\input ./style/arxiv-general.cfg
\documentclass[MSNbibl,number,citesort,dvips]{arxbj}
\makeatletter
   \@ifpackageloaded{graphicx}{}{\usepackage{graphicx}}
\makeatother
\usepackage{mathbh}
\usepackage{algorithm,algorithmic}
\usepackage{upgreek}
\usepackage{mathrsfs}

\volume{22}
\issue{2}
\pubyear{2016}
\firstpage{794}
\lastpage{856}
\doi{10.3150/14-BEJ676} % kopijuoti is laisko
\docsubty{FLA}

\makeatletter

\newcommand{\rrvert}{\vert}

\newcommand{\llvert}{\vert}
\newtheorem{corollary}{Corollary}
\newproclaim{defn}{Definition}
\newtheorem{theorem}{Theorem}
\newproclaim{cond}{Condition}
\newproclaim{res}{Result}
\newproclaim{prin}{Principle}
\newremark{remark}{Remark}
\makeatother

\begin{document}
\begin{frontmatter}

\title{On the exact and $\varepsilon$-strong simulation of (jump) diffusions}
\runtitle{Exact Simulation of Jump Diffusions}

\begin{aug}
%%%% inicialai - be tarpu
% Corresponding author: Murray Pollock - m.pollock@warwick.ac.uk% Updated by VTEXPTS2LaTeX.exe, 18.11.2014 08:31
%Updated by VTEXPTS2LaTeX.exe, 17.11.2014 15:53
\author[A]{\inits{M.}\fnms{Murray}~\snm{Pollock}\corref{}\thanksref{e1}\ead[label=e1,mark]{m.pollock@warwick.ac.uk}},
\author[A]{\inits{A.M.}\fnms{Adam M.}~\snm{Johansen}\thanksref{e2}\ead[label=e2,mark]{a.m.johansen@warwick.ac.uk}}
\and
\author[A]{\inits{G.O.}\fnms{Gareth O.}~\snm{Roberts}\thanksref{e3}\ead[label=e3,mark]{gareth.o.roberts@warwick.ac.uk}}
%%\runauthor{} %% auto
%\dedicated{}
\address[A]{Department of Statistics, University of Warwick, Coventry, CV4
7AL, United Kingdom.\\ \printead{e1}; \printead*{e2};\\ \printead*{e3}}
%\address[]{. \printead{}}
\end{aug}

% HISTORY:
\received{\smonth{7} \syear{2013}}
\revised{\smonth{4} \syear{2014}}

% ABSTRACT
%
\begin{abstract}
This paper introduces a framework for simulating finite dimensional
representations of (jump) diffusion sample paths over finite intervals,
without discretisation error (\textit{exactly}), in such a way that the
sample path can be restored at any finite collection of time points.
Within this framework we extend existing exact algorithms and introduce
novel adaptive approaches. We consider an application of the
methodology developed within this paper which allows the simulation of
upper and lower bounding processes which almost surely constrain (jump)
diffusion sample paths to any specified tolerance. We demonstrate the
efficacy of our approach by showing that with finite computation it is
possible to determine whether or not sample paths cross various
irregular barriers, simulate to any specified tolerance the first
hitting time of the irregular barrier and simulate killed diffusion
sample paths.
\end{abstract}

% KEYWORDS
% visi is mazosios raides ir pagal abecele
%
\end{frontmatter}

%s1 #&#
\section{Introduction}\label{sintroduction}

Diffusions and jump diffusions are widely used across a number of
application areas. An extensive literature exists in economics and
finance, spanning from the seminal Black--Scholes model \cite{JPEBS73,BJEMSM73,JFEM76} to the present \cite{JFEJP03,JFEBS04}. Other
applications can be easily found within the physical \cite{SJSPGD09}
and life sciences \cite{SCGW06,CSDAGW08} to name but a few. A jump
diffusion $V\dvtx  \mathbb{R} \to \mathbb{R}$ is a Markov process. In
this paper, we consider jump diffusions defined as the solution to a
stochastic differential equation (SDE) of the form (denoting $V_{t-} :=
\lim_{s \uparrow t }V_s$),
%
%e1 #&#
\begin{equation}\label{eqjumpdiffusion}
\mathrm{d}V_t  = \beta(V_{t-}) \,\mathrm{d}t +
\sigma(V_{t-}) \,\mathrm{d}W_t + \mathrm{d}J^{\lambda,\mu
}_t,\qquad
V_0 = v \in\mathbb{R}, t\in[0,T],
\end{equation}
where $\beta\dvtx  \mathbb{R} \to \mathbb{R}$ and $\sigma\dvtx
\mathbb{R} \to \mathbb{R}_+$ denote the (instantaneous) drift and diffusion
coefficients, respectively, $W_t$ is a standard Brownian motion and
$J^{\lambda,\mu}_t$ denotes a compound Poisson process.
$J^{\lambda,\mu}_t$ is parameterised with (finite) jump intensity $\lambda\dvtx  \mathbb{R} \to \mathbb{R}_+$ and jump size coefficient
$\mu\dvtx  \mathbb{R}
\to \mathbb{R}$ with jumps distributed with density $f_{\mu}$. All
coefficients are themselves (typically) dependent on $V_t$. Regularity
conditions are assumed to hold to ensure the existence of a unique
non-explosive weak solution (see, e.g., \cite{BKASCJD,BKNSSDEJF}).
To apply the methodology developed within this
paper we primarily restrict our attention to univariate diffusions and
require a number of additional conditions on the coefficients of
(\ref{eqjumpdiffusion}), details and a discussion of which can be found in
Section~\ref{spreliminaries}.

Motivated by the wide range of possible applications we are typically
interested (directly or indirectly) in the measure of $V$ on the path
space induced by (\ref{eqjumpdiffusion}), denoted $\mathbb{T}^v_{0,T}$.
As $\mathbb{T}^v_{0,T}$ is typically not explicitly
known, in order to compute expected values $\mathbb{E}_{\mathbb{T}^v_{0,T}} [h(V) ]$ for various test functions
$h \dvtx \mathbb{R} \to \mathbb{R}$, we can construct a Monte Carlo estimator. In
particular, if it is possible to draw independently $V^{(1)}, V^{(2)},
\ldots, V^{(n)} \sim\mathbb{T}^v_{0,T}$ then by applying the strong
law of large numbers we can construct a consistent estimator of the
expectation (unbiasedness follows directly by linearity),
%
%e2 #&#
\begin{equation}
\label{eqmcest}
\mbox{w.p. 1:}\qquad\lim_{n \to\infty} \frac{1}{n} \sum
^n_{i=1} h\bigl(V^{(i)}\bigr)  =
\mathbb{E}_{\mathbb{T}^v_{0,T}} \bigl[h(V) \bigr].
\end{equation}
Furthermore, provided $\mathbb{V}_{\mathbb{T}^v_{0,T}}
[h(V)
] =: \sigma^2_h < \infty$, by application of the central limit theorem
we have,
%
%e3 #&#
\begin{equation} \label{eqmcvar}
\lim_{n \to\infty} \sqrt{n} \Biggl[\mathbb{E}_{\mathbb{T}^v_{0,T}}
\bigl[h(V) \bigr] - \frac{1}{n} \sum^n_{i=1}
h\bigl(V^{(i)}\bigr) \Biggr]  \stackrel{\mathcal{D}}{=} \xi, \qquad \mbox{where }\xi\sim\mathrm{N}\bigl(0,\sigma^2_h\bigr).
\end{equation}
Unfortunately, as diffusion sample paths are infinite dimensional
random variables it isn't possible to draw an entire sample path from
$\mathbb{T}^v_{0,T}$ -- at best we can hope to simulate some finite
dimensional subset of the sample path, denoted ${V}^{\mathrm{fin}}$
(we further
denote the remainder of the sample path by
${V}^{\mathrm{rem}}:=V\setminus{V}^{\mathrm{fin}}$). Careful
consideration has to be taken as to how to simulate ${V}^{\mathrm{fin}}$ as any numerical approximation impacts the unbiasedness and
convergence of the resulting Monte Carlo estimator (\ref{eqmcest}), (\ref{eqmcvar}). Equally, consideration has to be given to the form of the
test function, $h$, to ensure it's possible to evaluate it given
${V}^{\mathrm{fin}}$.
%
%
%f1 #&#
\begin{figure}[t]
\begin{tabular}{@{}c@{\quad}c@{\quad}c@{}}

\includegraphics{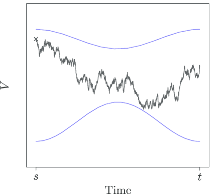}
 & \includegraphics{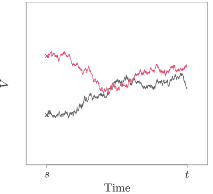}& \includegraphics{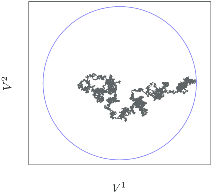}\\
{\fontsize{9}{11}\selectfont{(a) Non-linear two sided barrier}} &
{\fontsize{9}{11}\selectfont{(b) Diffusion barrier}} &
{\fontsize{9}{11}\selectfont{(c) 2-D circular barrier}}
\end{tabular}
\caption{Examples of test functions in which evaluation requires the
characterisation of an entire sample path.}\label{figbarrier}
\end{figure}

To illustrate this point we consider some possible applications. In
Figure~\ref{figbarrier}(a), (b) and (c) we are interested in whether a simulated sample path
$V\sim\mathbb{T}^v_{0,T}$, crosses some barrier (i.e., for some set $A$
we have $h(V) := \mathbh{1}(V\in A)$). Note that in all three cases in
order to evaluate $h$ we would require some characterisation of the
entire sample path (or some further approximation) and even for
diffusions with constant coefficients and simple barriers this is
difficult. For instance, as illustrated in Figure~\ref{figbarrier}(c),
even in the case where $\mathbb{T}^v_{0,T}$ is known (e.g., when $\mathbb{T}^v_{0,T}$ is
Wiener measure) and the barrier is known in advance and has a simple
form, there may still not exist any exact approach to evaluate whether
or not a sample path has crossed the barrier.

\begin{figure}%[b]

\includegraphics{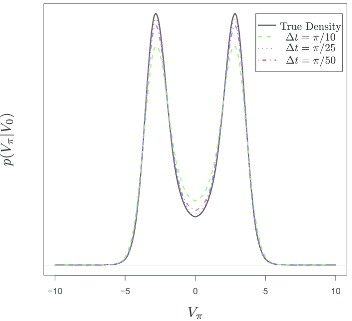}

\caption{Density of $V_\pi$ (obtained using 1\,000\,000 exact samples)
and approximations given by an Euler discretisation with various mesh
sizes, given $V_0 = 0$ where $\mathrm{d}V_t = \sin(V_t)\,
\mathrm{d}t +  \mathrm{d}W_t$.} \label{figeuler}
\end{figure}

Diffusion sample paths can be simulated approximately at a finite
collection of time points by \textit{discretisation} \cite{BKDP,BKNSSDE,BKNSSDEJF}, noting that as Brownian motion has a
Gaussian transition density then over short intervals the transition
density of (\ref{eqjumpdiffusion}) can be approximated by that of an
SDE with fixed coefficients (by a continuity argument). This can be
achieved by breaking the interval the sample path is to be simulated
over into a fine mesh (e.g., of size $\Delta t$), then
iteratively (at each mesh point) fixing the coefficients and simulating
the sample path to the next mesh point. For instance, in an \textit
{Euler} discretisation \cite{BKDP} of (\ref{eqjumpdiffusion}), the
sample path is propagated between mesh points as follows (where $\xi
\sim\mathrm{N}(0,\Delta t)$ and $\mu_t\sim f_{\mu}(\cdot;V_t)$),
%
%e4 #&#
\begin{equation}\label{eqeulerdiscretisation}
V_{t+\Delta t}  = \cases{V_t +
\beta (V_t )\Delta t + \sigma (V_t ) \xi, &$\quad \mbox{w.p. }
\exp \bigl\{-\lambda(V_t)\Delta t \bigr\}$,\vspace*{3pt}
\cr
V_t + \beta (V_t )\Delta t + \sigma (V_t
) \xi + \mu _t, & $\quad\mbox{w.p. } 1-\exp \bigl\{-\lambda(V_t)
\Delta t \bigr\}$.}
\end{equation}
It is hoped the simulated sample path (generated approximately at a
finite collection of mesh points) can be used as a proxy for an entire
sample path drawn exactly from $\mathbb{T}^v_{0,T}$. More complex
discretisation schemes exist (e.g., by exploiting It\^{o}'s lemma
to make higher order approximations or by local linearisation of the
coefficients \cite{BKNSSDE,BKNSSDEJF}), but all suffer from common
problems. In particular, minimising the approximation error (by
increasing the mesh density) comes at the expense of increased
computational cost, and further approximation or interpolation is
needed to obtain the sample path at non-mesh points (which can be
non-trivial). As illustrated in Figure~\ref{figeuler}, even when our test
function $h$ only requires the simulation of sample paths at a single
time point, discretisation introduces approximation error resulting in
the loss of unbiasedness of our Monte Carlo estimator (\ref{eqmcest}).
If $\mathbb{T}^v_{0,T}$ has a highly non-linear drift or includes a
compound jump process or $h$ requires simulation of sample paths at a
collection of time points then this problem is exacerbated. In the case
of the examples in Figure~\ref{figbarrier}, mesh based discretisation
schemes don't sufficiently characterise simulated sample paths for the
evaluation of $h$.

%
%
%f2 #&#

Recently, a new class of \textit{Exact Algorithms} for simulating
sample paths at finite collections of time points without approximation
error have been developed for both diffusions \cite{AAPBR05,BBPR06,MCAPBPR08,MORCH13} and jump diffusions
\cite{MCAPCR10,ORGS12,MCAPGR13}. These algorithms are based on rejection
sampling, noting that sample paths can be drawn from the (target)
measure $\mathbb{T}^v_{0,T}$ by instead drawing sample paths from an
equivalent proposal measure $\mathbb{P}^v_{0,T}$, and accepting or
rejecting them with probability proportional to the Radon--Nikod{\'y}m
derivative of $\mathbb{T}^v_{0,T}$ with respect to $\mathbb{P}^v_{0,T}$. However, as with
discretisation schemes, given a simulated sample path at a finite
collection of time points subsequent simulation of the sample path at
any other intermediate point may require approximation or interpolation
and may not be exact. Furthermore, we are again unable to evaluate test
functions of the type illustrated in Figure~\ref{figbarrier}.

The key contribution of this paper is the introduction of a novel
mathematical framework for constructing exact algorithms which
addresses this problem. In particular, instead of exactly simulating
sample paths at finite collections of time points, we focus on the
extended notion of simulating \textit{skeletons} which in addition
characterise the entire sample path.
% Skeleton
%

%de1 #&#
\begin{defn}[(Skeleton)] \label{defnskeleton}
A skeleton $(\mathcal{S})$ is a finite dimensional representation of a
diffusion sample path $(V\sim\mathbb{T}^v_{0,T})$, that can be
simulated without any approximation error by means of a proposal sample
path drawn from an equivalent proposal measure $(\mathbb{P}^v_{0,T})$
and accepted with probability proportional to $\frac{\mathrm{d}\mathbb{T}^v_{0,T}}{ \mathrm{d}\mathbb{P}^v_{0,T}}(V)$, which is
sufficient to
restore the sample path at any finite collection of time points exactly
with finite computation where $V|\mathcal{S}\sim \mathbb{P}^v_{0,T}|\mathcal{S}$. A skeleton typically comprises
information regarding the sample path at a finite collection of time
points and path space information which ensures the sample path is
almost surely constrained to some compact interval.
\end{defn}

Methodology for simulating skeletons (the size and structure
of which is dependent on exogenous randomness) is driven by both
computational and mathematical considerations (i.e., we need to ensure
the required computation is finite and the skeleton is exact). Central
to both notions is that the path space of the proposal measure
$\mathbb{P}^v_{0,T}$ can be partitioned (into a set of \textit{layers}), and
that the layer to which any sample path belongs to can be simulated.
% Layer
%

%de2 #&#
\begin{defn}[(Layer)] \label{defnlayer}
A layer $R(V)$, is a function of a diffusion sample path $V\sim
\mathbb{P}^v_{0,T}$ which determines the compact interval to which any
particular sample path $V(\omega)$ is constrained.
\end{defn}

To illustrate the concept of a layer and skeleton,
we could, for instance, have $R(V)=\inf\{i\in\mathbb{N}\dvt  \forall u\in
[0,T], V_u\in[v-i,v+i]\}$ and $\mathcal{S}=\{V_0=v,V_T=w,R(V)=1\}$.

We show that a valid exact algorithm can be constructed if it is
possible to partition the proposal path space into layers, simulate
unbiasedly to which layer a proposal sample path belongs and then,
conditional on that layer, simulate a skeleton. Our exact algorithm
framework for simulating skeletons is based on three principles for
choosing a proposal measure and simulating a path space layer:
% Principles
%

%pr1 #&#
\begin{prin}[(Layer construction)] \label{prinlayer}
The path space of the process of interest, can be partitioned and the
layer to which a proposal sample path belongs can be unbiasedly
simulated, $R(V)\sim\mathcal{R} := \mathbb{P}^v_{0,T} \circ R^{-1}$.
\end{prin}

%
%pr2 #&#
\begin{prin}[(Proposal exactness)] \label{prinprop}
Conditional on $V_0$, $V_T$ and $R(V)$, we can simulate any finite
collection of intermediate points of the trajectory of the proposal
diffusion exactly, $V\sim\mathbb{P}^v_{0,T}|_{R^{-1}(R(V))}$.
\end{prin}

Together Principles \ref{prinlayer} and \ref{prinprop}
ensure it is possible to simulate a skeleton. However, in addition we
want to characterise the entire sample path and so we construct exact
algorithms with the following additional principle.
%

%pr3 #&#
\begin{prin}[(Path restoration)] \label{prinrest}
Any finite collection of intermediate (inference) points, conditional
on the skeleton, can be simulated exactly, $V_{t_1},\ldots,V_{t_n}\sim
\mathbb{P}^v_{0,T}|\mathcal{S}$.
\end{prin}

In developing a methodological framework for simulating exact
skeletons of (jump) diffusion sample paths we are able to present a
number of significant extensions to the existing literature on exact
algorithms \cite{BBPR06,MCAPBPR08,MCAPCR10,MCAPGR13}. In
particular, we present novel exact algorithm constructions requiring
the simulation of fewer points of the proposal sample path in order to
evaluate whether to accept or reject (in effect a Rao--Blackwellisation
of EA3 for diffusions \cite{MCAPBPR08}, which we term the Unbounded
Exact Algorithm (UEA), and a Rao--Blackwellisation of the Jump Exact
Algorithms (JEAs) for jump diffusions \cite{MCAPCR10,MCAPGR13}, which
we term the Bounded Jump Exact Algorithm (BJEA) and Unbounded Jump
Exact Algorithm (UJEA), resp.) -- all of which we recommend are
adopted in practice instead of the existing equivalent exact algorithm.
Furthermore, we extend both existing and novel exact algorithms to
satisfy Principle~\ref{prinrest}, enabling the further simulation of a
proposed sample path \textit{after} it has been accepted, which
hitherto has not been possible except under the limiting conditions
imposed in EA1 \cite{BBPR06}.

Although in the context of a particular application we do not
necessarily require the ability to further simulate a proposal sample
path after acceptance (e.g., particle filtering for partially
observed jump diffusions \cite{PhDP13}, in which only the end point of
each sample path is required), to tackle the type of problem in which
we do require that Principle~\ref{prinrest} is satisfied (e.g., in
the examples outlined Figure~\ref{figbarrier}) we introduce a novel class
of \textit{Adaptive Exact Algorithms (AEAs)} for both diffusions and
jump diffusions (which are again in effect a Rao--Blackwellisation of
the Unbounded Exact Algorithm (UEA) requiring fewer points of the
proposal sample path in order
to evaluate whether to accept or reject).

By direct application of the methodology we develop for Adaptive Exact
Algorithms (AEA), we
significantly extend \textit{$\varepsilon$-Strong Simulation} methodology
\cite{BBPR12} (which allows the simulation of upper and lower bounding
processes which almost surely constrain stochastic process sample paths
to any specified tolerance), from Brownian motion sample paths to a
general class of jump diffusions, and introduce novel results to ensure
the methodology in \cite{BBPR12} can be implemented exactly. Finally,
we highlight a number of possible applications of the methodology
developed in this paper by returning to the examples introduced in
Figure~\ref{figbarrier}. We demonstrate that it is possible not only to
simulate skeletons exactly from the correct target measure but also to
evaluate exactly whether or not non-trivial barriers have been crossed
and so construct Monte Carlo estimators for computing barrier crossing
probabilities. It should be noted that there are a wide range of other
possible direct applications of the methodology in this paper, for
instance, the evaluation of path space integrals and barrier hitting
times to arbitrary precision, among many others.

In summary, the main contributions of this paper are as follows:
\begin{itemize}[--]
\item[--] A mathematical framework for constructing exact algorithms for
both diffusions and jump diffusions, enabling improvements to existing
exact algorithms, extension of existing exact algorithms to satisfy
Principle~\ref{prinrest} and a new class of adaptive exact algorithms (see
Sections~\ref{sea} and \ref{sjea}).
\item[--] Methodology for the $\varepsilon$-strong simulation of diffusion and
jump diffusion sample paths, along with a novel exact algorithm based
on this construction (see Sections~\ref{sepss} and \ref{snlbb}).
\item[--] New methodology for constructing Monte Carlo estimators to
compute irregular barrier crossing probabilities, simulating first
hitting times to any specified tolerance and simulating killed
diffusion sample path skeletons (see Sections~\ref{sepss} and \ref{sexampfinal}).
This work is reliant on the methodological extensions
of Sections~\ref{sea}--\ref{sepss} and is presented along with
examples based on the illustrations in Figure~\ref{figbarrier}.
\end{itemize}
We also make a number of other contributions which are necessary for
the implementation of our methodology. In particular, we detail how to
simulate unbiasedly events of probability corresponding to various
Brownian path space probabilities (see Section~\ref{sbpss}); and, we make
significant extensions to existing $\varepsilon$-strong methodology
enabling the initialisation of the algorithm and ensuring exactness
(see Sections~\ref{sepss} and \ref{snlbb}).

This paper is organised as follows: in Section~\ref{spreliminaries}, we
detail conditions sufficient to establish results necessary for
applying the methodology in this paper. In Sections~\ref{sea} and
\ref{sjea}, we outline our exact algorithm framework for diffusions and
jump diffusions, respectively, presenting the resulting
UEA, Bounded Jump Exact Algorithm (BJEA),
Unbounded Jump Exact Algorithm (UJEA) and
AEAs. In Section~\ref{sepss}, we apply our methodology
enabling the $\varepsilon$-strong simulation of (jump) diffusions. We
extend existing layered Brownian bridge constructions in Section~\ref{slbb},
introducing novel constructions for the AEAs in Section~\ref{snlbb} (both of which rely on novel Brownian path space simulation
results which are summarised in Section~\ref{sbpss}). Finally in
Section~\ref{sexampfinal}, we revisit the examples outlined in
Figure~\ref{figbarrier} to which we apply our methodology.

%s2 #&#
\section{Preliminaries}\label{spreliminaries}

In order to apply the methodology in this paper, we require conditions
in order to establish Results \ref{reslamp}--\ref{rescompact} which
we introduce below. To present our work in some generality we assume
Conditions \ref{condexis}--\ref{condphi} hold (see below) along with
some indication of why each is required. However, these conditions can
be difficult to check and so in Section~\ref{spraccond} we discuss
verifiable sufficient conditions under which Results \ref{reslamp}--\ref{rescompact} hold.
%

%co1 #&#
\begin{cond}[(Solutions)] \label{condexis}
The coefficients of (\ref{eqjumpdiffusion}) are sufficiently regular
to ensure the existence of a unique, non-explosive, weak solution.
\end{cond}
%

%co2 #&#
\begin{cond}[(Continuity)] \label{condcont}
The drift coefficient $\beta\in C^1$. The volatility coefficient
$\sigma\in C^2$ and is strictly positive.
\end{cond}

%
%co3 #&#
\begin{cond}[(Growth bound)] \label{condgrow}
We have that $\exists K>0$ such that $|\beta(x)|^2 + |\sigma(x)|^2
\leq K(1+|x|^2)$ $\forall x\in\mathbb{R}$.
\end{cond}
%

%co4 #&#
\begin{cond}[(Jump rate)] \label{condjump}
$\lambda$ is non-negative and locally bounded.
\end{cond}

Conditions \ref{condcont} and \ref{condgrow} are sufficient
to allow us to transform our SDE in (\ref{eqjumpdiffusion}) into one
with unit volatility,
%

%re1 #&#
\begin{res}[(Lamperti transform \protect\cite{BKNSSDE}, Chapter~4.4)]
\label{reslamp}
Let $\eta(V_t) =: X_t$ be a transformed process, where $\eta(V_t) :=
\int^{V_t}_{v^*} 1/\sigma(u) \,\mathrm{d}u$ (where $v^*$ is an arbitrary
element in the state space of $V$). We denote by $\psi_1,\ldots,\psi
_{N_T}$ as the jump times in the interval $[0,T]$, $\psi_0:=0$, $\psi
_{N_T+1}-:=\psi_{N_T+1}:=T$ and $N_t := \sum_{i \geq1} \mathbh{1}\{
\psi_i \leq t\}$ a Poisson jump counting process. Further denoting by
${V}^{\mathrm{cts}}$ the continuous component of $V$ and applying It\^{o}'s formula
for jump diffusions to find $ \mathrm{d}X_t$ we have (where $\mu
_t\sim f_{\mu
}(\cdot;V_{t-}) = f_{\mu}(\cdot;\eta^{-1}(X_{t-}))$),
%
%e5 #&#
\begin{eqnarray}
\mathrm{d}X_t & =& \bigl[\eta'\,\mathrm{d}
{V}^{\mathrm{cts}}_t + \eta'' \bigl(\mathrm{d} {V}^{\mathrm{cts}}_t \bigr)^2/2 \bigr] +
\bigl(\eta [V_{t-}+ \mu_t ] - \eta (V_{t-} )
\bigr)\,\mathrm{d}N_t
\nonumber
\\[-8pt]
\label{eqdXt}
\\[-8pt]
\nonumber
& = & \underbrace{ \biggl[\frac{\beta (\eta^{-1}
(X_{t-}
) )}{\sigma (\eta^{-1}  (X_{t-} ) )} - \frac
{\sigma' (\eta^{-1}  (X_{t-} ) )}{2}
\biggr]}_{\alpha
 (X_{t-} )} \,\mathrm{d}t + \mathrm{d}W_t +
 \underbrace{\bigl(\eta \bigl[\eta^{-1} (X_{t-} ) + \mu_t
\bigr] - X_{t-} \bigr)\, \mathrm{d} N_t}_{ \mathrm{d}J^{\lambda,\nu}_t}.
\end{eqnarray}
\end{res}

This transformation is typically possible for univariate
diffusions and for many multivariate diffusions \cite{ASA08}. A
significant class of multivariate diffusions can be simulated by direct
application of our methodology, and ongoing work is aimed at extending
these methodologies more broadly to multivariate diffusions (see
\cite{SJSSPRF12}).

In the remainder of this paper, we assume (without loss of generality)
that we are dealing with transformed SDEs with unit volatility
coefficient as in (\ref{eqdXt}). As such, we introduce the following
simplifying notation. In particular, we denote by $\mathbb{Q}^x_{0,T}$
the measure induced by (\ref{eqdXt}) and by $\mathbb{W}^x_{0,T}$ the
measure induced by the driftless version of (\ref{eqdXt}). We define
$A(u) := \int^u_0 \alpha(y)\,\mathrm{d}y$ and set $\phi(X_s) :=
\alpha^2(X_s)/2
+ \alpha'(X_s)/2$. If $\lambda=0$ in (\ref{eqdXt}) then $\mathbb{W}^x_{0,T}$ is Wiener measure. Furthermore, we impose the following
final condition.
%

%co5 #&#
\begin{cond}[($\Phi$)] \label{condphi}
There exists a constant $\Phi> -\infty$ such that $\Phi\leq\phi$.
\end{cond}

It is necessary for this paper to establish that the
Radon--Nikod{\'y}m derivative
of $\mathbb{Q}^x_{0,T}$ with respect to $\mathbb{W}^x_{0,T}$ exists
(Result~\ref{resradon}) and can be bounded on compact sets (Results
\ref{resquad} and \ref{rescompact}) under Conditions \ref{condexis}--\ref{condphi}.

%re2 #&#
\begin{res}[(Radon--Nikod{\'y}m derivative \cite{BKASCJD,BKNSSDEJF})]
\label{resradon}
Under Conditions \ref{condexis}--\ref{condjump}, the Radon--Nikod{\'y}m derivative of
$\mathbb{Q}^x_{0,T}$ with respect to $\mathbb{W}^x_{0,T}$ exists and
is given by Girsanov's formula:
%
%e6 #&#
\begin{equation}
\frac{ \mathrm{d}\mathbb{Q}^x_{0,T}}{ \mathrm{d}\mathbb{W}^x_{0,T}}(X) = \exp \biggl\{\int^{T}_{0}
\alpha (X_{s} )\,\mathrm {d}W_s - \frac
{1}{2}\int
^T_0\alpha^2(X_s)\,
\mathrm{d}s \biggr\}.
\end{equation}
As a consequence of Condition~\ref{condcont}, we have $A \in C^2$ and
so we
can apply It\^{o}'s formula to remove the stochastic integral,
%
%e7 #&#
\begin{equation}\label{eqjumprnd}
\frac{\mathrm{d}\mathbb{Q}^x_{0,T}}{ \mathrm{d}\mathbb{W}^x_{0,T}}(X) = \exp \Biggl\{A(X_T)-A(x)-\int
^T_0 \phi(X_s)\,\mathrm{d}s -
\sum^{N_T}_{i=1} \bigl[A(X_{\psi
_i})-A(X_{\psi_{i}-})
\bigr] \Biggr\}.
\end{equation}
In the particular case where we have a diffusion $(\lambda=0)$,
%
%e8 #&#
\begin{equation}\label{eqrnd}
\frac{ \mathrm{d}\mathbb{Q}^x_{0,T}}{ \mathrm{d}\mathbb{W}^x_{0,T}}(X) = \exp \biggl\{A(X_T)-A(x)-\int
^T_0 \phi(X_s)\,\mathrm{d}s
\biggr\}.
\end{equation}
\end{res}
%

%re3 #&#
\begin{res}[(Quadratic growth)]\label{resquad}
As a consequence of Condition~\ref{condgrow}, we have that $A$ has a
quadratic growth bound and so there exists some $T_0<\infty$ such that
$\forall T \leq T_0$:
%
%e9 #&#
\begin{equation}
c(x,T) := \int_{\mathbb{R}}\exp \biggl\{A(y) - \frac
{(y-x)^2}{2T}
\biggr\}\,\mathrm{d}y < \infty.
\end{equation}
\end{res}

Throughout this paper, we rely on the fact that upon
simulating a path space layer (see Definition~\ref{defnlayer}) then
$\forall
s \in[0,T]$ $\phi(X_s)$ is bounded, however this follows directly from
the following result.
%

%re4 #&#
\begin{res}[(Local boundedness)] \label{rescompact}
By Condition~\ref{condcont}, $\alpha$ and $\alpha'$ are bounded on compact
sets. In particular, suppose $\exists \ell, \upsilon\in\mathbb{R}$
such that $\forall t\in[0,T]$, $X_t(\omega)\in[\ell,\upsilon]$
$\exists L_X:=L (X(\omega) )\in\mathbb{R},U_X:=U
(X(\omega) )\in\mathbb{R}$ such that $\forall  t\in[0,T]$,
$\phi
 (X_t(\omega) )\in[L_X,U_X]$.
\end{res}

%s2.1 #&#
\subsection{Verifiable Sufficient Conditions} \label{spraccond}
As discussed in \cite{BKASCJD}, Theorem~1.19, to ensure Condition~\ref{condexis} it is
sufficient to assume that the coefficients of (\ref{eqjumpdiffusion}) satisfy the following linear growth and Lipschitz
continuity conditions for some constants $C_1,C_2<\infty$ (recalling
that $f_{\mu}$ is the density of the jump sizes),
%
%e10 #&#
%e11 #&#
\begin{eqnarray}\label{equlin}
&&\bigl|\beta(x)\bigr|^2 + \bigl|\sigma(x)\bigr|^2 + \int
_\mathbb{R} \bigl|f_\mu(z;x)\bigr|^2 \lambda (
\mathrm{d}z)  \leq   C_1\bigl(1+|x|^2\bigr), \qquad\forall x\in
\mathbb{R},
\\
&& \bigl|\beta(x)-\beta(y)\bigr|^2 + \bigl|\sigma(x)-\sigma(y)\bigr|^2
 + \int_\mathbb{R} \bigl|f_\mu(z;x)-f_\mu(z;y)\bigr|^2
\lambda( \mathrm{d}z)
\nonumber
\\[-8pt]
\label{equlip}
\\[-8pt]
\nonumber
&&\quad\leq C_2|x-y|^2,\qquad \forall x,y\in
\mathbb{R}.
\end{eqnarray}
(\ref{equlin}) and (\ref{equlip}) together with Condition~\ref{condcont}
are sufficient for the purposes of implementing the methodology in this
paper (in particular, Conditions \ref{condexis}, \ref{condgrow},
\ref{condjump} and \ref{condphi} will hold in this situation) but are not
necessary. Although easy to verify, (\ref{equlin}) and (\ref{equlip})
are somewhat stronger than necessary for our purposes and so we impose
Condition~\ref{condexis} instead.

It is of interest to note that if we have a diffusion (i.e., in (\ref{eqjumpdiffusion}) we have $\lambda=0$), then, by application of the
Mean Value Theorem, Condition~\ref{condcont} ensures $\beta$ and
$\sigma$ are
locally Lipschitz and so (\ref{eqjumpdiffusion}) admits a unique weak
solution \cite{BKSDE} and so Condition~\ref{condexis} holds. In particular,
the methodology in this paper will hold under Conditions \ref{condcont}, \ref{condgrow} and \ref{condphi}.

%s3 #&#
\section{Exact Simulation of Diffusions} \label{sea}
In this section, we outline how to simulate skeletons for diffusion
sample paths (we will return to jump diffusions in Section~\ref{sjea})
which can be represented (under the conditions in Section~\ref{spreliminaries} and following the transformation in (\ref{eqdXt})),
as the solution to SDEs with unit volatility,
%
%e12 #&#
\begin{equation}\label{eqdiffusion}
\mathrm{d}X_t  = \alpha(X_t) \,\mathrm{d}t +
\mathrm{d}W_t,\qquad  X_0 = x \in \mathbb{R}, t\in[0,T].
\end{equation}
As discussed in Section~\ref{sintroduction}, exact algorithms are a class
of rejection samplers operating on diffusion path space. In this
section, we begin by reviewing rejection sampling and outline an
(idealised) rejection sampler originally proposed in \cite{AAPBR05}
for simulating entire diffusion sample paths. However, for
computational reasons this idealised rejection sampler can't be
implemented so instead, with the aid of new results and algorithmic
step reordering, we address this issue and construct a rejection
sampler for simulating sample path skeletons which only requires finite
computation. A number of existing exact algorithms exist based on this
approach \cite{AAPBR05,BBPR06,MCAPBPR08}, however, in this paper we
present two novel algorithmic interpretations of this rejection
sampler. In Section~\ref{sbuea}, we present the Unbounded Exact
Algorithm (UEA), which is a
methodological extension of existing exact algorithms \cite{MCAPBPR08}, requiring less of the proposed sample path to be
simulated in order to evaluate acceptance or rejection. Finally, in
Section~\ref{sauea} we introduce the novel Adaptive Unbounded Exact
Algorithm (AUEA), which as noted in the
\hyperref[sintroduction]{Introduction}, is well suited to problems in which further simulation of
a proposed sample path \textit{after} it has been accepted is
required.

\textit{Rejection sampling} \cite{BKMCSM} is a standard Monte Carlo
method in which we can sample from some (inaccessible) target
distribution $\pi$ by means of an accessible dominating distribution
$g$ with respect to which $\pi$ is absolutely continuous with bounded
Radon--Nikod{\'y}m derivative. In particular, if we can find a bound
$M$ such that $\sup_x
\frac{ \mathrm{d}\pi}{ \mathrm{d}g}(x) \leq M < \infty$, then
drawing $X\sim g$ and
accepting the draw (setting $I=1$) with probability $P_g(X):=\frac
{1}{M}\frac{ \mathrm{d}\pi}{ \mathrm{d}g}(X)\in[0,1]$ then
\mbox{$(X|I=1)\sim\pi$}.

Similarly, we could simulate sample paths from our target measure (the
measure induced by~(\ref{eqdiffusion}) and denoted $\mathbb{Q}^x_{0,T}$) by means of a proposal measure which we can simulate
proposal sample paths from, provided a bound for the Radon--Nikod{\'y}m
derivative can be
found. A natural equivalent measure to choose as a proposal is Wiener
measure ($\mathbb{W}^x_{0,T}$, as (\ref{eqdiffusion}) has unit
volatility). In particular, drawing $X\sim\mathbb{W}^x_{0,T}$ and
accepting the sample path $(I=1)$ with probability $P_{\mathbb{W}^x_{0,T}}(X) := \frac{1}{M}\frac{ \mathrm{d}
\mathbb{Q}^x_{0,T}}{ \mathrm{d}
\mathbb{W}^x_{0,T}}(X)\in[0,1]$ (where $\frac{ \mathrm{d}\mathbb{Q}^x_{0,T}}{\mathrm{d}\mathbb{W}^x_{0,T}}(X)$ is as given in
(\ref{eqrnd})) then $(X|I=1)\sim\mathbb{Q}^x_{0,T}$. On average sample
paths would be accepted with probability $P_{\mathbb{W}^x_{0,T}} :=
\mathbb{E}_{\mathbb{W}^x_{0,T}}[P_{\mathbb{W}^x_{0,T}}(X)]$.
However, the function $A(X_T)$ in (\ref{eqrnd}) only has a quadratic
growth bound (see Result~\ref{resquad}), so typically no appropriate bound
($M<\infty$) exists.

To remove the unbounded function $A(X_T)$ from the acceptance
probability, one can use Biased Brownian motion (BBM) \cite{AAPBR05}
as the proposal measure and consider the resulting modification to the
acceptance probability.
%

%de3 #&#
\begin{defn}\label{dfnbbm}
Biased Brownian motion is the process $Z_t  \stackrel{\mathcal{D}}{=} (W_t|W_0=x, W_T := y \sim  h )$ with
measure $\mathbb{Z}^x_{0,T}$, where $x,y\in\mathbb{R}$, $t\in[0,T]$
and $h$ is defined as (by Result~\ref{resquad} we have $\forall T
\leq
T_0$, $h(y;x,T)$ is integrable),
%
%e13 #&#
\begin{equation}
h(y;x,T) := \frac{1}{c(x,T)}\exp \biggl\{A(y) -
\frac{(y-x)^2}{2T} \biggr\}.
\end{equation}
\end{defn}

%
%th1 #&#
\begin{theorem}[(Biased Brownian motion \protect\cite{AAPBR05}, Proposition~3)]\label{thmbbm}
$\mathbb{Q}^x_{0,T}$ is equivalent to $\mathbb{Z}^x_{0,T}$
with Radon--Nikod{\'y}m derivative:
%
%e14 #&#
\begin{equation}\label{eqbdrnd}
\frac{ \mathrm{d}\mathbb{Q}^x_{0,T}}{ \mathrm{d}\mathbb{Z}^x_{0,T}}(X) \propto \exp \biggl\{-\int^T_0
\phi(X_s)\, \mathrm{d}s \biggr\}.
\end{equation}
\end{theorem}

Sample paths can be drawn from $\mathbb{Z}^x_{0,T}$ in two
steps by first simulating the end point $X_T=:y\sim h$ (although $h$
doesn't have a tractable form, a rejection sampler with Gaussian
proposal can typically be constructed) and then simulating the
remainder of the sample path in $(0,T)$ from the law of a Brownian
bridge, $(X|X_0=x,X_T=y)\sim\mathbb{W}^{x,y}_{0,T}$. We can now
construct an (idealised) rejection sampler to draw sample paths from
$\mathbb{Q}^x_{0,T}$ as outlined in Algorithm~\ref{algidrsI},
noting that as
$\inf_{u\in[0,T]}\phi(X_u)\geq\Phi$ (see Condition~\ref{condphi}) we can
identify $\Phi$ and choose $M:=\exp\{-\Phi T\}$ to ensure
$P_{\mathbb{Z}^x_{0,T}}(X)\in[0,1]$.
\begin{algorithm}[t]
\caption{Idealised Rejection Sampler \cite{AAPBR05}} \label{algidrsI}
\begin{enumerate}
\item Simulate $X \sim\mathbb{Z}^x_{0,T}$, \label{algidrsIstart}
\begin{enumerate}[(a)]
\item[(a)] Simulate $X_T=:y\sim h$.
\item[(b)] Simulate $X_{(0,T)}\sim\mathbb{W}^{x,y}_{0,T}$. \label{stepirdsIinf}
\end{enumerate}
\item With probability ${{P}}_{\mathbb{Z}^x_{0,T}}(X) = \exp \{
-\int^T_0 \phi(X_s)\,\mathrm{d}s \}\cdot\exp\{\Phi T\}$
accept, else
reject and return to Step~1. \label{stepirdsIint}
\end{enumerate}
\end{algorithm}

\begin{algorithm}[b]
\caption{Implementable Exact Algorithm \protect\cite{AAPBR05,BBPR06}} \label{algea}
\begin{enumerate}
\item Simulate $X_T=:y\sim h$. \label{algeastart}
\item Simulate $F\sim\mathbb{F}$. \label{algeavar}
\item Simulate ${X}^{\mathrm{fin}}\sim\mathbb{W}^{x,y}_{0,T}|F$. \label{algeacond}
\item With probability ${{P}}_{\mathbb{Z}^x_{0,T}| F} (X )$
accept, else reject and return to Step~1. \label{algeaprob}\\[-3pt]
%\vspace{0.25cm}
\hrule
\vspace{0.15cm}
\item\textit{${}^*$Simulate ${X}^{\mathrm{rem}}\sim\mathbb{W}^{x,y}_{0,T}|({X}^{\mathrm{fin}},F) $.} \label{algeainf}
\end{enumerate}
\end{algorithm}

Unfortunately, Algorithm~\ref{algidrsI} can't be implemented directly
as it
isn't possible to draw entire sample paths from $\mathbb{W}^{x,y}_{0,T}$ in Step~1(b) (they're infinite
dimensional random variables) and it isn't possible to evaluate the
integral expression in the acceptance probability in Step~2.

The key to constructing an implementable algorithm (which requires only
finite computation), is to note that by first simulating some finite
dimensional auxiliary random variable $F\sim\mathbb{F}$ (the details
of which are in Sections~\ref{sbuea} and \ref{sauea}), an unbiased
estimator of the acceptance probability can be constructed which can be
evaluated using only a finite dimensional subset of the proposal sample
path. As such, we can use the simulation of $F$ to inform us as to what
finite dimensional subset of the proposal sample path to simulate
(${X}^{\mathrm{fin}}\sim\mathbb{W}^{x,y}_{0,T}|F$) in Step~1(b) in
order to evaluate the acceptance probability. The rest of the sample
path can be simulated as required \textit{after} the acceptance of the
sample path from the proposal measure conditional on the simulations
performed,
${X}^{\mathrm{rem}}\sim\mathbb{W}^{x,y}_{0,T}|({X}^{\mathrm
{fin}},F)$ (noting
that $X={X}^{\mathrm{fin}}\cup {X}^{\mathrm{rem}}$).
The synthesis of this argument is
presented in Algorithm~\ref{algea}.

Note that Algorithm~\ref{algea} Step~5 is separated
from the rest of
the algorithm and asterisked. This convention is used within this paper
to indicate the final step within an exact algorithm, which cannot be
conducted in its entirety as it involves simulating an infinite
dimensional random variable. However, as noted in the introductory
remarks to this section, our objective is to simulate a finite
dimensional sample path skeleton, with which we can simulate the
accepted sample path at any other finite collection of time points
without error. This final step simply indicates how this subsequent
simulation may be conducted.

In conclusion, although it isn't possible to simulate entire sample
paths from $\mathbb{Q}^x_{0,T}$, it is possible to simulate exactly a
finite dimensional subset of the sample path, characterised by its
\textit{skeleton} $\mathcal{S}(X) :=\{X_0,{X}^{\mathrm{fin}},X_T,F\}$. Careful
consideration has to be taken to construct $\mathbb{F}$ which existing
exact algorithms \cite{AAPBR05,BBPR06,MCAPBPR08} achieve by applying
Principles \ref{prinlayer} and \ref{prinprop}. However, no existing
exact algorithm addresses how to construct $\mathbb{F}$ under the
conditions in Section~\ref{spreliminaries} to additionally perform
Algorithm~\ref{algea} Step~5. We address this in
Sections~\ref{sbuea} and \ref{sauea}.

In the next two sections, we present two distinct, novel
interpretations of Algorithm~\ref{algea}. In Section~\ref{sbuea},
we present the UEA which is a methodological extension
of existing exact algorithms
and direct interpretation of Algorithm~\ref{algea}. In Section~\ref{sauea}, we
introduce the AUEA which takes a
recursive approach to Algorithm~\ref{algea} Steps 2,
3 and 4.

%s3.1 #&#
\subsection{Bounded and
Unbounded Exact Algorithms} \label{sbuea}

In this section, we present the Unbounded Exact Algorithm (UEA) along
with the Bounded Exact Algorithm (BEA) (which can
viewed as a special case of the UEA) by
revisiting Algorithm~\ref{algea}
and considering how to construct a suitable finite dimensional random
variable $F\sim\mathbb{F}$.

As first noted in \cite{BBPR06}, it is possible to construct and
simulate the random variable $F$ required in Algorithm~\ref{algea}, provided
$\phi(X_{[0,T]})$ can be bounded above and below. It was further noted
in \cite{MCAPBPR08} that if a Brownian bridge proposal sample path was
simulated along with an interval in which is was contained, and that
conditional on this interval $\phi(X_{[0,T]})$ was bounded above and
below, then $F$ could similarly be constructed and simulated. Finding a
suitable set of information that establishes an interval in which $\phi
(X_{[0,T]})$ is contained (by means of finding and mapping an interval
in which the sample path $X_{[0,T]}$ is contained), is the primary
motivation behind the notion of a sample path layer (see
Definition~\ref{defnlayer}). In this paper, we discuss more than one layer
construction (see Sections~\ref{slbb} and \ref{snlbb}), both of which
complicate the key ideas behind the UEA
and so layers are only
discussed in abstract terms at this stage.

Further to \cite{MCAPBPR08}, we note that $\phi(X_{[0,T]})$ is bounded
on compact sets (see Result~\ref{rescompact}) and so if, after simulating
the end point from Biased Brownian motion (BBM), we partition the path
space of $\mathbb{Z}^{x}|X_T$ into disjoint layers and simulate the
layer to which our proposal sample path belongs (see Principle~\ref
{prinlayer}, denoting $R:=R(X)\sim\mathcal{R}$ as the simulated layer
the precise details of which are given in Section~\ref{slbb}), then an
upper and lower bound for $\phi(X_{[0,T]})$ can always be found
conditional on this layer ($U_X\in\mathbb{R}$ and $L_X\in\mathbb{R}$,
resp.). As such, we have for all test functions $H\in\mathcal{C}_{\mathrm{b}}$,
%
%e15 #&#
\begin{eqnarray}
\mathbb{E}_{\mathbb{Z}^x_{0,T}} \bigl[P_{\mathbb{Z}^x_{0,T}}(X)\cdot H(X) \bigr] &=&
\mathbb{E}_{h}\mathbb{E}_{\mathbb{W}^{x,y}_{0,T}} \bigl[{{P}}_{\mathbb{Z}^x_{0,T}}(X)
\cdot H(X) \bigr]
\nonumber
\\[-8pt]
\\[-8pt]
\nonumber
& =& \mathbb{E}_{h}\mathbb{E}_\mathcal{R}
\mathbb{E}_{\mathbb{W}^{x,y}_{0,T}|R } \bigl[{{P}}_{\mathbb{Z}^x_{0,T}}(X)\cdot H(X) \bigr],
\end{eqnarray}
recalling that,
%
%e16 #&#
\begin{equation}\label{eqzaccprob}
P_{\mathbb{Z}^x_{0,T}}(X) = \exp \biggl\{-\int^T_0
\phi(X_s)\, \mathrm{d} s \biggr\}\cdot \mathrm{e}^{\Phi T}.
\end{equation}
Proceeding in a similar manner to \cite{JRSSBBPRF06} to construct our
finite dimensional estimator we consider a Taylor series expansion of
the exponential function in (\ref{eqzaccprob}),
%
%e17 #&#
\begin{eqnarray}
P_{\mathbb{Z}^x_{0,T}}(X) &=& \mathrm{e}^{-(L_X-\Phi)T}\cdot \mathrm{e}^{-(U_X-L_X)T}\exp \biggl\{
\int^T_0 U_X - \phi
(X_s)\,\mathrm{d}s \biggr\}
\nonumber
\\[-8pt]
\\[-8pt]
\nonumber
& =&  \mathrm{e}^{-(L_X-\Phi)T}\cdot \Biggl[\sum^\infty_{j=0}
\frac
{\mathrm{e}^{-(U_X-L_X)T} [(U_X - L_X) T ]^j}{j!} \biggl\{\int^T_0
\frac
{U_X - \phi(X_s)}{(U_X - L_X) T} \,\mathrm{d}s \biggr\}^j \Biggr],\hspace*{8pt}\quad
\end{eqnarray}
again employing methods found in \cite{JRSSBBPRF06}, we note that if
we let $\mathbb{K}_R$ be the law of $\kappa \sim \operatorname{Poi}((U_X-L_X)
T)$, $\mathbb{U}\kappa$ the distribution of $(\xi_1,\ldots,\xi
_\kappa
)\stackrel{\mathrm{iid}}{\sim} \mathrm{U}[0,T]$ we have,
%
%e18 #&#
\begin{eqnarray}
P_{\mathbb{Z}^x_{0,T}}(X) & =&  \mathrm{e}^{-(L_X-\Phi)T}\cdot\mathbb{E}_{\mathbb{K}_{R}}
\biggl[ \biggl(\int^T_0 \frac{U_X - \phi(X_s)}{(U_X - L_X)T} \,
\mathrm{d}s \biggr)^\kappa \Big| X \biggr]
\nonumber
\\[-8pt]
\label{eqbpe}
\\[-8pt]
\nonumber
& =& \mathrm{e}^{-(L_X-\Phi)T}\cdot\mathbb{E}_{\mathbb{K}_{R}} \Biggl[\mathbb{E}_{\mathbb{U}\kappa} \Biggl[\prod^\kappa_{i=1}
\biggl(\frac{U_X
- \phi
(X_{\xi_i})}{U_X - L_X} \biggr) \Big| X \Biggr]\Big| X \Biggr].
\end{eqnarray}
The key observation to make from (\ref{eqbpe}) is that the acceptance
probability of a sample path $X\sim\mathbb{Z}^{x}_{0,T}$ can be
evaluated without using the entire sample path, and can instead be
evaluated using a finite dimensional realisation, $X^{\mathrm{fin}}$. Simulating
a finite dimensional proposal as suggested by (\ref{eqbpe}) and
incorporating it within Algorithm~\ref{algea} results directly in the
UEA
presented in Algorithm~\ref{alguea}. A number of alternate methods for
simulating unbiasedly layer information (Step~2),
layered Brownian bridges (Step~4), and the sample path
at further times after acceptance (Step~6), are given in
Section~\ref{slbb}.
\begin{algorithm}[b]
\caption{Unbounded Exact Algorithm (UEA)} \label{alguea}
\begin{enumerate}[5.]
\item[1.] Simulate skeleton end point $X_T=:y \sim h$. \label{algueastart}
\item[2.] Simulate layer information $R\sim\mathcal{R}$. \label{alguealayer}
\item[3.] With probability $ (1-\exp \{-(L_X-\Phi)T \}
 )$
reject path and return to Step~1. \label{algueapre}
\item[4.] Simulate\vspace*{-2pt} skeleton points $ (X_{\xi_1},\ldots,X_{\xi
_\kappa} )|R $, \label{alguealbb}
\begin{enumerate}[(a)]
\item[(a)] Simulate\vspace*{1pt} $\kappa\sim\operatorname{Poi} ((U_X-L_X) T )$ and
skeleton times $\xi_1,\ldots,\xi_\kappa\stackrel{\mathrm{iid}}{\sim} \mathrm{U}
[0,T]$. \label{algueapoisskel}
\item[(b)] Simulate sample path at skeleton times $X_{\xi_1},\ldots,X_{\xi
_\kappa}\sim\mathbb{W}^{x,y}_{0,T}|R $.
\end{enumerate}
\item[5.] With probability $\prod^\kappa_{i=1} [ (U_X-\phi
(X_{\xi
_i}) )/ (U_X-L_X ) ]$, accept entire path, else
reject and return to Step~1. \label{algueaacpr}\\[-5pt]
%\vspace{0.15cm}
\hrule\vspace{0.15cm}
\item[6.] \textit{${}^*$Simulate ${X}^{\mathrm{rem}}\sim  (\bigotimes^{\kappa
+1}_{i=1} \mathbb{W}^{X_{\xi_{i-1}},X_{\xi_i}}_{\xi_{i-1},\xi
_i}
)|R $}. \label{algueainf}
\end{enumerate}
\end{algorithm}

The UEA can be viewed as a multi-step
rejection sampler in which the
acceptance probability is broken into a computational inexpensive step
(Step~3), and a computationally expensive step
(Step~5) which to evaluate requires partial
simulation of the
proposal sample path (Step~4). Unlike existing exact
algorithms (EA3 in \cite{MCAPBPR08}), the UEA\ conducts early
rejection to avoid any further unnecessary simulation of the rejected
sample path. In particular, the UEA
requires simulation of fewer
points of the proposal sample path in order to evaluate whether to
accept or reject.

The skeleton of an accepted sample path includes any information
simulated for the purpose of evaluating the acceptance probability (any
subsequent simulation must be consistent with the skeleton). As such,
the skeleton is composed of terminal points, skeletal points ($X_{\xi
_1},\ldots,X_{\xi_\kappa}$) and layer $R$ (denoting $\xi_0:=0$ and
$\xi_{\kappa+1}:=T$),
%
%e19 #&#
\begin{equation}\label{dfnueas}
\mathcal{S}_{\mathrm{UEA}} (X ):= \bigl\{ (\xi _i,X_{\xi_i}
)^{\kappa+1}_{i=0}, R \bigr\}.
\end{equation}
After simulating an accepted sample path skeleton, we may want to
simulate the sample path at further intermediate points. In the
particular case in which $\phi(X_{[0,T]})$ is almost surely bounded,
there is no need to simulate layer information in Algorithm~\ref{alguea}, the
skeleton can be simulated from the law of a Brownian bridge and given
the skeleton we can simulate further intermediate points of the sample
path from the law of a Brownian bridge (so we satisfy Principle~\ref{prinrest}). This leads to the \textit{Exact Algorithm}~1 (EA1)
proposed in \cite{BBPR06}, which we term the BEA,
%
%e20 #&#
\begin{equation}\label{dfnbeas}
\mathcal{S}_{\mathrm{BEA}} (X ):= \bigl\{ (\xi _i,X_{\xi
_i}
)^{\kappa+1}_{i=0} \bigr\}.
\end{equation}
A second exact algorithm (EA2) was also proposed in \cite{BBPR06} (the
details of which we omit from this paper), in which simulating the
sample path at further intermediate points after accepting the sample
path skeleton was possible by simulating from the law of two
independent Bessel bridges. However, EA1 (BEA) and EA2 both have very
limited applicability and are the only existing exact algorithms which
directly satisfy Principle~\ref{prinrest}.

Unlike existing exact algorithms \cite{AAPBR05,BBPR06,MCAPBPR08},
after accepting a sample path skeleton using the UEA, it is possible
to simulate the sample path at further finite collections of time
points without approximation under the full generality of the
conditions outlined in Section~\ref{spreliminaries} (so satisfying
Principle~\ref{prinrest}). Algorithm~\ref{alguea} Step~6 can't be
conducted in
existing exact algorithms as the layer imparts information across the
entire interval. However, in Section~\ref{slbb} we show that
Step~6 is possible (with additional computation), by augmenting
the skeleton with sub-interval layer information (denoting
$R^{[a,b]}_X$ as the layer for the sub-interval $[a,b] \subseteq[0,T]$),
%
%e21 #&#
\begin{eqnarray}
\mathcal{S}'_{\mathrm{UEA}} (X ) & :=& \underbrace{ \bigl\{ (
\xi_i,X_{\xi_i} )^{\kappa+1}_{i=0}, R,
\bigl(R^{[\xi_{i-1},\xi_i]}_X \bigr)^{\kappa+1}_{i=1} \bigr\}}_{A}
\nonumber
\\[-8pt]
\label{extendedskeleton}
\\[-8pt]
\nonumber
& \equiv& \underbrace{ \bigl\{ (\xi_i,X_{\xi_i}
)^{\kappa
+1}_{i=0}, \bigl(R^{[\xi_{i-1},\xi_i]}_X
\bigr)^{\kappa
+1}_{i=1} \bigr\} }_{B}.
\end{eqnarray}
The augmented skeleton allows the sample path to be decomposed into
conditionally independent paths between each of the skeletal points and
so the layer $R$ no longer imparts information across the entire
interval $[0,T]$. In particular, the sets in (\ref{extendedskeleton})
are equivalent in the sense that $\mathbb{W}^{x,y}_{0,T}|A = \mathbb{W}^{x,y}_{0,T}|B$.
As such, simulating the sample path at further
times after acceptance as in Algorithm~\ref{alguea} Step~6 is direct,
%
%e22 #&#
\begin{equation}
{X}^{\mathrm{rem}} \sim \mathbb{W}^{x,y}_{0,T}|
\mathcal {S}'_{\mathrm{UEA}} = \bigotimes^{\kappa+1}_{i=1}
\bigl(\mathbb{W}^{X_{\xi
_{i-1}},X_{\xi_i}}_{\xi_{i-1},\xi_i}|R^{[\xi_{i-1},\xi_i]}_X
\bigr).
\end{equation}

%s3.1.1 #&#
\subsubsection{Implementational Considerations --
Interval Length} \label{sbueaimp}
It transpires that the computational cost of simulating a sample path
scales worse than linearly with interval length. However, this scaling
problem can be addressed by exploiting the fact that sample paths can
be simulated by successive simulation of sample paths of shorter length
over the required interval by applying the strong Markov property,
noting the Radon--Nikod{\'y}m derivative in (\ref{eqrnd}) decomposes
as follows (for any $t\in[0,T]$),
%
%e23 #&#
\begin{eqnarray}
\frac{ \mathrm{d}\mathbb{Q}^x_{0,T}}{ \mathrm{d}\mathbb{W}^x_{0,T}}(X) & =& \exp \biggl\{A(X_t)-A(x)-\int
^t_0 \phi(X_s)\,\mathrm{d}s
\biggr\}
\nonumber
\\[-8pt]
\label{eqdrnd}\\[-8pt]
\nonumber
&& {}\times\exp \biggl\{A(X_T)-A(X_t)-\int
^T_t \phi (X_s) \,\mathrm{d} s
\biggr\}.
\end{eqnarray}

%s3.2 #&#
\subsection{Adaptive Unbounded Exact Algorithm} \label{sauea}
\begin{figure}[t]

\includegraphics{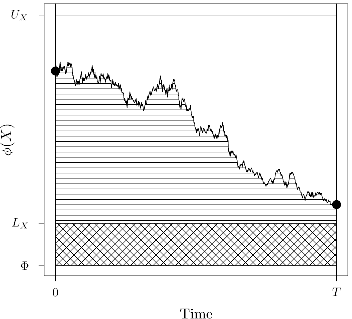}
\caption{Example trajectory of $\phi(X)$ where $X\sim \mathbb{W}^{x,y}_{0,T}|R(X)$.}
\label{figphil}
\end{figure}

Within this section, we outline a novel Adaptive Unbounded Exact
Algorithm (AUEA). To motivate this, we
revisit Algorithm~\ref{algea} noting that the acceptance probability
(\ref{eqzaccprob}) of a sample path $X\sim\mathbb{Z}^x_{0,T}$ can be
rewritten as follows,
%
%e24 #&#
\begin{equation}
P_{\mathbb{Z}^x_{0,T}}(X)  = \exp \biggl\{-\int^T_0
\bigl(\phi (X_s) - L_X \bigr)\, \mathrm{d}s \biggr\}
\cdot \mathrm{e}^{-(L_X - \Phi)T} =: {{\tilde {P}}}_{\mathbb{Z}^x_{0,T}|R}(X)\cdot \mathrm{e}^{-(L_X-\Phi)T}.
\end{equation}
Now following Algorithm~\ref{alguea}, after simulating layer information
(Step~2) and conditionally accepting the proposal
sample path in the first (inexpensive) part of the multi-step rejection
sampler (Step~3) the probability of accepting the sample
path is,
%
%e25 #&#
\begin{equation}\label{eqpoisap}
{\tilde{P}}_{\mathbb{Z}^x_{0,T}|R}(X)\in \bigl[\mathrm{e}^{-
(U_X-L_X
)T},1 \bigr]
\subseteq(0,1].
\end{equation}
Reinterpreting the estimator in (\ref{eqbpe}) in light of (\ref{eqpoisap}) and with the aid of Figure~\ref{figphil}, we are exploiting
the fact that ${\tilde{P}}_{\mathbb{Z}^x_{0,T}|R}(X)$ is precisely the
probability a Poisson process of intensity $1$ on the graph $\mathcal{G}_A:=
\{(x,y)\in[0,T]\times[L_X,\infty)\dvt y\leq\phi(x)\}$ contains no
points. As this is a difficult set in which to simulate a Poisson
process (we don't even know the entire trajectory of $X$), we are
instead simulating a Poisson process of intensity $1$ on the larger
graph $\mathcal{G}_P:=[0,T]\times[L_X,U_X]\supseteq\mathcal{G}_A$
(which is easier as $U_X-L_X$ is a constant), and then conducting
Poisson thinning by first computing $\phi(X)$ at the finite collection
of time points of the simulated Poisson process and then determining
whether or not any of the points lie in $\mathcal{G}_A$ (accepting the
entire sample path if there are no Poisson points in $\mathcal
{G}_A\subseteq\mathcal{G}_P$). This idea was first presented in \cite
{BBPR06} and formed the basis of the Bounded Exact Algorithm (BEA)
discussed in Section~\ref{sbuea}.

As an aside, it should be noted that conditional acceptance of the
proposal sample path with probability $\mathrm{e}^{-(L_X-\Phi)T}$ in
Algorithm~\ref{alguea} Step~3 is simply the
probability that a Poisson process
of intensity $1$ has no points on the graph $\mathcal{G}_R:=[0,T]\times
[\Phi,L_X]$ (the crosshatched region in Figure~\ref{figphil}).
%
%
%f3 #&#

In some settings $\mathcal{G}_P$ can be much larger than $\mathcal{G}_A$ and the resulting exact algorithm can be inefficient and
computationally expensive. In this section, we propose an \textit{adaptive} scheme which exploits the simulation of intermediate
skeletal points of the proposal sample path in Algorithm~\ref{alguea}
Step~4. In particular, note that each simulated
skeletal point implicitly provides information regarding the layer the
sample path is contained within in both the sub-interval before and
after it. As such, by simulating each point separately we can use this
information to construct a modified bounding region $\mathcal{G}^M_P$
such that $\mathcal{G}_A\subseteq\mathcal{G}^M_P\subseteq\mathcal
{G}_P$, composed of a Poisson process with piecewise constant
intensity, for the simulation of the remaining points.

In Algorithm~\ref{alguea} Step~4(a), we simulate a
Poisson process
of intensity $\Delta_XT:= (U_X-L_X)T$ on the interval $[0,T]$ to
determine the skeletal points ($\xi_1,\ldots,\xi_\kappa$).
Alternatively we can exploit the \textit{exponential waiting time}
property between successive events \cite{BKPP}. In particular,
denoting $T_1,\ldots,T_\kappa$ as the time between each event $\xi
_1,\ldots,\xi_\kappa$, then the timing of the events can be simulated
by successive $\operatorname{Exp}(\Delta_X)$ waiting times while $\sum_i T_i
\leq T$.

The independence of arrival times of the points of a homogeneous
Poisson process allows us to simulate them in any convenient order. In
our case,\ it is likely the sample path at points closer to the
mid-point of the interval will contain more information about the layer
structure of the entire sample path. As such, there is an advantage in
simulating these points first. If we begin at the interval mid-point
($T/2$), we can find the skeletal point closest to it by simulating an
$\operatorname{Exp}(2\Delta_X)$ random variable, $\tau$ (we are simulating the
first point at \textit{either} side of the mid-point). As such, the
simulated point (denoted $\xi$) will be with equal probability at
either $T/2-\tau$ or $T/2+\tau$. Considering this in the context of
(\ref{eqpoisap}), upon simulating $\xi$ we have simply broken the
acceptance probability into the product of three probabilities
associated with three disjoint sub-intervals, the realisation of the
sample path at $X_\xi$ providing a binary unbiased estimate of the
probability corresponding to the central sub-interval (where the
expectation is with respect to $u\sim\mathrm{U}[0,1]$),
%
%e26 #&#
\begin{eqnarray}
\tilde{P}_{\mathbb{Z}^x_{0,T}|R,X_\xi}(X) & =& \exp \biggl\{-\int^{T/2-\tau}_0
\bigl[\phi(X_s)-L_X \bigr]\, \mathrm{d}s-\int
^T_{T/2+\tau} \bigl[\phi(X_s)-L_X
\bigr]\,\mathrm{d}s \biggr\}
\nonumber
\\[-8pt]
\label{eqaueacondec}
\\[-8pt]
\nonumber
&&{}\times\mathbb{E} \biggl(\mathbh{1} \biggl[u\leq \frac
{U_X-\phi(X_\xi)}{U_X-L_X}
\biggr]\Big|X_\xi \biggr).
\end{eqnarray}
If the central sub-interval is rejected, the entire sample path can be
discarded. However, if it is accepted then the acceptance of the entire
sample path is conditional on the acceptance of \textit{both} the left-
and right-hand sub-intervals in (\ref{eqaueacondec}), each of which
has the same structural form as we originally had in (\ref{eqpoisap}). As such, for each we can simply iterate the above process until
we have exhausted the entire interval $[0,T]$.

As outlined above, our approach is an algorithmic reinterpretation, but
otherwise identical, to Algorithm~\ref{alguea}. However, we now have the
flexibility to exploit the simulated skeletal point $X_\xi$, to
simulate new layer information for the remaining sub-intervals
conditional on the existing layer $R_X$ (which we detail in
Section~\ref{snlbb}). In particular, considering the left-hand
sub-interval in
(\ref{eqaueacondec}), we can find new layer information (denoted
$R^{[0,\xi]}_X$) which will contain tighter bound information regarding
the sample path ($\ell_X\leq\ell^{[0,\xi]}_X\leq X_{[0,\xi
]}(\omega)\leq
\upsilon^{[0,\xi]}_X\leq\upsilon_X$) and so (as a consequence of
Result~\ref{rescompact}) can be used to compute tighter bounds for
$\phi(X_{[0,\xi
]})$ (denoted $U_X^{[0,\xi]} (\leq U_X)$ and $L_X^{[0,\xi]} (\geq L_X)$),
%
%e27 #&#
\begin{eqnarray}
 &&  {\tilde{P}}^{[0,\xi]}_{\mathbb{Z}^x_{0,T}|R_X^{[0,\xi]},X_\xi} (X)\nonumber\\
 &&\label{eqlhsdec}\quad=  \exp \biggl\{- \int
^{{T}/{2}-\tau}_0 \bigl[\phi (X_s) -
L_X \bigr] \,\mathrm{d}s \biggr\}
\\
&&\quad =  \exp \biggl\{ - \bigl(L^{[0,\xi]}_X - L_X
\bigr)\cdot \biggl(\frac{T}{2} - \tau \biggr) \biggr\}\cdot \exp \biggl\{-
\int^{{T}/{2}-\tau}_0 \bigl[\phi(X_s) -
L^{[0,\xi
]}_X \bigr] \,\mathrm{d}s \biggr\}.\nonumber
\end{eqnarray}
The left-hand exponential in (\ref{eqlhsdec}) is a constant and it is
trivial to immediately reject the entire path with the complement of
this probability. The right-hand exponential of (\ref{eqlhsdec}) has
the same form as~(\ref{eqpoisap}) and so the same approach as
outlined\vspace*{1.5pt} above can be employed, but over the shorter interval $[0,T/2
- \tau]$ and with the lower rate \smash{$\Delta^{[0,\xi]}_X (:=
U_X^{[0,\xi]} -L_X^{[0,\xi]} \leq\Delta_X)$}. As a consequence, the~expected number of intermediary points required in order to evaluate
the acceptance probability in~(\ref{eqpoisap}) is lower than the
Unbounded Exact Algorithm (UEA)
in Algorithm~\ref{alguea}.
\begin{algorithm}[t]
\caption{Adaptive Unbounded Exact Algorithm (AUEA)} \label{algauea}
\begin{enumerate}[6.]
\item[1.] Simulate skeleton end point $X_T=:y \sim h$. \label{algaueastart}
\item[2.] Simulate initial layer information $R_X\sim\mathcal{R}$, setting
$\Pi:=  \{\Xi \} :=  \{ \{[0,T], X_0, X_T,
R_X \}
 \}$ and $\kappa=0$. \label{algauealayer}
\item[3.] With probability $ (1-\exp \{-(L_X-\Phi)T \}
 )$
reject path and return to Step~1. \label{algaueaprelim}
\item[4.] While $ |\Pi |\neq0$, \label{algauealoop}
\begin{enumerate}[(a)]
\item[(a)] Set $\Xi=\Pi(1)$.
\item[(b)] Simulate $\tau\sim\operatorname{Exp} (2\Delta^{\Xi}_X )$. If
$\tau> d(\Xi)$ then set $\Pi:=\Pi\setminus\Xi$ else,
\begin{enumerate}[iii.]
\item[i.] Set $\kappa=\kappa+1$ and with probability $1/2$ set $\xi
'_\kappa
=m(\Xi)-\tau$ else $\xi'_\kappa=m(\Xi)+\tau$.
\item[ii.] Simulate $X_{\xi'_\kappa}\sim \mathbb{W}^{x(\Xi),
y(\Xi
)}_{{\fontsize{7.6}{10.6}\selectfont{\overleftarrow{s}}}(\Xi),{\fontsize{7.6}{10.6}\selectfont{\overrightarrow{t}}}(\Xi)}|R^{\Xi
}_X $. \label{algaueaconlayer}
\item[iii.] With probability $ (1- [U^{\Xi}_X-\phi (X_{\xi
'_\kappa
} ) ]/\Delta^{\Xi}_X )$ reject path and return to
Step~1.
\item[iv.] Simulate new layer information $R_X^{ [{\fontsize{7.6}{10.6}\selectfont{\overleftarrow
{s}}}(\Xi
),\xi'_\kappa ]}$ and $R_X^{ [\xi'_\kappa
,{\fontsize{7.6}{10.6}\selectfont{\overrightarrow
{t}}}(\Xi) ]}$ conditional on $R^{\Xi}_X$. \label{algaueasimlayer}
\item[v.] With probability $ (1-\exp \{- [L^{
[{\fontsize{7.6}{10.6}\selectfont{\overleftarrow{s}}}(\Xi),\xi'_\kappa ]}_X+L^{ [\xi
'_\kappa
,{\fontsize{7.6}{10.6}\selectfont{\overrightarrow{t}}}(\Xi) ]}_X-2L^{\Xi}_X ] [d(\Xi
)-\tau
 ] \} )$ reject path and return to Step~1.
\item[vi.] Set $\Pi:=\Pi\cup \{ [s(\Xi),m(\Xi) - \tau
 ],
X^{\Xi}_{{\fontsize{7.6}{10.6}\selectfont{\overleftarrow{s}}}}, X_{\xi'_\kappa}, R^{
[{\fontsize{7.6}{10.6}\selectfont{\overleftarrow
{s}}}(\Xi),\xi'_\kappa ]}_X \}
%\\
%\hphantom{ \{X^{\Xi}_{{\fontsize{7.6}{10.6}\selectfont{\overleftarrow{s}}}},
%X_{\xi'_\kappa}, R^{ [\overleftarrow{s}(\Xi),\xi'_\kappa
%]}_X \}}
\cup \{ [m(\Xi) + \tau,t(\Xi) ],
X_{\xi
'_\kappa},  X^\Xi_{{\fontsize{7.6}{10.6}\selectfont{\overrightarrow{t}}}},  R^{ [\xi'_\kappa
,{\fontsize{7.6}{10.6}\selectfont{\overrightarrow{t}}}(\Xi) ]}_X \}\setminus\Xi$.
\end{enumerate}
\end{enumerate}
\item[5.] Define skeletal points $\xi_1,\ldots,\xi_\kappa$ as the order
statistics of the set $ \{\xi'_1,\ldots,\xi'_\kappa \}
$.\\[-4pt]
%\vspace{0.25cm}
\hrule\vspace{0.15cm}
\item[6.]\textit{${}^*$Simulate ${X}^{\mathrm{rem}} \sim (\bigotimes^{\kappa+1}_{i=1}
\mathbb{W}^{X_{\xi_{i-1}},X_{\xi_i}}_{\xi_{i-1},\xi_i}|R^{[\xi
_{i-1},\xi_i]}_X )$}. %\label{algaueainf}
\end{enumerate}\vspace*{-3pt}
\end{algorithm}

\begin{figure}[t]
\begin{tabular}{@{}p{170pt}p{170pt}@{}}

\includegraphics{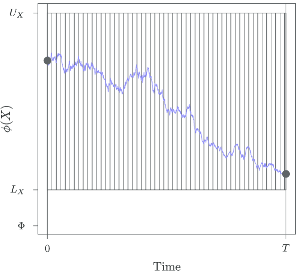}
 & \includegraphics{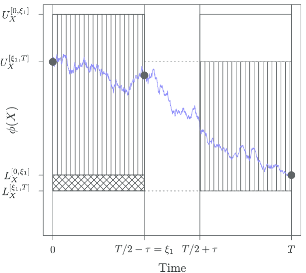}\\
{\fontsize{9}{11}\selectfont{(a) After preliminary acceptance (Algorithm~\ref{algauea}
Step~3)}}\vspace*{11pt} &
\multicolumn{1}{c@{}}{{\fontsize{9}{11}\selectfont{(b) After simulating $\xi_1$ (Algorithm~\ref{algauea}
Step~4)}}}\\[5pt]

\includegraphics{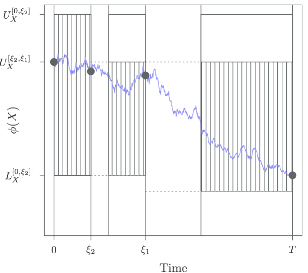}
 & \includegraphics{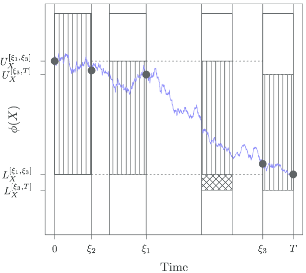}\\
\multicolumn{1}{c}{{\fontsize{9}{11}\selectfont{(c) After simulating $\xi_2$}}} &
\multicolumn{1}{c@{}}{{\fontsize{9}{11}\selectfont{(d) After simulating $\xi_3$}}}
\end{tabular}
\caption{AUEA applied to the trajectory of $\phi(X)$ in Figure~\protect\ref{figphil} (where $X\sim \mathbb{W}^{x,y}_{0,T}|R(X)$).}\vspace*{-6pt}\label{figauea}
\end{figure}

This leads to the novel AUEA
detailed in Algorithm~\ref{algauea}, the
recursive nature of the algorithm being illustrated in Figure~\ref{figauea} which is an extension to the example in Figure~\ref{figphil}.
We outline how to simulate (unbiasedly) layer information (Step~2), intermediate skeletal points (Step~4(b)ii) and new layer information
(Step~4(b)iv) in a variety of ways in Section~\ref{snlbb}. Our
iterative scheme outputs a skeleton comprising skeletal points and
layer information for the intervals between consecutive skeletal points,
%
%e28 #&#
\begin{equation}\label{eqaueaskel}
\mathcal{S}_{\mathrm{AUEA}} (X ) := \bigl\{ (\xi _i,X_{\xi
_i}
)^{\kappa+1}_{i=0}, \bigl(R^{[\xi_{i-1},\xi_i]}_X
\bigr)^{\kappa+1}_{i=1} \bigr\}.
\end{equation}
The AUEA with this skeleton has
the distinct advantage that
Principles \ref{prinlayer}, \ref{prinprop} and \ref{prinrest} are
satisfied directly. In particular, any finite collection of
intermediate points required after the skeleton has been simulated can
be simulated directly (by application of Algorithm~\ref{algauea}
Step~4(b)ii and Step~4(b)iv), without
any augmentation of the skeleton (as in Algorithm~\ref{alguea}). If
necessary, further refinement of the layers given the additionally
simulated points can be performed as outlined in Section~\ref{snlbb}.
%
%
%f4 #&#

\begin{figure}
\begin{tabular}{@{}p{185pt}p{185pt}@{}}

\includegraphics{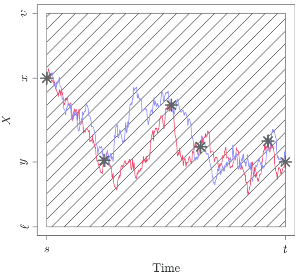}
 & \includegraphics{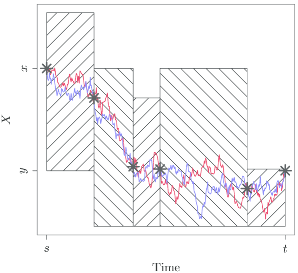}\\
{\fontsize{9}{11}\selectfont{(a) Example sample path skeleton output from the Unbounded Exact
Algorithm (UEA; Algorithm~\ref{alguea}), $\mathcal{S}_{\mathrm{UEA}}
(X )$, overlaid with two possible example sample path trajectories
consistent with the skeleton}}%\label{figilluea}
& {\fontsize{9}{11}\selectfont{(b) Example sample path skeleton output from the Adaptive
Unbounded Exact Algorithm (AUEA; Algorithm~\ref{algauea}),
$\mathcal{S}_{\mathrm{AUEA}} (X)$, overlaid with two possible example sample path
trajectories consistent with the skeleton}}%\label{figillauea}
\end{tabular}
\caption{Comparison of UEA and AUEA skeleton output. Hatched regions
indicate layer information, whereas the asterisks indicate skeletal
points.} \label{figillueaauea}
\end{figure}

It was noted in Section~\ref{sbuea} that the skeleton from the
UEA
(Algorithm~\ref{alguea}) could be augmented so that it satisfies
Principle~\ref{prinrest}. However, the augmentation requires the
application of the
same methodological steps as that developed for the
AUEA (see Section~\ref{snlbb}), but without the
advantage of reducing the number of points
required in the proposal sample path in order to evaluate whether to
accept or reject. As such the AUEA can be viewed as a
Rao--Blackwellised version of the UEA
which should be adopted
whenever further simulation of the proposal sample path is required
after acceptance (e.g., in the settings outlined in Sections~\ref{sepss} and
\ref{sexampfinal}). To emphasise this, in Figure~\ref{figillueaauea} we contrast the skeleton output from the
UEA
(Algorithm~\ref{alguea}, prior to augmentation) and the
AUEA (Algorithm~\ref{algauea}).

In Algorithm~\ref{algauea}, we introduce simplifying notation,
motivated by
the algorithm's recursive nature in which (as shown in (\ref{eqaueacondec})) the acceptance probability is iteratively decomposed over
intervals which have been estimated and are yet to be estimated. $\Pi$
%
%f5 #&#
denotes the set comprising information required to evaluate the
acceptance probability for each of the\vadjust{\goodbreak} intervals still to be estimated,
$\Pi:= \{\Pi(i) \}^{|\Pi|}_{i=1}$. Each $\Pi(i)$ contains
information regarding the time interval it applies to, the sample path
at known points at either side of this interval (which \textit{do not}
necessarily align with the end points of the sub-intervals
corresponding to the remaining probabilities requiring simulation (as
illustrated in Figure~\ref{figauea})) and the associated simulated layer
information, which we denote $[s(\Pi(i)),t(\Pi(i)) ]$,
$x(\Pi
(i)):=X^{\Pi(i)}_{{\fontsize{7.6}{10.6}\selectfont{\overleftarrow{s}}}}$, $y(\Pi(i)):=X^{\Pi
(i)}_{\fontsize{7.6}{10.6}\selectfont{\overrightarrow{t}}}$ and $R^{\Pi(i)}_X$, respectively (where
$\overleftarrow{s}(\Pi(i))\leq s(\Pi(i))<t(\Pi(i))\leq
\overrightarrow{t}(\Pi(i))$). As before, $R^{\Pi(i)}_X$ can be used to directly
compute bounds for $\phi$ for this specific sample path over the
interval $ [s(\Pi(i)),t(\Pi(i)) ]$ (namely $L^{\Pi(i)}_X$,
$U^{\Pi(i)}_X$ and $\Delta^{\Pi(i)}_X$). We further denote $m(\Pi(i))
:=  (s(\Pi(i))+t(\Pi(i)) )/2$, $d(\Pi(i)) :=  (t(\Pi
(i))-s(\Pi(i)) )/2$ and $\Xi:=\Pi(1)$.
%
%

%s3.2.1 #&#
\subsubsection{Implementational Considerations -- Known Intermediate Points} \label{saueaimp}
It should be noted that if a number of intermediate points of a sample
path are required and the time points at which they occur are known in
advance, then rather than\vadjust{\goodbreak} simulating them after the acceptance of the
sample path skeleton in Algorithm~\ref{algauea} Step~6, their
simulation can be incorporated into Algorithm~\ref{algauea}. In particular,
if these points are simulated immediately after Algorithm~\ref{algauea} Step~3 (this can be performed using
Algorithm~\ref{alglayer} as detailed in Section~\ref{sMEA}), then we have additional
layer information regarding the sample path which can be used to
compute tighter bounds for $\phi(X_{[0,T]})$ leading to a more
efficient algorithm (as in Section~\ref{sauea}). A drawback of this
approach is that these additional points of the sample path constitute
part of the resulting skeleton.

%s4 #&#
\section{Exact Simulation of Jump Diffusions} \label{sjea}
In this section, we extend the methodology of Section~\ref{sea},
constructing exact algorithms for simulating skeletons of \textit{jump
diffusion} sample paths which can be represented as the solution to the
following SDE,
%
%e29 #&#
\begin{equation}\label{eqjeajumpsde}
\mathrm{d}X_t  = \alpha (X_{t-} ) \,\mathrm{d}t +
\mathrm{d}W_t + \mathrm{d}J^{\lambda,\nu
}_t,\qquad
X_0 = x \in\mathbb{R}, t\in[0,T].
\end{equation}
Denoting by $\mathbb{Q}^x_{0,T}$ the measure induced by (\ref{eqjeajumpsde}), we can draw sample paths from $\mathbb{Q}^x_{0,T}$ by
instead drawing sample paths from an equivalent proposal measure
$\mathbb{W}^x_{0,T}$ (a natural choice being a driftless version of
(\ref{eqjeajumpsde}), which will no longer coincide with Wiener
measure), and accepting them with probability proportional to the
Radon--Nikod{\'y}m derivative
 of $\mathbb{Q}^x_{0,T}$ with respect to $\mathbb{W}^x_{0,T}$. The
resulting Radon--Nikod{\'y}m derivative (\ref{eqjumprnd}) differs
from that for diffusions
(\ref{eqrnd}) with the inclusion of an additional term, so the
methodology of Section~\ref{sea} can't be applied. However, (\ref{eqjumprnd})~can be re-expressed in a product form similar to (\ref{eqdrnd})
(with $\psi_1,\ldots,\psi_{N_T}$ denoting the jump times in the
interval $[0,T]$, $\psi_0:=0$, $\psi_{N_T+1}:=T$ and $N_t := \sum_{i
\geq1} \mathbh{1}\{\psi_i \leq t\}$),
%
%e30 #&#
\begin{equation}\label{eqjrnd}
\frac{ \mathrm{d}\mathbb{Q}^x_{0,T}}{ \mathrm{d}\mathbb{W}^x_{0,T}}(X) = \prod^{N_T+1}_{i=1}
\biggl[\exp \biggl\{A(X_{\psi_i-})-A(X_{\psi
_{i-1}})-\int
^{\psi_i-}_{\psi_{i-1}} \phi(X_s)\,\mathrm{d}s
\biggr\} \biggr].
\end{equation}
This form of the Radon--Nikod{\'y}m derivative is the key to
constructing \textit{Jump Exact
Algorithms} (JEAs). Recall that in Section~\ref{sbueaimp},
decomposing the
Radon--Nikod{\'y}m derivative for diffusions justified the simulation
of sample paths by
successive simulation of sample paths of shorter length over the
required interval (see (\ref{eqdrnd})). Similarly, jump diffusion
sample paths can be simulated by simulating diffusion sample paths of
shorter length between consecutive jumps.

In this section, we present three novel Jump Exact Algorithms (JEAs).
In contrast with
existing algorithms \cite{MCAPCR10,ORGS12,MCAPGR13}, we note that
the Bounded, Unbounded and Adaptive Unbounded Exact Algorithms in
Section~\ref{sea} can all be incorporated (with an appropriate choice of
layered Brownian bridge construction) within any of the JEAs we
develop. In Section~\ref{sbjea}, we present the Bounded Jump Exact
Algorithm (BJEA), which is a
reinterpretation and methodological extension of \cite{MCAPCR10},
addressing the case where there exists an explicit bound for the
intensity of the jump process. In Section~\ref{sujea}, we present the
Unbounded Jump Exact Algorithm (UJEA)
which is an extension to existing exact algorithms \cite{ORGS12,MCAPGR13} in which the jump intensity is only locally
bounded. Finally, in Section~\ref{saujea} we introduce an entirely novel
Adaptive Unbounded Jump Exact Algorithm (AUJEA) based on the adaptive
approach of Section~\ref{sauea}, which
should be adopted in the case where further simulation of a proposed
sample path is required after acceptance.

%s4.1 #&#
\subsection{Bounded Jump Intensity Jump Exact Algorithm} \label{sbjea}
The case where the jump diffusion we want to simulate (\ref{eqjeajumpsde}) has an explicit jump intensity bound ($\sup_{u\in[0,T]}\lambda
(X_u)\leq\Lambda<\infty$) is of specific interest as an especially
efficient exact algorithm can be implemented in this context. In
particular, proposal jump times, $\Psi_1,\ldots,\Psi_{N^{\Lambda}_T}$
can be simulated according to a Poisson process with the homogeneous
intensity $\Lambda$ over the interval $[0,T]$ (where $N^\Lambda_T$
denotes the number of events in the interval $[0,T]$ of a Poisson
process of intensity $\Lambda$). A simple Poisson thinning argument
\cite{BKPP} can be used to accept proposal jump times with probability
$\lambda(X_{\Psi_i})/\Lambda$. As noted in \cite{MCAPCR10}, this
approach allows the construction of a highly efficient algorithmic
interpretation of the decomposition in (\ref{eqjrnd}). The interval
can be broken into segments corresponding precisely to the intervals
between proposal jump times, then iteratively between successive times,
an exact algorithm (as outlined in Section~\ref{sea}) can be used to
simulate a diffusion sample path skeleton. The terminal point of each
skeleton can be used to determine whether the associated proposal jump
time is accepted (and if so a jump is simulated).

\begin{algorithm}[t]
\caption{Bounded Jump Exact Algorithm (BJEA) \cite{MCAPCR10}}
\label{algbjea}
\begin{enumerate}
\item Set $j=0$. While $\Psi_j<T$,
\begin{enumerate}[(a)]
\item[(a)] Simulate $\tau\sim\operatorname{Exp}(\Lambda)$. Set $j=j+1$ and $\Psi
_j=\Psi_{j-1}+\tau$.
\item[(b)] Apply an exact algorithm to the interval $ [\Psi
_{j-1},
(\Psi_j \wedge T ) )$, obtaining skeleton $\mathcal
{S}^j_{\mathrm{EA}}$.
\item[(c)] If $\Psi_j>T$ then set $X_T=X_{T-}$ else,
\begin{enumerate}[i.]
\item[i.] With probability $\lambda(X_{\Psi_i})/\Lambda$ set $X_{\Psi
_{j}}:=X_{\Psi_{j}-}+\mu_{\Psi_{j}}$ where $\mu_{\Psi_{j}} \sim
f_\nu
 (\cdot;X_{\Psi_{j}-} )$, else set $X_{\Psi_{j}}:=X_{\Psi_{j}-}$.
\end{enumerate}
\end{enumerate}
\vspace{0.15cm}\hrule\vspace{0.15cm}
\item\textit{${}^*$Simulate ${X}^{\mathrm{rem}}\sim\bigotimes^{N^{\Lambda
}_T+1}_{j=1} (\bigotimes^{\kappa_j+1}_{i=1} \mathbb{W}^{X_{\xi
_{j,i-1}},X_{\xi_{j,i}}}_{\xi_{j,i-1},\xi_{j,i}}|R^{\mathrm{EA}}_{X[\xi_{j,0},\xi_{j,\kappa_j+1}]} )$}. \label{algbjeainf}
\end{enumerate}
\end{algorithm}

The Bounded Jump Exact Algorithm (BJEA) we outline in Algorithm~\ref
{algbjea} is a modification of that
originally proposed in \cite{MCAPCR10} (where we define $\Psi_0:=0$,
$\Psi_{N^\Lambda_T+1}:=T$ and $X[s,t]$ to be the trajectory of $X$ in
the interval $[s,t]\subseteq[0,T]$). In particular, we simulate the
proposal jump times iteratively (exploiting the exponential waiting
time property of Poisson processes \cite{BKPP} as in Section~\ref{sauea}),
noting that the best proposal distribution may be different for each
sub-interval. Furthermore, we note that any of the exact algorithms we
introduced in Section~\ref{sea} can be incorporated within
the BJEA (and
so the BJEA  will satisfy at least
Principles~\ref{prinlayer} and~\ref{prinprop}). In
particular, the BJEA skeleton is a
concatenation of
exact algorithm skeletons for the intervals between each proposal jump
time, so to satisfy Principle~\ref{prinrest} and simulate the sample
path at
further intermediate time points (Step~2), we either
augment the skeleton if the exact algorithm chosen is the Unbounded
Exact Algorithm (UEA) (as
discussed in Sections~\ref{sbuea} and \ref{slbb}), or, if the exact
algorithm chosen is the Adaptive Unbounded Exact Algorithm (AUEA)
then simulate them directly (as
discussed in Sections~\ref{sauea} and \ref{snlbb}),
%
%e31 #&#
\begin{equation}\label{BJEAskeleton}
\mathcal{S}_{\mathrm{BJEA}} (X ) := \bigcup^{N^{\Lambda
}_T+1}_{j=1}
\mathcal{S}^j_{\mathrm{EA}}(X).
\end{equation}
%

%s4.2 #&#
\subsection{Unbounded Jump Intensity Jump Exact Algorithm} \label{sujea}
Considering the construction of a Jump Exact Algorithm (JEA) under
the weaker Condition~\ref{condjump} (in which we assume only that the
jump intensity in (\ref{eqjeajumpsde}) is locally bounded), it is not possible to first
simulate the jump times as in Section~\ref{sbjea}. However (as in
Section~\ref{sea} and as noted in \cite{ORGS12,MCAPGR13}), it is
possible to
simulate a layer $R(X)\sim\mathcal{R}$, and then compute a jump
intensity bound $(\lambda\leq\Lambda_X<\infty)$ conditional on this
layer. As such, we can construct a JEA in this
case by simply
incorporating the jump intensity bound simulation within the layer
framework of the Unbounded Exact Algorithm (UEA) and Adaptive
Unbounded Exact Algorithm (AUEA).
%al6
\begin{algorithm}[t]
\caption{Unbounded Jump Exact Algorithm (UJEA) \cite{MCAPGR13}}
\label{algujea}
\begin{enumerate}[2.]
\item[1.] Set $j=0$, $\psi_j=0$ and $N^\lambda_T=0$,
\begin{enumerate}[(a)]
\item[(a)] Simulate skeleton end point $X_T=:y \sim h(y;X_{\psi_j},T-\psi
_j)$. \label{algujeastart}
\item[(b)] Simulate layer information $R^j_{X[\psi_j,T]}\sim\mathcal{R}$ and
compute $\Lambda^j_{X[\psi_j,T]}$.
\item[(c)] Simulate proposal jump times $N^{\Lambda,j}_T\sim\operatorname{Poi}
(\Lambda^j_{X[\psi_j,T]}(T-\psi_j) )$ and $\Psi^j_1,\ldots,\Psi
^j_{N^{\Lambda,j}_T}\stackrel{\mathrm{iid}}{\sim} \mathrm{U}[\psi_j,T]$.
\item[(d)] Simulate skeleton points and diffusion at proposal jump times
$ (X_{\xi^j_1},\ldots,X_{\xi^j_\kappa}, X_{\Psi^j_1},\ldots,X_{\Psi^j_{N({\Lambda,j},T)}} )$,
\begin{enumerate}[ii.]
\item[i.] Simulate $\kappa_j\sim\operatorname{Poi} ( [U^j_{X[\psi
_j,T]}-L^j_{X[\psi_j,T]} ] \cdot(T-\psi_j) )$ and skeleton
times $\xi^j_1,\ldots,\xi^j_\kappa\stackrel{\mathrm{iid}}{\sim}
\mathrm{U}[\psi_j,T]$.
\item[ii.] Simulate sample path at $X_{\xi^j_1},\ldots,X_{\xi^j_\kappa
},X_{\Psi^j_1},\ldots,X_{\Psi^j_{N({\Lambda,j},T)}}\sim
\mathbb{W}^{x,y}_{\psi_j,T}|R^j_{X[\psi_j,T]} $. \label{algujeatimes}
\end{enumerate}
\item[(e)] With probability $ (1-\prod^{\kappa_j}_{i=1} [
(U^j_{X[\psi_j,T]}-\phi (X_{\xi^j_i} ) ) /
(U^j_{X[\psi_j,T]}-L^j_{X[\psi_j,T]} ) ] )$, reject and
return to Step~1(a).
\item[(f)] For $i$ in $1$ to $N^{\Lambda,j}_T$,
\begin{enumerate}[i.]
\item[i.] With probability $\lambda(X_{\Psi^j_i})/\Lambda^j_{X[\psi_j,T]}$
set $X_{\Psi^j_{i}-}=X_{\Psi^j_{i}}$, $X_{\Psi^j_{i}}:=X_{\Psi
^j_{i}-}+\mu_{\Psi^j_{i}}$ where $\mu_{\Psi^j_{i}}\sim f_\nu
(\cdot
;X_{\Psi^j_{i}} )$, $\psi_{j+1}:=\Psi^j_i$, $j=j+1$,
$N^\lambda
_T=j$, and return to Step~1(a).
\end{enumerate}
\end{enumerate}
\vspace{0.15cm}\hrule\vspace{0.15cm}
\item[2.]\textit{${}^*$Simulate ${X}^{\mathrm{rem}}\sim\bigotimes^{N^{\lambda}_T}_{j=0}
[ (\bigotimes^{\kappa_j+1}_{i=1} \mathbb{W}^{X_{\xi
_{j,i-1}},X_{\xi
_{j,i}}}_{\xi_{j,i-1},\xi_{j,i}} )|R^{j}_{X[\psi_j,T]}
]$}. \label{algujeainf}
\end{enumerate}
\end{algorithm}

The Unbounded Jump Exact Algorithm (UJEA) which we present in
Algorithm~\ref{algujea} is a JEA
construction based on the UEA and
extended from \cite{MCAPGR13}.
The UJEA is necessarily more
complicated than the Bounded Jump Exact Algorithm (BJEA) as
simulating a layer in the UEA requires
first simulating an end
point. Ideally we would like to segment the interval the jump diffusion
is to be simulated over into sub-intervals according to the length of
time until the next jump (as in the BJEA), however, as we have
simulated the end point in order to find a jump intensity bound then
this is not possible. Instead we need to simulate a diffusion sample
path skeleton over the entire interval (along with all proposal jump
times) and then determine the time of the first accepted jump (if any)
and simulate it. If a jump is accepted another diffusion sample path
has to be proposed from the time of that jump to the end of the
interval. This process is then iterated until no further jumps are
accepted. The resulting UJEA
satisfies Principles~\ref{prinlayer} and~\ref{prinprop}, however, as a consequence of the layer
construction, the jump diffusion skeleton is composed of the \textit{entirety} of each proposed diffusion sample path skeleton. In
particular, we can't apply the strong Markov property to discard the
sample path skeleton after an accepted jump because of the interaction
between the layer and the sample path before and after the time of that jump.
%
%e32 #&#
\begin{equation}\label{equjeaskel}
\mathcal{S}_{\mathrm{UJEA}} (X ) := \bigcup^{N^{\lambda
}_T}_{j=0}
\bigl\{ \bigl(\xi^j_i,X_{\xi^j_i}
\bigr)^{\kappa
_j+1}_{i=0}, \bigl(\Psi^j_1,X_{\Psi^j_1}
\bigr)^{N^{\Lambda,j}_T}_{i=1}, R^j_{X[\psi_j,T]} \bigr\}.
\end{equation}
The UJEA doesn't satisfy
Principle~\ref{prinrest} unless the skeleton is
augmented (as with the UEA outlined in
Sections~\ref{sbuea} and \ref
{slbb}). As this is computationally expensive, it is not recommended
in practice. Alternatively we could use the AUEA within the UJEA
to directly satisfy Principle~\ref{prinrest}, however it is more efficient
in this case to implement the Adaptive Unbounded Jump Exact Algorithm
(AUJEA) which will be described in
Section~\ref{saujea} (for the same reasons in which the AUEA is more
efficient than the UEA, as detailed in
Section~\ref{sauea}).
%

%s4.3 #&#
\subsection{Adaptive Unbounded Jump Intensity Jump Exact Algorithm}
\label{saujea}
The novel Adaptive Unbounded Jump Exact Algorithm (AUJEA) which we
present in Algorithm~\ref{algaujea} is based on
the Adaptive Unbounded Exact Algorithm (AUEA) and a reinterpretation
of the Unbounded Jump Exact Algorithm (UJEA). Considering the
UJEA, note that if we simulate diffusion sample path skeletons using the
AUEA  then, as the AUEA satisfies Principle~\ref{prinrest}
directly, we
\begin{algorithm}[b]
\caption{Adaptive Unbounded Jump Exact Algorithm (AUJEA)} \label{algaujea}
\begin{enumerate}[5.]
\item[1.] Set $j=0$ and $\psi_j=0$.
\item[2.] Apply AUEA to interval $ [\psi_{j},T ]$, obtaining
skeleton $\mathcal{S}^{[\psi_j,T]}_{\mathrm{AUEA}}$. \label{algaujeastart}
\item[3.] Set $k=0$ and $\Psi^j_k=\psi_j$. While $\Psi^j_k<T$,
\begin{enumerate}[(a)]
\item[(a)] Compute $\Lambda^j_{X[\Psi^j_k,T]}$.
\item[(b)] Simulate $\tau\sim\operatorname{Exp} (\Lambda^j_{X[\Psi
^j_k,T]}
)$. Set $k=k+1$ and $\Psi^j_k=\Psi^j_{k-1}+\tau$.
\item[(c)] If $\Psi^j_k\leq T$,
\begin{enumerate}[iii.]
\item[i.] Simulate $X_{\Psi^j_k}\sim \mathbb{W}^{X_{\psi
_j},X_T}_{\psi
_j,T}|\mathcal{S}^{[\psi_j,T]}_{\mathrm{AUEA}} $.
\item[ii.] Simulate $R^{[\xi^j_-,\Psi^j_k]}_X$ and $R^{[\Psi^j_k,\xi_+]}_X$
and set $\mathcal{S}^{[\psi_j,T]}_{\mathrm{AUEA}}:= \mathcal
{S}^{[\psi
_j,T]}_{\mathrm{AUEA}}\cup \{X_{\Psi^j_k},R^{[\xi^j_-,\Psi
^j_k]}_X,  R^{[\Psi^j_k,\xi_+]}_X \}\setminus R^{[\xi^j_-,\xi_+]}_X$.
\item[iii.] With probability $\lambda(X_{\Psi^j_k})/\Lambda^j_{X[\Psi
^j_{k-1},T]}$ set $X_{\Psi^j_{k}-}=X_{\Psi^j_{k}}$, $X_{\Psi
^j_{k}}:=X_{\Psi^j_{k}-}+\mu_{\Psi^j_{k}}$ where $\mu_{\Psi
^j_{k}}\sim
f_\nu (\cdot;X_{\Psi^j_{k}} )$, $\psi_{j+1}:=\Psi^j_k$, retain
$\mathcal{S}^{[\psi_j,\psi_{j+1})}_{\mathrm{AUEA}}$, discard
$\mathcal
{S}^{[\psi_{j+1},T]}_{\mathrm{AUEA}}$, set $j=j+1$ and return to
Step~2.
\end{enumerate}
\end{enumerate}
\item[4.] Let skeletal points $\chi_1,\ldots,\chi_m$ denote the order
statistics of the time points in $\mathcal{S}_{\mathrm{AUJEA}} :=
\bigcup^{j+1}_{i=1}\mathcal{S}^{[\psi_{i-1},\psi_i)}_{\mathrm{AUEA}}$.\\[-2pt]
%\vspace{0.15cm}
\hrule\vspace{0.15cm}
\item[5.]\textit{${}^*$Simulate ${X}^{\mathrm{rem}} \sim \bigotimes^{m+1}_{i=1}\mathbb{W}^{[X_{\chi_{i-1}},X_{\chi_{i}})}_{[\chi_{i-1},\chi_{i})}|
R^{[\chi_{i-1},\chi_i)}_{X} $}.
\end{enumerate}
\end{algorithm}
can simulate proposal jump times after proposing and accepting a
diffusion sample path as opposed to simulating the proposal times in
conjunction with the sample path (see Algorithm~\ref{algujea}
Step~1(d)ii). As such, we only need to simulate the next
proposal jump time (as opposed to all of the jump times), which (as
argued in Section~\ref{sauea}), provides further information about the
sample path. In particular, the proposal jump time necessarily lies
between two existing skeletal times, $\xi_-\leq\Psi\leq\xi_+$, so the
layer information for that interval, $R^{[\xi_-,\xi_+]}_X$ can be
updated with layer information for each sub-interval $R^{[\xi_-,\Psi
]}_X$ and $R^{[\Psi,\xi_+]}_X$ (the mechanism is detailed in
Section~\ref{sSLBB}). Furthermore, upon accepting a proposal jump
time $\Psi$, the
sample path skeleton in the sub-interval after $\Psi$ contains no
information regarding the skeleton preceding $\Psi$ (so it can be
discarded). As such, the AUJEA satisfies Principles \ref{prinlayer}, \ref{prinprop} and \ref{prinrest} and the skeleton is
composed of only the accepted segments of each
AUEA skeleton,
%
%e33 #&#
\begin{equation} \label{eqaujeaskel}
\mathcal{S}_{\mathrm{AUJEA}} (X ) := \bigcup^{N^{\lambda
}_T+1}_{j=1}
\mathcal{S}^{[\psi_{j-1},\psi_j)}_{\mathrm{AUEA}} (X ).
\end{equation}
%

%s4.4 #&#
\subsection{An Extension to the Unbounded
and  Adaptive Unbounded Jump Exact Algorithms}\label{seuaujea}
In both the UJEA  and AUJEA, we are unable to segment the interval
the jump diffusion is to be simulated over into sub-intervals according
to the length of time until the next jump (in contrast to the Bounded
Jump Exact Algorithm (BJEA)).
As a consequence, we simulate diffusion sample paths which are longer
than necessary (so computationally more expensive), then (wastefully)
partially discard them. To avoid this problem we could break the
interval into segments and iteratively simulate diffusion sample paths
of shorter length over the interval (as in (\ref{eqdrnd})), thereby
minimising the length of discarded segments beyond an accepted jump.
However, the computational cost of simulating a sample path does not
scale linearly with the interval it has to be simulated over, so the
optimal length to decompose the interval is unknown.

It is possible to extend the UJEA
and AUJEA based on this
decomposition and Poisson superposition \cite{BKPP}. In particular, if
it is possible to find a \textit{lower} bound for the jump intensity
$\lambda{\downarrow}\in(0,\lambda)$, then we can consider the target
jump process as being the superposition of two jump processes (one of
homogeneous intensity $\lambda{\downarrow}$ and the other with
inhomogeneous intensity $\lambda-\lambda{\downarrow}$). As such we can
simulate the timing of an accepted jump in the jump diffusion sample
path under the homogeneous jump intensity $\lambda{\downarrow}$ by
means of a $\tau\sim\operatorname{Exp}(\lambda{\downarrow})$ random variable.
If $\tau\in[0,T]$ then there is no need to simulate proposal diffusion
skeletons over the entire interval $[0,T]$, instead we can simulate
them over $[0,\tau]$. Furthermore, we can modify the bounding jump
intensity in the UJEA and AUJEA for generating the proposal jump
times in the proposal diffusion sample path skeletons from $\Lambda_X$
to $\Lambda_X-\lambda{\downarrow}$.

%s5 #&#
\section{\texorpdfstring{$\varepsilon$}{$varepsilon$}-Strong Simulation of (Jump) Diffusions} \label{sepss}
In this section, we outline a novel approach for simulating upper and
lower bounding processes which almost surely constrain (jump) diffusion
sample paths to any specified tolerance. We do this by means of a
significant extension of the \textit{$\varepsilon$-Strong Simulation}
algorithm proposed in \cite{BBPR12}, based upon the adaptive exact
algorithms we developed in Sections~\ref{sauea} and \ref{saujea}.

As originally proposed in \cite{BBPR12} and presented in
Algorithm~\ref{algesamod}, $\varepsilon$-strong simulation is an
algorithm which
simulates upper and lower convergent bounding processes ($X^{\uparrow}$
and $X^{\downarrow}$) which enfold, almost surely, Brownian motion
sample paths over some finite interval $[0,T]$. In particular, we have
$\forall u\in[0,T]$ and some counter $n$,
%
%e34 #&#
\begin{equation} \label{eqesuplow}
X^{\downarrow}_u(n) \leq X^{\downarrow}_u(n+1)
\leq X_u \leq X^{\uparrow
}_u(n+1) \leq
X^{\uparrow}_u(n), \qquad \mbox{w.p. } 1.
\end{equation}
The details on how to simulate initial layer information
(Algorithm~\ref{algesamod} Step~2) can be found in
Section~\ref{sil}
(Algorithm~\ref{algint}), simulate intermediate points
(Algorithm~\ref{algesamod} Step~3(b)ii) can be found in
Section~\ref{silip}, and
simulate new layer information (Algorithm~\ref{algesamod}
Step~3(b)iii,
also know as \textit{Bisection}) can be found in Section~\ref{silb}
(Algorithm~\ref{algbis}) -- note that each of these steps can also be
found within
the Adaptive Unbounded Exact Algorithm (AUEA; Algorithm~\ref
{algauea}). The
additional \textit{Refinement} Step 3(b)iv in
Algorithm~\ref{algesamod} is a method by which a tighter interval in
which the
sample path is constrained can be found and, as shown in \cite
{BBPR12}, ensures a rate of convergence of $\mathcal{O}(2^{-n/2})$ of
the upper and lower bounding processes ($X^{\uparrow}$ and
$X^{\downarrow}$) to $X$. Details on how to conduct Algorithm~\ref
{algesamod} Step~3(b)iv can be found in Section~\ref{silr}
(Algorithm~\ref{algref}).

Note that the $\varepsilon$-strong simulation algorithm presented in
Algorithm~\ref{algesamod} differs from that presented in \cite
{BBPR12}. In
particular, in contrast to \cite{BBPR12}, we simulate initial layer
information unbiasedly (Algorithm~\ref{algesamod} Step~2) and
simulate intermediate points exactly (Algorithm~\ref{algesamod}
Step~3(b)ii). It should be further noted that
Algorithm~\ref{algesamod} doesn't guarantee any particular tolerance
in the
difference between the integrals of the upper and lower bounding
processes over the interval $[0,T]$ can be achieved as currently
written (we address this later in this section), but does have
controlled computational cost.
\begin{algorithm}[t]
\caption{$\varepsilon$-Strong Simulation of Brownian Motion sample paths
($n$ bisections)} \label{algesamod}
\begin{enumerate}[3.]
\item[1.] Simulate $X_T=:y\sim\mathrm{N}(0,T)$.
\item[2.] Simulate initial layer information $R_X\sim\mathcal{R}$, setting
$\mathcal{S} :=  \{\Xi \} :=  \{ \{[0,T], X_0, X_T,
R_X \} \}$. \label{algesamods1}
\item[3.] While $ |\mathcal{S} |\leq2^{n-1}$,
\begin{enumerate}[(b)]
\item[(a)] Set $\Gamma=\varnothing$.
\item[(b)] For $i$ in $1$ to $ |\mathcal{S} |$,
\begin{enumerate}[iii.]
\item[i.] Set $\Xi=\mathcal{S}_i$ and $q:= (s(\Xi)+t(\Xi) )/2$.
\item[ii.] Simulate $X_{q}\sim \mathbb{W}^{x(\Xi), y(\Xi)}_{s(\Xi
),t(\Xi)}|R^{\Xi}_X $. \label{algesamods2}
\item[iii.] Simulate new layer information $R_X^{[s(\Xi),q]}$ and
$R_X^{[q,t(\Xi)]}$ conditional on $R^{\Xi}_X$. \label{algesamods3}
\item[iv.] Refine layer information $R_X^{[s(\Xi),q]}$ and $R_X^{[q,t(\Xi
)]}$. \label{algesamods4}
\item[v.] Set $\Gamma:=\Gamma\cup \{ [s(\Xi),q ],
X^{\Xi
}_{s}, X_{q}, R^{ [s(\Xi),q ]}_X \}\cup \{
[q,t(\Xi) ], X_{q}, X^\Xi_{t}, R^{ [q,t(\Xi)
]}_X \}$.
\end{enumerate}
\item[(c)] Set $\mathcal{S}=\Gamma$.
\end{enumerate}
\end{enumerate}
\end{algorithm}

The upper and lower convergent bounding processes ($X^{\uparrow}$ and
$X^{\downarrow}$) can be found as a function of the layer information
for each interval in Algorithm~\ref{algesamod}, and as shown in \cite{BBPR12}
Proposition~3.1,  convergence in the supremum norm holds (denoting by $s_i$
and $t_i$ the left and right hand time points of $\mathcal{S}_i$,
respectively, and $\ell^i_{s,t}$ and $\upsilon^i_{s,t}$ as the infimum
and supremum of $R^{\mathcal{S}(i)}_X$, resp.),
%
%e35 #&#
\begin{equation}\label{eqesconv}
\mbox{w.p. 1:}\qquad\lim_{n \to\infty} \sup_u
\bigl|X^{\uparrow}_u(n) - X^{\downarrow}_u(n)\bigr|
\rightarrow0,
\end{equation}
where
%
%e36 #&#
\begin{equation}\label{eqXupdown}
X^{\uparrow}_u(n) := \inf\bigl\{\upsilon^i_{s,t}\dvt u
\in[s_i,t_i]\bigr\}, \qquad X^{\downarrow}_u(n)
:= \sup\bigl\{\ell^i_{s,t}\dvt u\in[s_i,t_i]
\bigr\}.
\end{equation}
Furthermore, it was shown in \cite{BBPR12}, Proposition~3.2,  that dominating
processes can be constructed which converge in the $L_1$ norm with rate
of the order $\mathcal{O}(2^{-n/2})$,
%
%e37 #&#
\begin{equation}\label{eqesl1}
2^{n/2}\times\mathbb{E} \bigl[\bigl\llvert X^{\uparrow}-X^{\downarrow
}
\bigr\rrvert _1 \bigr] =\mathcal{O}(1),
\end{equation}
where
%
%e38 #&#
\begin{equation}\label{eqXupdownA}
X^{\uparrow}(n) := \sum^{2^{n-1}}_{i=1}
\upsilon^{i}_{s,t}\cdot (t_{i}-s_{i}),\qquad
X^{\downarrow}(n) := \sum^{2^{n-1}}_{i=1}
\ell^{i}_{s,t}\cdot(t_{i}-s_{i}).
\end{equation}
Now, considering the \textit{$\varepsilon$-Strong Simulation of Jump
Diffusions}, note that upon simulating a jump diffusion sample path
skeleton (as per the AUJEA), it has a form (see (\ref{eqaueaskel})
and (\ref{eqaujeaskel})) that can be used in Algorithm~\ref{algesamod}. As
such, Algorithm~\ref{algesamod} can be extended to jump diffusions
(Algorithm~\ref{algesb}), and (\ref{eqesconv}) and (\ref{eqesl1})
still hold.
\begin{algorithm}[b]
\caption{$\varepsilon$-Strong Simulation of Jump Diffusion sample paths
($n$ bisections)} \label{algesb}
\begin{enumerate}[2.]
\item[1.] Simulate jump diffusion skeleton as per Algorithm~\ref{algauea} to
obtain initial intersection layer.
\item[2.] Simulate further intersection layers as required ($n$ bisections)
as per Algorithm~\ref{algesamod}.
\end{enumerate}
\end{algorithm}
%
%
%f6 #&#
\begin{figure}[t]
\begin{tabular}{@{}ccc@{}}

\includegraphics{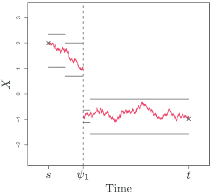}
 & \includegraphics{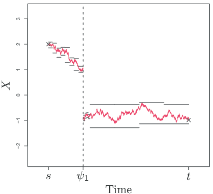} &\includegraphics{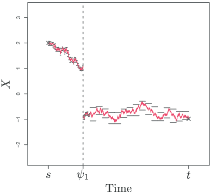} \\
{\fontsize{9}{11}\selectfont{(a) Sample path skeleton}} &
{\fontsize{9}{11}\selectfont{(b) After $n=2$ bisections}}  &
{\fontsize{9}{11}\selectfont{(c) After $n=4$ bisections}}
\end{tabular}
\caption{Illustration of standard $\varepsilon$-strong simulation of a
jump diffusion sample path, overlaid with sample path.} \label{figsesex1}
\end{figure}

\begin{figure}[b]
\begin{tabular}{@{}ccc@{}}

\includegraphics{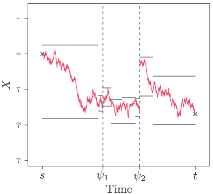}
 & \includegraphics{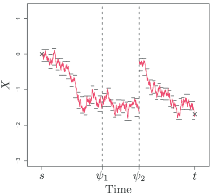} & \includegraphics{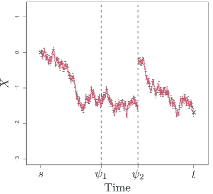}\\
{\fontsize{9}{11}\selectfont{(a) Sample path skeleton}}&
{\fontsize{9}{11}\selectfont{(b) $\llvert X^{\uparrow}-X^{\downarrow}\rrvert \leq0.5$}}&
{\fontsize{9}{11}\selectfont{(c) $\llvert X^{\uparrow}-X^{\downarrow}\rrvert \leq0.2$}}
\end{tabular}
\caption{Illustration of modified tolerance-based $\varepsilon$-strong
simulation of a jump diffusion sample path, overlaid with sample path.}
\label{figsesex2}
\end{figure}

As far as we are aware, there are no existing methods for the $\varepsilon
$-strong simulation of jump diffusions. The class of jump diffusions to
which this methodology can be applied is broad (the conditions outlined
in Section~\ref{spreliminaries} are sufficient) and motivate a number of
avenues for future research. In particular, non-trivial characteristics
of the diffusion path can be simulated (e.g., extrema, hitting
times, integrals) and can be applied to areas such as option pricing,
rare event simulation and the simulation of stochastic volatility
models (which are currently being explored in related work). The
precise implementation of Algorithm~\ref{algesb} can be tailored to the
specific application. For instance, in Figure~\ref{figsesex1} we present
the $\varepsilon$-strong simulation of a jump diffusion sample path as
detailed in Algorithm~\ref{algesb}, whereas in Figure~\ref{figsesex2} we instead
consider an alternate tolerance-based $\varepsilon$-strong simulation of a
jump diffusion sample path in which we are instead interested in
minimising (for any given computational budget) the $L_1$ distance. In
particular, in this example an iterative procedure was established (in
which the intersection layer with the greatest contribution to the
$L_1$ distance was selected and split into two), until such point that
the difference between the integrals of the upper and lower bounding
processes over the entire interval reached a predetermined tolerance.
%
%f7 #&#

%s5.1 #&#
\subsection{An \texorpdfstring{$\varepsilon$}{$varepsilon$}-Strong Exact Algorithm for Diffusions} \label{sesea}
Reconsidering our initial Implementable Exact Algorithm (Algorithm~\ref
{algea}), recall that after simulating the end point from biased
Brownian motion (Algorithm~\ref{algea} Step~1), the
remainder of the
proposal sample path can be simulated exactly from the law of a
Brownian bridge (see Theorem~\ref{thmbbm}). In order to determine whether
to accept or reject a sample path simulated from our proposal measure
($X\sim\mathbb{Z}^x_{0,T}$) as a sample path from our target measure
(denoted $\mathbb{Q}^x_{0,T}$), we accept the sample path with
probability ${{P}}_{\mathbb{Z}^x_{0,T}}(X)$. In Section~\ref{sea}, we
explored how to simulate an event of probability ${{P}}_{\mathbb{Z}^x_{0,T}}(X)$ using only a finite dimensional realisation of the
proposal sample path, however, it is interesting to note that if we
reconsider the simulation of the proposal sample path in light of
Algorithm~\ref{algesamod} we can find upper and lower convergent bounding
sequences for ${{P}}_{\mathbb{Z}^x_{0,T}}(X)$ (in analogous fashion to
(\ref{eqesuplow})) by directly mapping the upper and lower bounds of
the underlying proposal sample path $X$ obtained from $\varepsilon$-strong
simulation (recalling that ${{P}}_{\mathbb{Z}^x_{0,T}}(X) = \exp\{
-\int^T_0 (\phi(X_s)-\Phi)\,\mathrm{d}s\}$),
%
%e39 #&#
\begin{equation}\label{eqeseabound}
0\leq\cdots\leq\phi^{\downarrow}_n \leq\phi^{\downarrow}_{n+1}
\leq\cdots\leq{{P}}_{\mathbb{Z}^x_{0,T}}(X) \leq\cdots\leq \phi
^{\uparrow}_{n+1} \leq\phi^{\uparrow}_n \leq
\cdots\leq1,
\end{equation}
\begin{algorithm}[b]
\caption{$\varepsilon$-Strong Exact Algorithm ($\varepsilon$EA)} \label{algesea}
\begin{enumerate}[6.]
\item[1.] Simulate skeleton end point $X_T=:y \sim h$.\label{algeseastart}
\item[2.] Simulate initial layer information $R_X\sim\mathcal{R}$, setting
$\mathcal{S} :=  \{\Xi \} :=  \{ \{[0,T], X_0, X_T,
R_X \} \}$.
\item[3.] Simulate $u\sim\mathrm{U}[0,1]$, set $n=1$ and compute $\phi
^{\downarrow
}_n$ and $\phi^{\uparrow}_n$. \label{algeseainitlayer}
\item[4.] While $u \in(\phi^{\downarrow}_n,\phi^{\uparrow}_n)$,
\begin{enumerate}[(d)]
\item[(a)] Set $\Xi=\mathcal{S}_{i^*}$ (where $i^*$ is as in (\ref
{eqeseaistar})) and $q:= (s(\Xi)+t(\Xi) )/2$.
\item[(b)] Simulate $X_{q_n}\sim \mathbb{W}^{x(\Xi), y(\Xi
)}_{s(\Xi
),t(\Xi)}|R^{\Xi}_X $.
\item[(c)] Simulate new layer information $R_X^{[s(\Xi),q(n)]}$ and
$R_X^{[q(n),t(\Xi)]}$ conditional on $R^{\Xi}_X$.
\item[(d)] Refine layer information $R_X^{[s(\Xi),q(n)]}$ and
$R_X^{[q(n),t(\Xi)]}$.
\item[(e)] Set $\mathcal{S}:=\mathcal{S}\cup \{ [s(\Xi
),q_n
], X^{\Xi}_{s}, X_{q_n}, R^{ [s(\Xi),q(n) ]}_X \}
\cup \{ [q_n,t(\Xi) ], X_{q_n}, X^\Xi_{t}, R^{
[q(n),t(\Xi
) ]}_X \} \setminus\Xi$, set $n=n+1$ and compute $\phi
^{\downarrow}_n$ and $\phi^{\uparrow}_n$.
\end{enumerate}
\item[5.] If $u \leq\phi^{\downarrow}_n$ accept skeleton, defining $\xi
_0,\xi_1,\ldots,\xi_{n+1}$ as the order statistics of the set
$ \{
s,q_1,\ldots,q_n,t \}$ else if $u\geq\phi^{\uparrow}_n$ reject
and return to Step~1.\\[-3pt]
%\vspace{0.15cm}
\hrule\vspace{0.15cm}
\item[6.]\textit{${}^*$Simulate ${X}^{\mathrm{rem}} \sim\bigotimes^{n+1}_{i=1}
(\mathbb{W}^{X_{\xi_{i-1}},X_{\xi_i}}_{\xi_{i-1},\xi_i}|R^{[\xi
_{i-1},\xi_i]}_X )$}. %\label{algaueainf}
\end{enumerate}
\end{algorithm}

\noindent 
where we define the bounding sequences as follows (recalling $\phi(X)$
is bounded on compact sets (Result~\ref{rescompact}), and that $s_i$ and
$t_i$ depend explicitly on $n$ (see Algorithm~\ref{algesamod})),
%
%e40 #&#
%e41 #&#
\begin{eqnarray}
\phi^{\downarrow}_n & :=&  \exp \Biggl\{-\sum
^{n}_{i=1} \Bigl(\sup_{u\in [\ell
^{i}_{s,t},\upsilon^{i}_{s,t} ]}
\bigl(\phi(u)-\Phi \bigr) \Bigr)\cdot(t_{i}-s_{i}) \Biggr\},
\\
\label{eqeseaphi}
\phi^{\uparrow}_n & :=&  \exp \Biggl\{-\sum
^{n}_{i=1} \Bigl(\inf_{u\in [\ell
^{i}_{s,t},\upsilon^{i}_{s,t} ]}
\bigl(\phi(u)-\Phi \bigr) \Bigr)\cdot(t_{i}-s_{i}) \Biggr\}.
\end{eqnarray}
As such we can simulate events of probability ${{P}}_{\mathbb{Z}^x_{0,T}}(X)$ by direct application of series sampling (see
Theorem~\ref{thmss} in Section~\ref{sbpss}) and hence construct an
\textit{$\varepsilon$-Strong Exact Algorithm} ($\varepsilon$EA) as outlined in Algorithm~\ref
{algesea}. The precise implementation of Algorithm~\ref{algesea} differs
from that of Algorithm~\ref{algesamod} as at each iteration of the algorithm
we want to select an interval to bisect and refine from the existing
finite dimensional realisation of the proposal sample path in order to
find bounds for $P_{\mathbb{Z}^x_{0,T}}(X)$ which are as tight as
possible (this is similar to the tolerance-based $\varepsilon$-strong
simulation illustrated in Figure~\ref{figsesex2}). More precisely, at step
$(n+1)$ we choose to bisect and refine the following interval,
%
%e42 #&#
\begin{equation}\label{eqeseaistar}
i^* := \mathop{\arg\max}_{i\in\{1,\ldots,n\}} \Bigl[ \Bigl(\sup_{u\in
[\ell
^{i}_{s,t},\upsilon^{i}_{s,t} ]}
\phi(u)-\inf_{u\in
[\ell
^{i}_{s,t},\upsilon^{i}_{s,t} ]} \phi(u) \Bigr)\cdot (t_{i}-s_{i})
\Bigr].
\end{equation}%

\noindent Algorithm~\ref{algesea} can be employed to simulate the same class of
diffusions as outlined in Section~\ref{spreliminaries} and furthermore
satisfies Principles \ref{prinlayer}, \ref{prinprop} and \ref
{prinrest}. The resulting skeleton comprises all simulated
intersection layers as shown in (\ref{eqepseaskel}) and admits the
further simulation of intermediate points by direct application of
Algorithm~\ref{alglayer} (as detailed in Section~\ref{sMEA}). It
should be noted
that extension of this exact algorithm to jump diffusions can be
straight forwardly performed in an analogous fashion to the extension
of the exact algorithms in Section~\ref{sea} to exact algorithms for jump
diffusions in Section~\ref{sjea}.
%
%e43 #&#
\begin{equation}\label{eqepseaskel}
\mathcal{S}_{\varepsilon\mathrm{EA}} (X ) := \bigl\{ (\xi
_i,X_{\xi_i} )^{\kappa+1}_{i=0},
\bigl(R^{[\xi_{i-1},\xi
_i]}_X \bigr)^{\kappa+1}_{i=1}
\bigr\}.
\end{equation}
The natural extension to Algorithm~\ref{algesea} is the
AUEA  presented in
Algorithm~\ref{algauea} of Section~\ref{sauea}, which on
implementation is far
more computationally efficient than Algorithm~\ref{algesea} due to
the slow
convergence of the bounding sequences enfolding ${P}_{\mathbb{Z}^x_{0,T}}(X)$ in (\ref{eqeseabound}). However, we have included
this algorithm here as in addition to providing a direct application of
$\varepsilon$-strong simulation as presented in Section~\ref{sepss}, it
is a
novel approach to the exact algorithm which opens up interesting
avenues to tackle related problems (which we are currently exploring in
related work).

%s6 #&#
\section[Brownian Path Space Simulation]{Brownian Path Space
Simulation} \label{sbpss}

In this section, we present key results which we use to construct
layered Brownian bridges in Sections~\ref{slbb} and \ref{snlbb}. In
Section~\ref{ssbb}, we outline a number of established results pertaining
to the simulation of a variety of aspects of Brownian bridge sample
paths. In Section~\ref{ssbpsp}, we consider known results (along with some
extensions) for simulating events corresponding to the probability that
Brownian and Bessel bridge sample paths are contained within particular
intervals. Finally, in Section~\ref{ssnbpsp} we present novel work in which
we consider simulating probabilities corresponding to a more complex
Brownian path space partitioning. Central to Sections \ref{ssbpsp}~and~\ref{ssnbpsp} are Theorem~\ref{thmss} and Corollaries \ref{corlt}
and \ref{corris}, which together form the basis for simulating events of
unknown probabilities $p$, which can be represented as alternating
Cauchy sequences of the following form,
%
%e44 #&#
\begin{equation}\label{eqristyp}
0 = S_0 \leq S_2 < S_4 <
S_6 < \cdots< p < \cdots < S_5 < S_3 <
S_1 \leq1.
\end{equation}
%
% Theorem - SS
%

%th2 #&#
\begin{theorem}[(Series sampling \protect\cite{BKNURVG}, Section~4.5)]
\label{thmss}
An event of (unknown) probability $p\in[0,1]$, where there exists
monotonically decreasing and increasing sequences, $(S^{+}_k  \dvt  k \in
\mathbb{Z}_{\geq0})$ and $(S^{-}_k  \dvt  k \in\mathbb{Z}_{\geq0})$
respectively, such that $\lim_{k\to\infty}S^+_k\downarrow p$ and
$\lim_{k\to\infty}S^+_k\uparrow p$, can be simulated unbiasedly. In
particular, a binary random variable $P:=\mathbh{1}(u\leq p)$ can be
simulated (where $u \sim\mathrm{U}[0,1]$), noting that as there
almost surely
exists a finite $K:=\inf\{k \dvt  u\notin(S^-_k,S^+_k)\}$ we have $\mathbh{1}(u\leq p)= \mathbh{1}(u\leq S^-_K)$ and $\mathbb{E}[\mathbh{1}(u\leq S^-_K)]=p$.
\end{theorem}
%
% Corollary - Linear Transform
%

%co1 #&#
\begin{corollary}[(Transformation)] \label{corlt}
Probabilities which are linear transformations or ratios of a
collection of probabilities, each of which have upper and lower
convergent sequences can be simulated by extension of Theorem~\ref{thmss}.
In particular, suppose $f\dvtx \mathbb{R}^m_+\to\mathbb{R}_+\in C^1$ such
that $\llvert   \mathrm{d}f/ \mathrm{d}u_i(u) \rrvert >0$ $\forall 1\leq i \leq m$
and $u\in\mathbb{R}^m_+$ and that the probability $p:=f(p_1,\ldots,p_m)$ then defining the sequences ($T^{i,-}_k  \dvt  k \in\mathbb{Z}_{\geq0}$) and
($T^{i,+}_k  \dvt  k \in\mathbb{Z}_{\geq0}$) as follows,
%
%e45 #&#
\begin{equation}
T^{i,-}_k  = \cases{
S^{i,-}_k, & \quad$\mbox{if } \mathrm{d}f/ \mathrm{d}u_i>0$,\vspace*{3pt}\cr
S^{i,+}_{k}, &\quad$\mbox{if } \mathrm{d}f/
\mathrm{d}u_i<0$,}\qquad
T^{i,+}_k
 = \cases{S^{i,+}_k,  &
\quad$\mbox{if } \mathrm{d}f/ \mathrm{d}u_i>0$,\vspace*{3pt}\cr
S^{i,-}_{k+1},  & \quad$\mbox{if } \mathrm{d}f/
\mathrm{d}u_i<0$.}
\end{equation}
We have that $S^-_k:=f(T^{1,-}_k,\ldots,T^{m,-}_k)$ is monotonically
increasing and converges to $p$ from below and
$S^+_k:=f(T^{1,+}_k,\ldots,T^{m,+}_k)$ is monotonically decreasing
and converges to $p$ from above.
\end{corollary}

% Corollary - RIS
%
%
%co2 #&#
\begin{corollary}[(Retrospective Bernoulli sampling  \cite{MCAPBPR08}, Proposition~1)] \label{corris}
If $p$ can be represented as the limit of an alternating Cauchy
sequence of the form of (\ref{eqristyp}) ($S_k  \dvt  k \in\mathbb{Z}_{\geq0}$), then splitting the sequence into subsequences composed
of the odd and even terms, respectively, each subsequence will converge
to $p$, one of which will be monotonically decreasing and the other
monotonically increasing, so events of probability $p$ can be simulated
by extension of Theorem~\ref{thmss}.
\end{corollary}

We conclude the introductory remarks to this section by
presenting Algorithm~\ref{algris}, in which we outline by application of
Corollary~\ref{corris} how an unknown probability which can be represented
as an alternating Cauchy sequence in which (without loss of generality)
the even terms converge from below and the odd terms from above, can be
simulated unbiasedly.
\begin{algorithm}[b]
\caption{Retrospective Bernoulli Sampling \cite{MCAPBPR08}} \label{algris}
\begin{enumerate}[3.]
\item[1.] Simulate $u \sim\mathrm{U}[0,1]$ and set $k=1$. \label{algrisunif}
\item[2.] While $u \in (S_{2k},S_{2k+1})$, $k=k+1$. \label{algrisstoc}
\item[3.] If $u \leq S_{2k}$ then $u<p$ so return $1$ else $u>p$ so return $0$.
\end{enumerate}
\end{algorithm}

Clearly as we have that the number of computations required to
implement retrospective Bernoulli sampling is stochastic (as a
consequence of Algorithm~\ref{algris} Step~2), the
efficiency of the
algorithm is dependent upon the expected number of iterations of that
step required (where $u\sim\mathrm{U}[0,1]$ as per Algorithm~\ref{algris} Step~1),
%
%e46 #&#
\begin{equation} \label{eqfinexp}
\mathbb{E}[K]  = \sum^\infty_{k=1}
\mathbb{P} (K\geq k ) = \sum^\infty _{k=0}
\mathbb{P} \bigl(u \in [S_{2k},S_{2k+1} ] \bigr) = \sum
^\infty_{k=0}\llvert S_{2k+1}-S_{2k}
\rrvert.
\end{equation}
At a minimum for any practical implementation, we require that the
$\mathbb{E}[K]<\infty$, which can't be ensured without imposing
further conditions. However, as we will encounter in Sections \ref{ssbpsp}~and~\ref{ssnbpsp}, the alternating Cauchy sequences which we
consider in this paper converge exponentially fast and so finiteness is ensured.

%s6.1 #&#
\subsection{Simulating Brownian Bridges and
Related Processes} \label{ssbb}
The density of a Brownian bridge sample path $W^{x,y}_{s,t}$, at an
intermediate time $q \in(s,t)$ is Gaussian with mean $\mu_w := x +
(q-s)(y-x)/(t-s)$ and variance $\sigma^2_w := (t-q)(q-s)/(t-s)$ (so can
be simulated directly). The joint distribution of the minimum value
reached by $W^{x,y}_{s,t}$, and the time at which it is attained ($\tau
,\hat{m}$), is given by \cite{BKBMSC},
%
%e47 #&#
\begin{eqnarray}
&& \mathbb{P} (\hat{m}\in \mathrm{d}w, \tau\in \mathrm {d}q|W_s=x,W_t=y
)
\nonumber
\\[-8pt]
\label{eqbrmindist}
\\[-8pt]
\nonumber
&&\quad \propto\frac{(w-x)(w-y)}{\sqrt
{(t-q)^3(q-s)^3}}\exp \biggl\{-\frac{(w-x)^2}{2(q-s)}-\frac{(w-y)^2}{2(t-q)}
\biggr\} \,\mathrm{d}w \,\mathrm{d}q.
\end{eqnarray}
Analogously the maximum ($\tau,\check{m}$), can be considered by reflection.
We can jointly draw $(\tau,\hat{m})$ (or $(\tau,\check{m})$) as
outlined in Algorithm~\ref{algbbm}, which is similar to the approach
taken in \cite{BBPR06},
noting that it is possible to condition the minimum to occur within a
particular interval. In particular, we can simulate $ (\tau,\hat{m}
)| (\hat{m}\in[a_1,a_2] ) $ where $a_1 < a_2 \leq
(x
\wedge  y)$.
\begin{algorithm}[t]
\caption{Brownian Bridge Simulation at its Minimum Point (constrained
to the interval $[a_1,a_2]$ where $a_1<a_2\leq x\wedge y$ and
conditional on $W_s =x$ and $W_t=y$ (denoting $\operatorname{IGau}(\mu
,\lambda
)$ as the inverse Gaussian distribution with mean $\mu$ and shape
parameter $\lambda$)} \label{algbbm}
\begin{enumerate}[4.]
\item[1.] Simulate $u_1 \sim\mathrm{U} [M(a_1),M(a_2) ]$ where
$M(a) :=
\exp \{-2(a-x)(a-y)/(t-s) \}$ and $u_2 \sim\mathrm{U}[0,1]$.
\item[2.] Set $\hat{m}:= x -  [\sqrt{(y-x)^2-2(t-s)\log(u_1)}-
(y-x) ]/2$.
\item[3.] If $u_2 \leq\frac{x-\hat{m}}{x+y-2\hat{m}}$ then $V\sim
\operatorname{IGau}
(\frac{y-\hat{m}}{x-\hat{m}},\frac{(y-\hat{m})^2}{t-s} )$
else $\frac{1}{V}\sim
\operatorname{IGau} (\frac{x-\hat{m}}{y-\hat{m}},\frac{(x-\hat
{m})^2}{t-s} )$.
\item[4.] Set $\tau:= \frac{sV+t}{1+V}$.
\end{enumerate}
\end{algorithm}
\begin{algorithm}[b]
\caption{(Minimum) Bessel Bridge Simulation (at time $q\in(s,t)$ given
$W_{s}=x,W_t=y$ and $W_\tau=\hat{m}$) \cite{AGPAAP95}} \label{algbsbm}
\begin{enumerate}[2.]
\item[1.] If $q<\tau$ then $r=s$ else $r=t$. Simulate $b_1,b_2,b_3\stackrel{\mathrm{iid}}{\sim} \mathrm{N} (0,\frac{|\tau-q|\cdot
|q-r|}{(\tau
-r)^2} )$.
\item[2.] Set $W_{q}:=\hat{m}+ \sqrt{|\tau-r|}\cdot\sqrt{ (\frac
{
(W_{r}-\hat{m} )\cdot|\tau-q|}{|\tau-r|^{3/2}}+b_1 )^2
+b^2_2+b^2_3}$.
\end{enumerate}
\end{algorithm}%

Conditional on a Brownian bridge sample path minimum (or maximum), the
law of the remainder of the trajectory is that of a \textit{Bessel
bridge}, which can be simulated by means of a 3-dimensional Brownian
bridge of unit length conditioned to start and end at zero as outlined
in~\cite{AGPAAP95} and Algorithm~\ref{algbsbm} (maximum by reflection).

%s6.2 #&#
\subsection{Simulating
Elementary Brownian Path Space Probabilities} \label{ssbpsp}
\begin{figure}[t]

\includegraphics{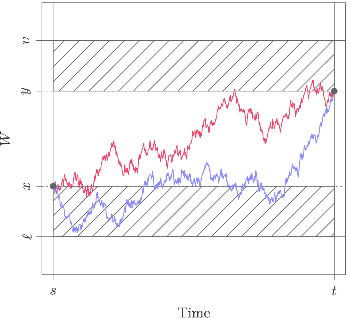}

\caption{Example sample path trajectories $W\sim \mathbb{W}^{x,y}_{s,t}|(W_{(s,t)}\in[\ell,\upsilon]) $.} \label{figgamimg}
\end{figure}

In this section, we briefly outline results pertaining to the
probability that a Brownian bridge sample path is contained within a
particular interval \cite{AMSA60,JAPPW01} (Theorem~\ref{thmgamma}) and
how to simulate events of this probability \cite{MCAPBPR08}
(Corollary~\ref{corgamma}). In Figure~\ref{figgamimg}, we show
example sample path
trajectories of a Brownian bridge $W\sim\mathbb{W}^{x,y}_{s,t}$, which
remain in the interval $[\ell,\upsilon]$. Similarly in Theorems \ref{thmdelta1} and \ref{thmdelta2} we outline a result (first shown in
\cite{MCAPBPR08}) that shows that the probability a Bessel bridge
sample path is contained within a particular interval can be
represented as an infinite series. We reproduce these results as they
are used and extended extensively throughout the remainder of this
paper. In the rest of this paper, with a slight abuse of notation, we
write $ \{W\in[\ell,\upsilon] \}$ to mean $\{W_u\dvt  s\leq
u\leq
t\}\subset[\ell,\upsilon]$. Further details on the results developed
in this section can be found in \cite{PhDP13}, Chapter~6.1.1.
%
%f8 #&#

Of particular importance for what follows is Corollary~\ref{cordelta2}, in
which we establish that it is possible to simulate events with a
probability corresponding to the probability that a Bessel bridge
sample path is contained within a particular interval (without
assumptions on the size of the interval), by application of
Corollary~\ref{corris}.

%
%
%th3 #&#
\begin{theorem}[(\protect\cite{JAPPW01}, Theorem~3)] \label{thmgamma}
The probability that a Brownian bridge sample path $W\sim\mathbb{W}^{x,y}_{s,t}$, remains in the interval $[\ell,\upsilon]$
(i.e.,
$\forall u\in[s,t]$ $W_u\in[\ell,\upsilon]$) can be represented as an
infinite series,
%
%e48 #&#
\begin{equation}
\gamma^{\ell,\upsilon}_{s,t}(x,y)  := \mathbb{P} \bigl(W\in[\ell,
\upsilon] \bigr) = 1 - \sum^{\infty
}_{j=1}
\bigl\{\varsigma^{\ell,\upsilon}_{s,t}(j;x,y) -\varphi
^{\ell
,\upsilon}_{s,t}(j;x,y) \bigr\},
\end{equation}
where $\varsigma^{\ell,\upsilon}_{s,t}(j;x,y) := \bar{\varsigma
}^{\ell
,\upsilon}_{s,t}(j;x,y) + \bar{\varsigma}^{-\ell,-\upsilon
}_{s,t}(j;-x,-y)$, $\varphi^{\ell,\upsilon}_{s,t}(j;x,y) := \bar
{\varphi
}^{\ell,\upsilon}_{s,t}(j;x,y) +\break \bar{\varphi}^{-\ell,-\upsilon
}_{s,t}(j;-x,-y)$ and,
%
%e49 #&#
%e50 #&#
\begin{eqnarray}\label{eqvarsigrep}
\bar{\varsigma}^{\ell,\upsilon}_{s,t}(j;x,y) & :=&  \exp \biggl\{-
\frac{2}{t-s} \bigl( |\upsilon-\ell | j+(\ell \wedge\upsilon)-x \bigr)\cdot
\bigl( |\upsilon-\ell | j+(\ell \wedge\upsilon)-y \bigr) \biggr\},
\\
\label{eqvarphirep}
\bar{\varphi}^{\ell,\upsilon}_{s,t}(j;x,y) & :=& \exp \biggl\{-
\frac{2j}{t-s} \bigl( |\upsilon-\ell |^2j+ |\upsilon-\ell | (x-y)
\bigr) \biggr\}.
\end{eqnarray}
\end{theorem}
%
% Corollary - Gamma is a Cauchy sequence
%

%co3 #&#
\begin{corollary}[(\protect\cite{MCAPBPR08}, Proposition~2)]\label{corgamma}
$\gamma^{\ell,\upsilon}_{s,t}(x,y)$ is an alternating Cauchy sequence,
so events of probability $\gamma^{\ell,\upsilon}_{s,t}(x,y)$ can be
simulated by retrospective Bernoulli sampling (Corollary~\ref{corris} and
Algorithm~\ref{algris}) using the following sequence,
%
%e51 #&#
\begin{equation}
S^\gamma_{2k} := 1 - \sum^{k}_{j=1}
\bigl\{\varsigma^{\ell,\upsilon}_{s,t}(j;x,y) -\varphi^{\ell,\upsilon}_{s,t}(j;x,y)
\bigr\}, \qquad  S^\gamma_{2k+1} := S^\gamma_{2k}
- \varsigma^{\ell,\upsilon}_{s,t}(k+1;x,y).
\end{equation}
\end{corollary}

As shown in \cite{MCAPBPR08}, Theorem~\ref{thmgamma} and
Corollary~\ref{corgamma} can be extended to consider simulating events
with a probability corresponding to the probability a Bessel bridge
sample path is contained within a particular interval. As indicated in
Definition~\ref{dfndelta} we have to consider two possible cases
where either
of the end points attain the sample path minimum (or maximum) or not.
% Definition - Delta
%

%de4 #&#
\begin{defn} \label{dfndelta}
We allow $\delta^{\hat{m},\upsilon}_{s,t}(x,y)$ to denote the probability
that a Bessel bridge sample path $W\sim \mathbb{W}^{x,y}_{s,t}|\hat{m} $ (with minimum $\hat{m}$) remains in the
interval $[\hat{m},\upsilon]$. We further denote $\delta^{\hat
{m},\upsilon
}_{s,t}(1;x,y) := \mathbb{P} (W \in[\hat{m},\upsilon]|W\geq
\hat{m},
(x \wedge y)>\hat{m} )$ and $\delta^{\hat{m},\upsilon
}_{s,t}(2;x,y):=
\mathbb{P} (W\in[\hat{m},\upsilon]|W \geq\hat{m}, (x
\wedge y)=\hat{m}
 )$ noting that $\delta^{\hat{m},\upsilon}_{s,t}(x,y)
=\mathbh{1}\{\hat{m}
< (x\wedge y)\}\cdot\delta^{\hat{m},\upsilon}_{s,t}(1;x,y)$
$+\mathbh{1}\{
\hat{m} = (x \wedge y)\}\cdot\delta^{\hat{m},\upsilon}_{s,t}(2;x,y)$.
\end{defn}

% Remark - Delta 1 by reflection
Note that we can similarly consider the probability that a
Bessel bridge sample path $W\sim \mathbb{W}^{x,y}_{s,t}|\check{m}
 $ (with maximum $\check{m}$) remains in the interval $[\ell
,\check{m}]$
($\forall u\in[s,t]$ $W_u\in[\ell,\check{m}]$) by a simple reflection
argument.

We first consider the case where neither end point attains the Bessel
bridge minimum.
% Theorem - Delta 1
%

%th4 #&#
\begin{theorem}[(\cite{MCAPBPR08}, Proposition 3)] \label{thmdelta1}
The probability that a Bessel bridge sample path $W\sim \mathbb{W}^{x,y}_{s,t}|\hat{m} $, (with minimum $\hat{m}<(x \wedge y)$)
remains in the interval $[\hat{m},\upsilon]$ ($\forall u\in[s,t]$
$W_u\in[\hat{m}
,\upsilon]$) can be represented as an infinite series,
%
%e52 #&#
\begin{eqnarray}
\delta^{\hat{m},\upsilon}_{s,t}(1;x,y) & :=&  \mathbb{P} \bigl(W \in [
\hat{m} ,\upsilon]|W\geq\hat{m}, (x\wedge y)>\hat{m} \bigr)
\nonumber
\\[-9pt]
\\[-9pt]
\nonumber
& =&  \frac{\gamma^{\hat{m},\upsilon}_{s,t}(x,y)}{1-\exp \{
-2(x-\hat{m})(y-\hat{m}
)/(t-s)  \}}.
\end{eqnarray}
\end{theorem}
%
% Corolloary - Delta 1 Cauchy

%
%co4 #&#
\begin{corollary}[(\protect\cite{MCAPBPR08}, Proposition~3)] \label{cordelta1}
Events of probability $\delta^{\hat{m},\upsilon}_{s,t}(1;x,y)$ can be
simulated by application of retrospective Bernoulli sampling (as per
Corollaries \ref{corlt}, \ref{corris} and Algorithm~\ref{algris})
using the
following sequence,
%
%e53 #&#
\begin{equation}
S^{\delta,1}_{k} := \frac{S^\gamma_{k}}{1-\exp \{-2(x-\hat{m})(y-\hat
{m})/(t-s)  \}}.
\end{equation}
\end{corollary}

  We now consider the case where either one of the end points
attains the Bessel bridge minimum.
% Theorem - Delta 2
%

%th5 #&#
\begin{theorem}[(\protect\cite{MCAPBPR08}, Proposition~3)] \label{thmdelta2}
The probability that a Bessel bridge sample path $W\sim \mathbb{W}^{x,y}_{s,t}|\hat{m}$ (with minimum $\hat{m}=x<y$)
remains in the
interval $[\hat{m},\upsilon]$ ($\forall u\in[s,t]$ $W_u\in[\hat
{m},\upsilon]$)
can be represented as an infinite series,
%
%e54 #&#
\begin{eqnarray}
\delta^{\hat{m},\upsilon}_{s,t}(2;x,y) & :=& \mathbb{P} \bigl(W\in[
\hat{m},\upsilon]|W \geq\hat{m} \bigr)
\nonumber
\\[-10pt]
\label{eqdel2deriv}\\[-10pt]
\nonumber
& =&  1 - \frac{1}{(y-\hat{m})}\sum^{\infty}_{j=1}
\bigl\{\psi ^{\hat{m},\upsilon
}_{s,t}(j;y) -\chi^{\hat{m},\upsilon}_{s,t}(j;y)
\bigr\},
\end{eqnarray}
where we denote,
%
%e55 #&#
%e56 #&#
\begin{eqnarray}
\psi^{\hat{m},\upsilon}_{s,t}(j;y)  & :=&  \bigl(2 |\upsilon-\hat{m} | j -
(y-\hat{m}) \bigr) \exp \biggl\{-\frac
{2 |\upsilon-\hat{m} |  j}{t-s} \bigl( |\upsilon-\hat {m} |
j- (y-\hat{m} ) \bigr) \biggr\},
\\[-2pt]
\chi^{\hat{m},\upsilon}_{s,t}(j;y) & :=&  \bigl(2 |\upsilon-\hat{m} | j +
(y-\hat{m}) \bigr) \exp \biggl\{-\frac
{2 |\upsilon-\hat{m} |  j}{t-s} \bigl( |\upsilon-\hat {m} |
j+ (y-\hat{m} ) \bigr) \biggr\}.
\end{eqnarray}
\end{theorem}

% Remark - Delta 2 Cauchy
%
%
%re1 #&#
\begin{remark}[(\cite{MCAPBPR08}, Proposition~3)] \label{remdelta2}
As before, we can consider the probability a Bessel bridge sample path
$W\sim \mathbb{W}^{x,y}_{s,t}|\hat{m}$ (with minimum
$\hat{m}
=y<x$) remains in the interval $[\hat{m},\upsilon]$ by a simple reflection
argument of Theorem~\ref{thmdelta2}.
\end{remark}

We conclude this section by showing that it is possible to
simulate events with probability corresponding to the probability a
Bessel bridge sample path is contained within a particular interval,
without any further assumption regarding the interval size (unlike
existing methods \cite{MCAPBPR08}, Proposition~3,  in which one requires that
$3(\upsilon-\hat{m})^2>(t-s)$). As we consider in this setting
$s,t,x,y,\hat{m}
,\upsilon$ fixed, for conciseness we denote $\psi_j:=\psi^{\hat
{m},\upsilon
}_{s,t}(j;y)$ and $\chi_j:=\chi^{\hat{m},\upsilon}_{s,t}(j;y)$.
% Lemma
%

%co5 #&#
\begin{corollary} \label{cordelta2}
After the inclusion of the first $\hat{k}:=\sqrt{(t-s)+|\upsilon
-\hat{m}
|^2}/ (2|\upsilon-\hat{m}| )$ terms, $\delta^{\hat
{m},\upsilon
}_{s,t}(2;x,y)$ is an alternating Cauchy sequence, so events of
probability $\delta^{\hat{m},\upsilon}_{s,t}(2;x,y)$ can be
simulated by
retrospective Bernoulli\vspace*{1pt} sampling (as per Corollary~\ref{corris} and
Algorithm~\ref{algris}) using the following sequence (where $k\in
\mathbb{N}$ such
that $k\geq\hat{k}$),
%
%e57 #&#
%e58 #&#
\begin{eqnarray}
S^{\delta,2}_{2k} & :=&  1 - \frac{1}{(y-\hat{m})}\sum
^{k}_{j=1} \bigl\{\psi^{\hat
{m},\upsilon
}_{s,t}(j;y)
-\chi^{\hat{m},\upsilon}_{s,t}(j;y) \bigr\},
\\[-6pt]
S^{\delta,2}_{2k+1} & :=&  S^{\delta,2}_{2k} -
\frac{1}{y-\hat{m}}\psi^{\hat
{m},\upsilon}_{s,t}(k+1;y).
\end{eqnarray}
\end{corollary}

\begin{pf}
As $(y-\hat{m}) \in(0, (\upsilon-\hat{m})]$ then $\forall j$ we
have $\psi_j, \chi
_j \geq0$. As such it is sufficient to show that $\forall j \geq\hat
{k}$ that $\psi_j \geq\chi_j \geq\psi_{j+1} \geq\chi_{j+1} \geq
\cdots$ which can be proved inductively by first showing that
$\forall j$ $\psi_j \geq\chi_j$ and then $\forall j$ $\chi_j \geq
\psi
_{j+1}$. Considering $\psi_j/\chi_j$ if $j\geq\hat{k}$ then this is
minimised when $y=\hat{m}$ and $\psi_j/\chi_j>1$. Similarly
considering $\chi
_j/\psi_{j+1}$ if $j\geq\hat{k}$ then this is minimised when
$y=\upsilon$ where $\chi_j/\psi_{j+1}>1$.
\end{pf}

%s6.3 #&#
\subsection{Simulating
Brownian Path Space Probabilities}\label{ssnbpsp}
In this section, we establish that the probability a Brownian bridge
sample path, conditioned on a number of intermediate points
($q_1,\ldots,q_n$), has a minimum in a lower interval and a maximum in an upper
interval (or in each sub-interval a minimum in a lower interval and a
maximum in an upper interval) can be represented as an infinite series
(Theorems \ref{thmrhon} and \ref{thmbetan}, resp.), and events
of this probability can be simulated (Corollaries \ref{corrhon} and
\ref{corbetan}, resp.). Further details on the results
developed in this section can be found in \cite{PhDP13}, Chapter~6.1.2.

In this section, we introduce the following simplifying notation,
$q_{1:n}:= \{q_1,\ldots,q_n \}$, $q_0:=s$ and $q_{n+1}:=t$.
We further denote $\hat{m}_{s,t} := \inf\{W_q; q \in[s,t]\}$,
$\check{m}_{s,t} :=
\sup\{W_q; q \in[s,t]\}$, $\mathcal{W}:= \{W_{q_1}=w_1,\ldots
,W_{q_n}=w_n \}$, $\mathcal{L}:= \{\hat{m}_{s,q_1}\in
[\ell
^{\downarrow}_{s,q_1},\ell^{\uparrow}_{s,q_1}],\ldots,\hat
{m}_{q_n,t}\in
[\ell^{\downarrow}_{q_n,t},\ell^{\uparrow}_{q_n,t}] \}$,
$\mathcal
{U}:= \{\check{m}_{s,q_1}\in[\upsilon^{\downarrow
}_{s,q_1},\upsilon
^{\uparrow}_{s,q_1}],\ldots,\check{m}_{q_n,t}\in[\upsilon
^{\downarrow
}_{q_n,t},\upsilon^{\uparrow}_{q_n,t}] \}$. We\vspace*{1pt} also use the
following abuse of notation $ \{W_{[s,t]}\in[\ell,\upsilon
] \}
:=\{W_u\in[\ell,\upsilon] \ \forall u\in[s,t]\}$, noting that $
\{
W_{[s,t]}\in[\ell,\upsilon] \}=\{\hat{m}_{s,t}\in[\ell,(x
\wedge
y)],\check{m}_{s,t}\in[(x \vee y),\upsilon]\}$.
% Theorem - Rho (n)
%

%th6 #&#
\begin{theorem} \label{thmrhon}
The probability a Brownian bridge sample path $W\sim \mathbb{W}^{x,y}_{s,t}|\mathcal{W} $ has a minimum $\hat{m}_{s,t}\in
[\ell
{\downarrow},\ell{\uparrow}]$ and a maximum $\check{m}_{s,t}\in
[\upsilon
{\downarrow},\upsilon{\uparrow}]$ can be represented as an infinite series,
%
%e59 #&#
\begin{eqnarray}
 && {}^{(n)}\rho^{\ell{\downarrow},\ell{\uparrow},\upsilon
{\downarrow},\upsilon{\uparrow}}_{s,t,x,y}(q_{1:n},
\mathcal{W}) \nonumber\\
&&\quad :=  \mathbb{P} \bigl(\hat{m}_{s,t} \in[\ell{\downarrow},
\ell {\uparrow}], \check{m} _{s,t} \in[\upsilon{\downarrow},\upsilon{
\uparrow}]|\mathcal {W} \bigr)
\\
&&\quad =  \Biggl[\prod^n_{i=0}
\gamma^{\ell{\downarrow
},\upsilon
{\uparrow}}_{q_i,q_{i+1}} \Biggr] - \Biggl[\prod
^n_{i=0}\gamma^{\ell
{\uparrow},\upsilon{\uparrow}}_{q_i,q_{i+1}}
\Biggr]- \Biggl[\prod^n_{i=0}
\gamma^{\ell{\downarrow},\upsilon{\downarrow
}}_{q_i,q_{i+1}} \Biggr] + \Biggl[\prod
^n_{i=0}\gamma^{\ell{\uparrow
},\upsilon{\downarrow}}_{q_i,q_{i+1}}
\Biggr].\nonumber
\end{eqnarray}
\end{theorem}

\begin{pf}
Follows by sample path inclusion--exclusion and the Markov property for
diffusions,
%
%e60 #&#
%e61 #&#
\begin{eqnarray}
&& \hspace*{-10pt}{}^{(n)}\rho^{\ell{\downarrow},\ell{\uparrow},\upsilon
{\downarrow},\upsilon{\uparrow}}_{s,t,x,y}(q_{1:n},
\mathcal{W})\nonumber \\
&&\hspace*{-10pt}\quad= \mathbb{P} \bigl(W \in[\ell{\downarrow},\upsilon{\uparrow}]|
\mathcal{W} \bigr) - \mathbb{P} \bigl(W \in[\ell{\uparrow },\upsilon {\uparrow}]|
\mathcal{W} \bigr)\nonumber
\\
&&\hspace*{-10pt}\qquad{}- \mathbb{P} \bigl(W \in[\ell{\downarrow},\upsilon {
\downarrow}]|\mathcal{W} \bigr) + \mathbb{P} \bigl(W \in[\ell {\uparrow},
\upsilon{\downarrow}]|\mathcal{W} \bigr)
\\
&&\hspace*{-10pt}\quad =  \prod^n_{i=0} \mathbb{P}
\bigl(W_{[q_i,q_{i+1}]} \in [\ell {\downarrow},\upsilon{\uparrow}]|W_{q_i},W_{q_{i+1}}
\bigr) - \prod^n_{i=0} \mathbb{P}
\bigl(W_{[q_i,q_{i+1}]} \in [\ell {\uparrow},\upsilon{\uparrow}]|W_{q_i},W_{q_{i+1}}
\bigr)
\nonumber
\\[-8pt]
\\[-8pt]
\nonumber
&&\hspace*{-10pt}\qquad{}- \prod^n_{i=0} \mathbb{P}
\bigl(W_{[q_i,q_{i+1}]} \in [\ell{\downarrow},\upsilon{\downarrow}]|W_{q_i},W_{q_{i+1}}
\bigr) + \prod^n_{i=0} \mathbb{P}
\bigl(W_{[q_i,q_{i+1}]} \in [\ell {\uparrow},\upsilon{\downarrow}]|W_{q_i},W_{q_{i+1}}
\bigr).\hspace*{12pt}
\end{eqnarray}
\upqed\end{pf}

\begin{figure}[t]
\begin{tabular}{@{}cc@{}}

\includegraphics{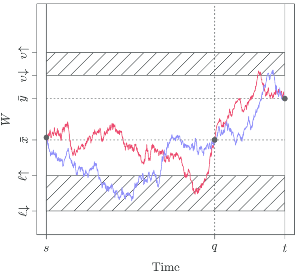}
 & \includegraphics{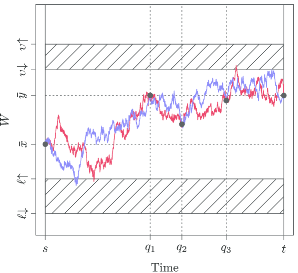}\\
{\fontsize{9}{11}\selectfont{(a) $n=1$}} %\label{figrho1}
& {\fontsize{9}{11}\selectfont{(b)   $n=3$}}
\end{tabular}
\caption{Example sample path trajectories $W\sim \mathbb{W}^{x,y}_{s,t}|(\hat{m}\in[\ell{\downarrow},\ell{\uparrow
}],\check{m}\in
[\upsilon{\downarrow},\upsilon{\uparrow}],q_{1:n},\mathcal
{W})$.}\label{figrhoimg}
\end{figure}

Intuitively ${}^{(n)}\rho^{\ell{\downarrow},\ell
{\uparrow},\upsilon{\downarrow},\upsilon{\uparrow
}}_{s,t,x,y}(q_{1:n},\mathcal{W})$ corresponds to the proportion of
Brownian bridge sample paths with the restriction $\mathcal{W}$, which
also have the restriction $ \{\hat{m}_{s,t} \in[\ell{\downarrow
},\ell
{\uparrow}]$, $\check{m}_{s,t} \in[\upsilon{\downarrow},\upsilon
{\uparrow
}] \}$ (paths of which are illustrated in Figure~\ref{figrhoimg}).
%
%
%f9 #&#

%co6 #&#
\begin{corollary} \label{corrhon}
Events of probability ${}^{(n)}\rho$ can be simulated by
retrospective Bernoulli sampling (as per Corollaries \ref{corlt},
\ref
{corris} and Algorithm~\ref{algris}), noting that ${}^{(n)}\rho$ is
a function of $\gamma$ probabilities, using the following sequence,
%
%e62 #&#
\begin{eqnarray}
S^{\rho(n)}_k & := & \Biggl[\prod
^n_{i=0} S^{\gamma(q_i,q_{i+1},\downarrow,\uparrow
)}_k \Biggr]
- \Biggl[\prod^n_{i=0}
S^{\gamma(q_i,q_{i+1},\uparrow
,\uparrow)}_{k+1} \Biggr]
\nonumber
\\[-8pt]
\\[-8pt]
\nonumber
&&{}- \Biggl[\prod^n_{i=0}
S^{\gamma
(q_i,q_{i+1},\downarrow
,\downarrow)}_{k+1} \Biggr] + \Biggl[\prod
^n_{i=0} S^{\gamma
(q_i,q_{i+1},\uparrow,\downarrow)}_k \Biggr].
\label{eqrho1de}
\end{eqnarray}
\end{corollary}

% Definition - Rho (1) = Rho
%

%de5 #&#
\begin{defn} \label{dfnrhon}
We define $\rho(s,q,t,x,w,y,\ell{\downarrow},\ell{\uparrow
},\upsilon
{\downarrow},\upsilon{\uparrow}):={}^{(1)}\rho^{\ell
{\downarrow
},\ell{\uparrow},\upsilon{\downarrow},\upsilon{\uparrow
}}_{s,t,x,y}( \{q \}, \{w \})$, which coincides with
$\rho$ in \cite{BBPR12}.
\end{defn}

% Theorem - Beta (n)
%
%
%th7 #&#
\begin{theorem} \label{thmbetan}
The probability that a Brownian bridge sample path $W\sim\mathbb{W}^{x,y}_{s,t} |\mathcal{W}$, has in the sub-intervals between
successive points in $\mathcal{W}$, a minimum and maximum in particular
intervals, $\mathcal{L}$~and $\mathcal{U}$ respectively, can be
represented as an infinite series,
%
%e63 #&#
\begin{equation}
{}^{(n)}\beta^{\mathcal{L},\mathcal
{U}}_{s,t,x,y}(q_{1:n},
\mathcal{W}) := \mathbb{P} (\mathcal {L},\mathcal{U}|\mathcal{W} ) = \prod
^n_{i=0} \bigl[\gamma ^{\ell{\downarrow},\upsilon{\uparrow}}_{q_i,q_{i+1}}
- \gamma ^{\ell
{\uparrow},\upsilon{\uparrow}}_{q_i,q_{i+1}} - \gamma^{\ell
{\downarrow
},\upsilon{\downarrow}}_{q_i,q_{i+1}}
+ \gamma^{\ell{\uparrow
},\upsilon
{\downarrow}}_{q_i,q_{i+1}} \bigr].
\end{equation}
\end{theorem}

\begin{pf}
Follows by the strong Markov property for diffusions and sample path
inclusion--exclusion,
%
%e64 #&#
%e65 #&#
%e66 #&#
\begin{eqnarray}
{}^{(n)}\beta^{\mathcal{L},\mathcal
{U}}_{s,t,x,y}(q_{1:n},
\mathcal{W}) & =&  \prod^n_{i=0} \bigl[
\mathbb{P} \bigl(\hat{m}_{q_i,q_{i+1}} \in \bigl[\ell ^{\downarrow}_{q_i,q_{i+1}},
\ell^{\uparrow}_{q_i,q_{i+1}} \bigr],
\nonumber
\\[-8pt]
\\[-8pt]
\nonumber
&&\hspace*{9pt}\qquad{}\check{m} _{q_i,q_{i+1}} \in \bigl[
\upsilon^{\downarrow
}_{q_i,q_{i+1}},\upsilon ^{\uparrow}_{q_i,q_{i+1}}
\bigr]
%\vphantom{ |
%W_{q_i}=w_i,W_{q_{i+1}}=w_{i+1} ) ]}
% \vphantom{= \prod^n_{i=0} [\mathbb{P} (\hat{m}
%_{q_i,q_{i+1}} \in [\ell^{\downarrow}_{q_i,q_{i+1}},\ell
%^{\uparrow
%}_{q_i,q_{i+1}} ], \check{m}_{q_i,q_{i+1}} \in [\upsilon
%^{\downarrow
%}_{q_i,q_{i+1}},\upsilon^{\uparrow}_{q_i,q_{i+1}} ]}
|W_{q_i}=w_i,W_{q_{i+1}}=w_{i+1}
\bigr) \bigr]\quad
\\
& =&  \prod^n_{i=0} \bigl[\mathbb{P}
\bigl(W_{[q_i,q_{i+1}]} \in \bigl[\ell ^{\downarrow}_{q_i,q_{i+1}},
\upsilon^{\uparrow}_{q_i,q_{i+1}} \bigr]|W_{q_i},W_{q_{i+1}}
\bigr)
%\vphantom{- \mathbb{P}
%(W_{[q_i,q_{i+1}]} \in [\ell^{\uparrow}_{q_i,q_{i+1}},\upsilon
%^{\uparrow}_{q_i,q_{i+1}} ]|W_{q_i},W_{q_{i+1}} ) -
%\mathbb{P} (W_{[q_i,q_{i+1}]} \in [\ell^{\downarrow
%}_{q_i,q_{i+1}},\upsilon^{\downarrow}_{q_i,q_{i+1}} ]|
%W_{q_i},W_{q_{i+1}} ) + \mathbb{P} (W_{[q_i,q_{i+1}]} \in
% [\ell^{\uparrow}_{q_i,q_{i+1}},\upsilon^{\downarrow
%}_{q_i,q_{i+1}} ]|W_{q_i},W_{q_{i+1}} )  ]}
\nonumber
\\
&&
%\vphantom{= \prod^n_{i=0} [\mathbb{P}
%(W_{[q_i,q_{i+1}]} \in [\ell^{\downarrow
%}_{q_i,q_{i+1}},\upsilon
%^{\uparrow}_{q_i,q_{i+1}} ]|W_{q_i},W_{q_{i+1}} )}
\hspace*{16pt}{}- \mathbb{P} \bigl(W_{[q_i,q_{i+1}]} \in \bigl[
\ell^{\uparrow
}_{q_i,q_{i+1}},\upsilon^{\uparrow}_{q_i,q_{i+1}}
\bigr]| W_{q_i},W_{q_{i+1}} \bigr)
%\vphantom{- \mathbb{P}
%(W_{[q_i,q_{i+1}]} \in [\ell^{\downarrow
%}_{q_i,q_{i+1}},\upsilon
%^{\downarrow}_{q_i,q_{i+1}} ]|W_{q_i},W_{q_{i+1}} ) +
%\mathbb{P} (W_{[q_i,q_{i+1}]} \in [\ell^{\uparrow
%}_{q_i,q_{i+1}},\upsilon^{\downarrow}_{q_i,q_{i+1}} ]|
%W_{q_i},W_{q_{i+1}} )  ]}
\nonumber
\\[-8pt]
\\[-8pt]
\nonumber
%\vphantom{= \prod^n_{i=0} [\mathbb{P}
%(W_{[q_i,q_{i+1}]} \in [\ell^{\downarrow
%}_{q_i,q_{i+1}},\upsilon
%^{\uparrow}_{q_i,q_{i+1}} ]|W_{q_i},W_{q_{i+1}} ) -
%\mathbb{P} (W_{[q_i,q_{i+1}]} \in [\ell^{\uparrow
%}_{q_i,q_{i+1}},\upsilon^{\uparrow}_{q_i,q_{i+1}} ]|
%W_{q_i},W_{q_{i+1}} )}
&&\hspace*{16pt}{}- \mathbb{P} \bigl(W_{[q_i,q_{i+1}]} \in \bigl[
\ell^{\downarrow}_{q_i,q_{i+1}},\upsilon^{\downarrow
}_{q_i,q_{i+1}}
\bigr]|W_{q_i},W_{q_{i+1}} \bigr)
%\vphantom{+
%\mathbb{P} (W_{[q_i,q_{i+1}]} \in [\ell^{\uparrow
%}_{q_i,q_{i+1}},\upsilon^{\downarrow}_{q_i,q_{i+1}} ]|
%W_{q_i},W_{q_{i+1}} )  ]}
\\
%\vphantom{= \vphantom{\prod^n_{i=0}}
%[\mathbb{P} (W_{[q_i,q_{i+1}]} \in [\ell^{\downarrow
%}_{q_i,q_{i+1}},\upsilon^{\uparrow}_{q_i,q_{i+1}} ]|
%W_{q_i},W_{q_{i+1}} ) - \mathbb{P} (W_{[q_i,q_{i+1}]} \in
% [\ell^{\uparrow}_{q_i,q_{i+1}},\upsilon^{\uparrow
%}_{q_i,q_{i+1}} ]|W_{q_i},W_{q_{i+1}} ) - \mathbb{P} (W_{[q_i,q_{i+1}]} \in [\ell^{\downarrow
%}_{q_i,q_{i+1}},\upsilon^{\downarrow}_{q_i,q_{i+1}} ]|
%W_{q_i},W_{q_{i+1}} )}
&&\hspace*{16pt}{}+ \mathbb{P} \bigl(W_{[q_i,q_{i+1}]} \in \bigl[
\ell^{\uparrow}_{q_i,q_{i+1}},\upsilon^{\downarrow
}_{q_i,q_{i+1}}
\bigr]|W_{q_i},W_{q_{i+1}} \bigr) \bigr].\nonumber
\end{eqnarray}
\upqed\end{pf}

\begin{figure}[b]
\begin{tabular}{@{}cc@{}}

\includegraphics{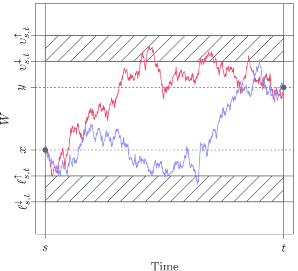}
 & \includegraphics{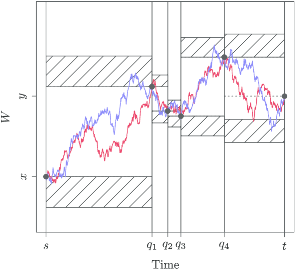}\\
{\fontsize{9}{11}\selectfont{(a) $n=0$}} &
{\fontsize{9}{11}\selectfont{(b) $n=4$}}
\end{tabular}
\caption{Example sample path trajectories $W\sim \mathbb{W}^{x,y}_{s,t}|(\mathcal{L},\mathcal{U},q_{1:n},\mathcal{W})$.}
\label{figbetaimg}
\end{figure}

As before, intuitively ${}^{(n)}\beta^{\mathcal
{L},\mathcal{U}}_{s,t,x,y}(q_{1:n},\mathcal{W})$ corresponds to the
proportion of Brownian bridge sample paths with the restriction
$\mathcal{W}$, which also have the restriction $ \{\mathcal{L},\mathcal{U} \}$ (paths of which are illustrated in
Figure~\ref{figbetaimg}).
%
%f10 #&#

%co7 #&#
\begin{corollary}\label{corbetan}
Events of probability ${}^{(n)}\beta$ can be simulated by
retrospective Bernoulli sampling (as per Corollaries \ref{corlt},
\ref
{corris} and Algorithm~\ref{algris}), noting that ${}^{(n)}\beta$ is
a function of $\gamma$ probabilities, using the following sequence,
%
%e67 #&#
\begin{equation}
S^{\beta(n)}_k  := \prod^n_{i=0}
\bigl[S^{\gamma(q_i,q_{i+1},\ell{\downarrow
},\upsilon
{\uparrow})}_k - S^{\gamma(q_i,q_{i+1},\ell{\uparrow},\upsilon
{\uparrow})}_{k+1} -
S^{\gamma(q_i,q_{i+1},\ell{\downarrow
},\upsilon
{\downarrow})}_{k+1} + S^{\gamma(q_i,q_{i+1},\ell{\uparrow
},\upsilon
{\downarrow})}_{k} \bigr].
\end{equation}
\end{corollary}

% Definition - Beta (0) = Beta
%
%
%de6 #&#
\begin{defn} \label{dfnbetan}
We define $\beta(s,t,x,y,\ell{\downarrow},\ell{\uparrow},\upsilon
{\downarrow},\upsilon{\uparrow}):={}^{(0)}\beta^{\mathcal
{L},\mathcal{U}}_{s,t,x,y}(\varnothing,\varnothing)$, which coincides with
$\beta$ in \cite{BBPR12} (where here we have $\mathcal{L}= \{
\hat{m}
_{s,t}\in[\ell{\downarrow},\ell{\uparrow}] \}$ and $\mathcal
{U}:= \{\check{m}_{s,t}\in[\upsilon{\downarrow},\upsilon
{\uparrow}]
\}$).
\end{defn}

%s7 #&#
\section{Layered Brownian Bridge Constructions} \label{slbb}
In this section, we outline how to construct and simulate finite
dimensional skeletons of layered Brownian bridges for use within the
Unbounded Exact Algorithm (UEA) (Algorithm~\ref{alguea}), which is
in turn used within the Bounded Jump Exact Algorithm (BJEA)
(Algorithm~\ref{algbjea}) and Unbounded Jump Exact Algorithm (UJEA)
(Algorithm~\ref{algujea}). In particular, we
address the simulation of layer information (Algorithm~\ref{alguea}
Step~2), intermediate skeletal points
(Algorithm~\ref{alguea} Step~4) and the process at
further times after
acceptance of the proposed sample path (Algorithm~\ref{alguea}
Step~6).

We present two alternate layered Brownian bridge constructions based on
extensions to existing exact algorithms. In Section~\ref{sea3}, we present
the \textit{Bessel Approach}, which is a reinterpretation of part of
the \textit{Exact Algorithm} 3 (EA3) proposed in \cite{MCAPBPR08}, in
which we incorporate the methodological improvements outlined in
Sections~\ref{sea} and \ref{sbpss} and introduce a novel approach for
conducting Algorithm~\ref{alguea} Step~6 (which
could not previously
be achieved). As a consequence, the resulting (complete) UEA, with
the inclusion of the Bessel approach, satisfies Principles \ref
{prinlayer}, \ref{prinprop} and \ref{prinrest} (as opposed to only
Principles \ref{prinlayer} and \ref{prinprop} in EA3 \cite{MCAPBPR08}). Finally, in Section~\ref{sloc} we briefly outline a
\textit
{Localised Approach} for constructing a layered Brownian bridge (based
on the \textit{Localised Exact Algorithm} (LEA) \cite
{MORCH13,ORGS12}), showing that the resulting UEA only satisfies
Principles \ref{prinlayer} and \ref{prinprop} and discussing the
difficulties in conducting Algorithm~\ref{alguea} Step~6 and
satisfying Principle~\ref{prinrest}.

In neither the Bessel nor the Localised approaches is it possible to
directly simulate intermediate points conditional on a simulated layer
(as required in Algorithm~\ref{alguea} Step~2).
Instead, in order
to simulate proposal sample path skeletons we employ other Monte Carlo
techniques, including rejection sampling (see Section~\ref{sea}) and
demarginalisation.

\textit{Demarginalisation} \cite{BKMCSM}, Section~5.3,  is a technique
whereby artificial extension of a density (with the incorporation of
auxiliary variables) simplifies sampling from it. To illustrate this,
consider the case where we want to draw a sample $g(X)$, but this is
not directly possible. However, suppose that with the introduction of
an auxiliary variable $Y$, sampling from $g(Y)$ and $g(X|Y)$ is
possible and $g(X,Y)$ admits $g(X)$ as a marginal,
%
%e68 #&#
\begin{equation}\label{eqdemarg}
g (X ) = \int_{\mathcal{Y}} g (X|Y )g (Y )\, \mathrm{d}Y.
\end{equation}
We can sample from $g(X)$ by first sampling $Y$ from $g(Y)$ and then
sampling from $g(X|Y)$. This algorithm can be viewed as a black box to
generate samples from $g(X)$ -- $Y$ can be simply marginalised out
(i.e.,  ``thrown'' away). Considering demarginalisation in the context of
the exact algorithms, we can simulate any (auxiliary) aspect of the
proposal diffusion sample path in addition to the skeleton to aid
sampling. Provided the auxiliary information does not influence the
acceptance probability then it is not part of the skeleton and doesn't
need to be retained.

%s7.1 #&#
\subsection{Bessel Approach} \label{sea3}
The central idea of the \textit{Bessel Approach} is that finite
dimensional subsets of Brownian bridge sample paths can be simulated
jointly with information regarding the interval in which they are
constrained (Algorithm~\ref{alguea} Step~2), by
means of a
partitioning of Brownian bridge path space with an (arbitrary)
increasing sequence, $\{a_\iota\}_{\iota\geq0}$, $a_0 =0$, which
radiates outwards from the interval $ [(x  \wedge  y),(x  \vee
y) ]$, demarcating layers (recalling Definition~\ref{defnlayer}).
We term
this particular layer construction the \textit{Bessel layer}. For
instance, the $\iota$th Bessel layer is defined as follows,
%
%e69 #&#
\begin{equation}
\mathcal{I}_\iota  = \bigl[(x \wedge y)-a_\iota,(x \vee
y)+a_\iota \bigr].
\end{equation}
The (smallest) Bessel layer, $\mathcal{I}=\iota$, in which a particular
Brownian bridge sample path is constrained can be simulated by
retrospective Bernoulli sampling and inversion sampling \cite{BKNURVG}, Section~2.1,  as detailed in Algorithm~\ref{algbbl} (where we denote by
$S^\gamma_k(s,t,x,y,\ell,\upsilon)$ as the alternating Cauchy sequence
whose limit as $k\to\infty$ is $\gamma^{\ell,\upsilon}_{s,t}(x,y)$).
The CDF of $\iota$ can be written as follows (with reference to
Theorem~\ref{thmgamma} and as shown in \cite{MCAPBPR08}),
%
%e70 #&#
\begin{equation}
\mathbb{P}(\mathcal{I}\leq\iota) = \mathbb{P} \bigl(W^{x,y}_{s,t}
\in \bigl[(x \wedge y)-a_\iota,(x \vee y)+a_\iota \bigr]
\bigr) = \gamma ^{(x\wedge y)-a_\iota,(x \vee y)+a_\iota}_{s,t}(x,y).
\end{equation}
\begin{algorithm}[b]
\caption{Simulation of a Brownian Bridge Bessel Layer \cite
{MCAPBPR08}}\label{algbbl}
\begin{enumerate}[3.]
\item[1.] Simulate $u\sim\mathrm{U}[0,1]$ and set $\iota=1$, $k=0$.
\item[2.] While $u \in (S^{\gamma}_{2k+1} (s,t,x,y,(x \wedge y)
- a_{\iota},(x \vee y) + a_{\iota} ),S^{\gamma}_{2k}
(s,t,x,y,(x \wedge y) - a_{\iota},(x \vee y) + a_{\iota}
) )$, $k=k+1$. \label{algbblloop}
\item[3.] If $u \geq S^\gamma_{2k}$ set $\iota= \iota+1$ and return to
Step~2 else set $\mathcal{I}=\iota$ and end.
\end{enumerate}
\end{algorithm}

Now, we require a method of simulating intermediate points
(Algorithm~\ref{alguea} Step~4) from a Brownian
bridge sample path restricted
to remain in the Bessel layer simulated in Algorithm~\ref{algbbl}. In
particular, denoting with $D_\iota$ the set of sample paths which are
contained in the $\iota$th Bessel layer we have,
%
%e71 #&#
\begin{equation}\label{eqdi}
D_\iota  = L_\iota\cup U_\iota,
\end{equation}
where
%
%e72 #&#
%e73 #&#
\begin{eqnarray}
L_\iota & :=&  \bigl\{W_{[s,t]}\dvt \hat{m}_{s,t} \in
\bigl[(x \wedge y) - a_{\iota}, (x \wedge y) -a_{\iota-1} \bigr) \bigr\}
\nonumber
\\[-8pt]
\label{eqdili}
\\[-8pt]
\nonumber
&&{}\cap \bigl\{ W_{[s,t]}\dvt \check{m} _{s,t} \in
\bigl[(x \vee y), (x \vee y) + a_{\iota} \bigr] \bigr\},
\\
U_\iota & := & \bigl\{W_{[s,t]}\dvt \hat{m}_{s,t} \in
\bigl[(x \wedge y) - a_{\iota
}, (x \wedge y) \bigr] \bigr\}
\nonumber
\\[-8pt]
\label{eqdiui}
\\[-8pt]
\nonumber
&&{}\cap \bigl\{ W_{[s,t]} \dvt  \check{m} _{s,t} \in \bigl((x
\vee y) + a_{\iota-1}, (x \vee y) + a_{\iota
} \bigr] \bigr\}.
\end{eqnarray}
Directly simulating intermediate points from a sample path
restricted to  $D_\iota$ (denoted $\mathbb{D}_{\iota}$) is not possible. Instead
(as proposed in \cite{MCAPBPR08}) we can propose sample paths from the
mixture measure $\mathbb{B}_{\iota} := \hat{\mathbb{M}}_{\iota
}/2 +
\check{\mathbb{M}}_{\iota}/2$ ($\hat{\mathbb{M}}_{\iota}$ and
$\check{\mathbb{M}}_{\iota}$ being the law induced by the restriction of
$\mathbb{W}^{x,y}_{s,t}$ to the sets $\hat{M}_\iota$ and $\check{M}_\iota$,
resp.)
and accept them with probability given by the Radon--Nikod{\'y}m
derivative of $\mathbb{D}_{\iota}$ with respect to $\mathbb{B}_{\iota}$, where
%
%e74 #&#
%e75 #&#
\begin{eqnarray}\label{eqhatMorvM1}
\hat{M}_\iota &=&  \bigl\{W_{[s,t]} \dvt  \hat{m}_{s,t} \in
\bigl[(x \wedge y) - a_{\iota
}, (x \wedge y) - a_{\iota-1} \bigr]
\bigr\},
\\
\label{eqhatMorvM2}
\check{M}_\iota &=&  \bigl\{W_{[s,t]}\dvt \check{m}_{s,t}
\in \bigl[(x \vee y) + a_{\iota
-1}, (x \vee y) + a_{\iota} \bigr]
\bigr\}.
\end{eqnarray}
It was shown in \cite{MCAPBPR08} that $\mathbb{D}_{\iota}$ is
absolutely continuous with respect to $\mathbb{B}_{\iota}$ with
Radon--Nikod{\'y}m derivative,
%
%e76 #&#
\begin{equation} \label{eqrnbbb}
\frac{ \mathrm{d}\mathbb{D}_{\iota}}{ \mathrm{d}\mathbb{B}_{\iota}}(x)  \propto\frac{\mathbh{1}(W\in D_\iota)}{1 + \mathbh{1}(W\in
\hat{M}
_\iota\cap\check{M}_\iota)}.
\end{equation}
Sample paths can be drawn from $\mathbb{D}_{\iota}$ by proposing them
from $\mathbb{B}_{\iota} := \hat{\mathbb{M}}_{\iota}/2 +
\check{\mathbb{M}}_{\iota}/2$ and then accepting them with probability given by (\ref
{eqrnbbb}). For instance, with probability $1/2$ we sample from~$\hat{\mathbb{M}}_{\iota}$ and accept with probability $1$ if the sample
path maximum is contained within the $(\iota-1)$th Bessel
layer or with probability $1/2$ if it is contained between the $(\iota
-1)$th and $\iota$th Bessel layer (and with
probability $0$ otherwise). In practice we first simulate the sample
path minimum $X_\tau=\hat{m}_{s,t}$ (or maximum $X_\tau=\check{m}_{s,t}$) as per
Algorithm~\ref{algris}, and subsequently simulate any required intermediate
points $\xi_1,\ldots,\xi_\kappa$ from a Bessel bridge as per
Algorithm~\ref{algbsbm}. As we can only simulate our sample path at a finite
collection of points we can't directly evaluate (\ref{eqrnbbb}).
However, we can obtain unbiased estimate and so simulate an event of
this probability by application of Corollaries \ref{thmdelta1} and
\ref
{thmdelta2} and Lemmata \ref{cordelta1} and \ref{cordelta2} (letting
$\chi_1,\ldots,\chi_{\kappa+3}$ be the order statistics of $
\{\xi
_1,\ldots,\xi_\kappa,s,\tau,t \}$),
%
%e77 #&#
%e78 #&#
\begin{eqnarray}
\mathbb{P}_{\hat{\mathbb{M}}_{\iota}} (X\in D_\iota |X_{\chi
_1},
\ldots,X_{\chi_{\kappa+3}} ) & =& \mathbb{P} \bigl(X\in\bigl[(x \wedge
y)-a_{\iota},(x \vee y)+a_{\iota
}\bigr]|X_{\chi_1},
\ldots,X_{\chi_{\kappa+3}} \bigr)\quad
\nonumber
\\[-8pt]
\label{eqPMinDi}
\\[-8pt]
\nonumber
& = & \prod^{\kappa+2}_{i=1}
\delta^{\hat{m},(x\vee y)+a_{\iota
}}_{\chi
_{i},\chi_{i+1}} (X_{\chi_{i}},X_{\chi_{i+1}} ),
\\
\hspace*{-10pt}\mathbb{P}_{\hat{\mathbb{M}}_{\iota}} (X\in\hat{M}_\iota \cap
\check{M}_\iota |X_{\chi_1},\ldots,X_{\chi_{\kappa+3}} ) & =&
\mathbb{P}_{\hat{\mathbb{M}}_{\iota}} (X\in D_\iota| X_{\chi_1},
\ldots,X_{\chi_{\kappa+3}} )
\nonumber
\\[-8pt]
\label{eqPMinvMM}
\\[-8pt]
\nonumber
&&{}- \prod^{\kappa+2}_{i=1}
\delta^{\hat
{m},(x\vee
y)+a_{\iota-1}}_{\chi_{i},\chi_{i+1}} (X_{\chi_{i}},X_{\chi
_{i+1}} ).
\end{eqnarray}
As both (\ref{eqPMinDi}) and (\ref{eqPMinvMM}) are probabilities
which can be represented as a function of $\delta$ probabilities,
events of this probability can be simulated by retrospective Bernoulli
sampling (as per Corollaries \ref{corlt}, \ref{corris} and
Algorithm~\ref{algris}). The synthesis of the above approach for
simulating a
Brownian bridge conditional on the Bessel layer simulated in
Algorithm~\ref{algbbl} (i.e., conducting Algorithm~\ref{alguea}
Step~4) leads to
Algorithm~\ref{algbessel}.
\begin{algorithm}[t]
\caption{Layered Brownian Bridge Simulation (Bessel Approach) --
Sampling $X$ at times $\xi_1,\ldots,\xi_\kappa$} \label{algbessel}
\begin{enumerate}[5.]
\item[1.] Simulate $u_1,u_2\sim\mathrm{U}[0,1]$, set $j=k=0$. \label
{algbesselstart}
\item[2.] Simulate Auxiliary Information (conditional on $I=\iota$),
\label{algbesselaux}
\begin{enumerate}[(a)]
\item[(a)] If $u_1\leq1/2$ simulate minimum point $(\tau,\hat{m}_{s,t})$
and set
$\ell_1=\ell_2=\hat{m}_{s,t}$, $\upsilon_1=(x \vee y) + a_{\iota
-1}$ and
$\upsilon_2=(x \vee y) + a_{\iota}$.
\item[(b)] If $u_1 >1/2$ simulate maximum $(\tau,\check{m}_{s,t})$ and set
$\ell
_1=(x \wedge y) - a_{\iota-1}$, $\ell_2=(x \wedge y) - a_{\iota-1}$
and $\upsilon_1=\upsilon_2=\check{m}_{s,t}$.
\end{enumerate}
\item[3.] Simulate intermediate times $X_{\xi_1},\ldots,X_{\xi_\kappa}$
from a Bessel Bridge conditional on $X_\tau$.
\item[4.] While $u_2 \in (\prod^{\kappa+2}_{i=1}S^{\delta
}_{2j+1}
(\ell_1,\upsilon_1 ),\prod^{\kappa+2}_{i=1}S^{\delta}_{2j}
(\ell
_1,\upsilon_1 ) )$, $j=j+1$,
\begin{enumerate}[(a)]
\item[(a)] If $u_2 \leq\prod^{\kappa+2}_{i=1}S^{\delta}_{2j+1} (\ell
_1,\upsilon_1 )$, then accept sample path.
\item[(b)] If $u_2 \geq\prod^{\kappa+2}_{i=1}S^{\delta}_{2j} (\ell
_1,\upsilon_1 )$ while $u_2 \in (\prod^{\kappa
+2}_{i=1}S^{\delta
}_{2k+1} (\ell_2,\upsilon_2 ),\prod^{\kappa
+2}_{i=1}S^{\delta
}_{2k} (\ell_2,\upsilon_2 ) )$, $k= k+1$,
\begin{enumerate}[ii.]
\item[i.] If $u_2 \leq\prod^{\kappa+2}_{i=1}S^{\delta}_{2k+1} (\ell
_2,\upsilon_2 )$, then with probability $1/2$ accept sample path,
else return to Step~1.
\item[ii.] If $u_2 \geq\prod^{\kappa+2}_{i=1}S^{\delta}_{2k} (\ell
_2,\upsilon_2 )$, then reject sample path and return to Step~1.
\end{enumerate}
\end{enumerate}
\item[5.] Discard or Retain Auxiliary Information.
\end{enumerate}
\end{algorithm}

Upon accepting a proposed sample path skeleton within the UEA (as
simulated by Algorithm~\ref{algbbl} and Algorithm~\ref{algbessel}
and so satisfying
Principles \ref{prinlayer} and \ref{prinprop}), we need to be able to
simulate the sample path at further times (Algorithm~\ref{alguea}
Step~2) in order to satisfy Principle~\ref
{prinrest}. Any
further simulation is conditional on information obtained constructing
the sample path skeleton. In particular, our sample path belongs to
$D_\iota$ (by Algorithm~\ref{algbbl}), the sample path minimum (or maximum)
belongs to a particular interval (w.p. $1$, as a consequence of the
mixture proposal in (\ref{eqhatMorvM1}), (\ref{eqhatMorvM2})), we
have simulated the sample path minimum (or maximum) (either $X_\tau
=\hat{m}
_{s,t}$ or $X_\tau=\check{m}_{s,t}$ by Algorithm~\ref{algris}) and
skeletal points
($X_{\xi_1},\ldots,X_{\xi_\kappa}$) and finally we have simulated
whether the sample path maximum (or minimum) is contained in the first
$(\iota-1)$ Bessel layers or in the $\iota$th Bessel layer
(by evaluating the Radon--Nikod{\'y}m derivative in (\ref{eqrnbbb})
by means of (\ref
{eqPMinDi}) and (\ref{eqPMinvMM})). In summary, we have four possible
sets of conditional information for our sample path,
%
%e79 #&#
%e80 #&#
%e81 #&#
%e82 #&#
\begin{eqnarray}
S_1 &:=& \bigl\{X_s,X_t,X\in
D_\iota,\hat{m}_{s,t} \in \bigl[(x \wedge y) -
a_{\iota}, (x \wedge y) - a_{\iota-1} \bigr],X_\tau=\hat
{m}_{s,t},
\nonumber
\\[-8pt]
\\[-8pt]
\nonumber
&&\hspace*{61pt}\qquad{} X_{\xi
_1},\ldots,X_{\xi_\kappa},\check{m}_{s,t} \in
\bigl[(x \vee y), (x \vee y) + a_{\iota-1} \bigr] \bigr\},
\\
S_2 &:=& \bigl\{X_s,X_t,X\in
D_\iota,\hat{m}_{s,t} \in \bigl[(x \wedge y) -
a_{\iota}, (x \wedge y) - a_{\iota-1} \bigr],X_\tau=\hat
{m}_{s,t},
\nonumber
\\[-8pt]
\\[-8pt]
\nonumber
&&\hspace*{42pt}\qquad{}X_{\xi
_1},\ldots,X_{\xi_\kappa},\check{m}_{s,t} \in
\bigl[(x \vee y) + a_{\iota
-1}, (x \vee y) + a_{\iota} \bigr] \bigr
\},
\end{eqnarray}
\begin{eqnarray}
S_3 &:=&  \bigl\{X_s,X_t,X\in
D_\iota,\check{m}_{s,t} \in \bigl[(x \vee y) +
a_{\iota-1}, (x \vee y) + a_{\iota} \bigr],X_\tau=\check
{m}_{s,t},
\nonumber
\\[-8pt]
\\[-8pt]
\nonumber
&&\hspace*{60pt}\qquad{}X_{\xi
_1},\ldots,X_{\xi_\kappa},\hat{m}_{s,t} \in
\bigl[(x \wedge y) - a_{\iota
-1},(x \wedge y) \bigr] \bigr\},
\\
S_4 &:=& \bigl\{X_s,X_t,X\in
D_\iota,\check{m}_{s,t} \in \bigl[(x \vee y) +
a_{\iota-1}, (x \vee y) + a_{\iota} \bigr],X_\tau=\check
{m}_{s,t},
\nonumber
\\[-8pt]
\\[-8pt]
\nonumber
&&\hspace*{42pt}\qquad{} X_{\xi
_1},\ldots,X_{\xi_\kappa},\hat{m}_{s,t} \in
\bigl[(x \wedge y) - a_{\iota
}, (x \wedge y) - a_{\iota-1} \bigr]
\bigr\}.
\end{eqnarray}
The difficulty in simulating the process at further intermediate times
conditional on the above is that information pertaining to the sample
path minimum and maximum induces a dependency between the sub-interval
in which we want to simulate an intermediate point, and all other
sub-intervals. An additional complication arises as we know precisely
the minimum (or maximum) of the sample path, so the law we need to
simulate further points from is that of a Bessel bridge conditioned to
remain in a given interval.

However, the minimum (or maximum) simulated in Algorithm~\ref
{algbessel} Step~2 is auxiliary sample path
information (as
in (\ref{eqdemarg})) and doesn't constitute an essential part of the
exact algorithm skeleton, so can be discarded. Furthermore, information
regarding the sample path minimum and maximum is sufficient in
determining an interval for the entire sample path. As such,
reconsidering $S_1$ ($S_2,S_3,S_4$ can be similarly considered) we have,
%
%e83 #&#
\begin{eqnarray}
\tilde{S}_1 &:=& \bigl\{X_s,X_t,X_{\xi_1},
\ldots,X_{\xi_\kappa
},\hat{m} _{s,t} \in \bigl[(x \wedge y) -
a_{\iota}, (x \wedge y) - a_{\iota
-1} \bigr],
\nonumber
\\[-8pt]
\\[-8pt]
\nonumber
&&\hspace*{95pt}\qquad\check{m}_{s,t} \in \bigl[(x \vee y), (x \vee y) + a_{\iota
-1}
\bigr] \bigr\}.
\end{eqnarray}
Now, to remove the induced dependency between sub-intervals of time we
can simulate, for each sub-interval of time, an interval of path space
in which the sample path minimum and maximum is constrained as outlined
in Section~\ref{silb} and Algorithm~\ref{algbis}. Further
intermediate points can
then be simulated as outlined in Section~\ref{sMEA}.

%s7.2 #&#
\subsection{Localised Approach} \label{sloc}
The \textit{Localised Approach} is based on the layered Brownian bridge
construction found in the \textit{Localised Exact Algorithm} (LEA)
originally proposed in \cite{MORCH13,ORGS12}. The LEA is a modified
construction of the exact algorithm based on the mathematical framework
of EA3 (see \cite{MCAPBPR08}). We outline the LEA only to highlight to
the reader the aspects of its construction which would lead to
computational challenges if implemented within the context which we
consider in this paper (in particular, significant computation is
required in order to satisfy Principle~\ref{prinrest}).

The key notion in the Localised approach is that rather than proposing
sample path skeletons from $\mathbb{Z}^x_{0,T}$ (where the end point
$X_T=:y\sim h$ is first simulated), the interval to be simulated
($[0,T]$) can be instead broken into a number of \textit{bounded}
segments (as in (\ref{eqdrnd})). Each segment is successively
simulated by means of simulating the first hitting time, $\tau$, of a
Brownian motion proposal sample path (as outlined in \cite{MaCiSBJ08})
of some user specified boundary symmetric around its start point (e.g., if $X_0=x$ with boundary $\theta$ then $\tau:=\inf\{
s:X_s\notin[x-\theta,x+\theta]\}$), and simulating and accepting a
sample path skeleton conditional on the simulated boundary (with a
suitable modification of the acceptance probability to account for the
modified proposal measure).

The benefit of the Localised approach is that simulating the first
hitting time of a boundary acts as a layer for the bounded segment
(i.e., $\forall u\in[0,\tau], X_u(\omega)\in[x-\theta,x+\theta]$)
and so
$\phi(X_{0,\tau})$ is conditionally bounded (as per Result~\ref{rescompact}) and a bound can be found for $A(X_\tau)$ in
(\ref{eqrnd}). As such it is possible to bound the Radon--Nikod{\'y}m
derivative without the need
for Condition~\ref{condphi}, however the acceptance rate of proposal sample
paths can be low as each component of the Radon--Nikod{\'y}m
derivative needs to be bounded
(the incongruity being that this can be particularly problematic in the
case where the diffusion doesn't satisfy Condition~\ref{condphi}). Moreover,
as with the UJEA and Adaptive
Unbounded Jump Exact Algorithm (AUJEA) this approach to simulating sample
path skeletons can result in simulating skeletons for intervals
exceeding that required (which is computationally wasteful), further
complicated by the need to specify the boundary $\theta$. Furthermore
(as discussed in \cite{MCAPGR13}), this methodology can't be used to
simulate \textit{conditioned} diffusion and jump diffusion sample path
skeletons (sample paths conditioned to hit some specified end point),
whereas the methodology developed elsewhere in this paper can be
directly extended to this setting (see \cite{PhDP13}, Chapter~5).
Finally, unlike the Bessel approach, the minimum or maximum that is
simulated forms part of the skeleton and so cannot be discarded. As
such, the demarginalisation strategy taken in Section~\ref{sea3} in order
to extend the UEA with the Bessel
approach for simulating layered
Brownian bridges to satisfy Principle~\ref{prinrest} can't be conducted.

%s8 #&#
\section{Adaptive Layered Brownian Bridge Constructions} \label{snlbb}
In Section~\ref{sauea}, we proposed the Adaptive Unbounded Exact
Algorithm (AUEA) (Algorithm~\ref{algauea}) as an
alternative to the UEA (Algorithm~\ref
{alguea}). In this section, we
outline how to simulate finite dimensional skeletons of layered
Brownian bridges for use within the AUEA (and by extension the BJEA
 (Algorithm~\ref{algbjea}) and  AUJEA (Algorithm~\ref{algaujea})). In
particular, we present new results for simulating an initial \textit
{intersection layer} (Algorithm~\ref{algauea} Step 2 -- Section~\ref{sil}),
intermediate points conditional on the layer (Algorithm~\ref{algauea}
Step~3.1.2 -- Section~\ref{silip}) and finally, new intersection layers
for each
sub-interval created by the intermediate point (Algorithm~\ref
{algauea} Step~3.1.4 -- Section~\ref{silb}).

We use the results we present in Sections~\ref{sil}--\ref{silb} to
outline novel layered Brownian bridge constructions in Section~\ref{sSLBB}
which can used within the AUEA,
all of which satisfy Principles \ref{prinlayer}, \ref{prinprop} and \ref{prinrest}.

%s8.1 #&#
\subsection{Simulating an Initial Intersection Layer} \label{sil}
Upon simulating a proposal Brownian bridge layer as per Algorithm~\ref
{algbbl} in Section~\ref{sea3}, we know that our entire Brownian bridge
sample path is contained within the $\iota$th Bessel layer,
but is not contained within the $(\iota-1)$th Bessel layer.
Simulating sample path intermediate points is complicated by this
conditional information (and as discussed in Section~\ref{slbb}, it
is not
possible to simulate intermediate points directly). The novel approach
we take in this paper is to simulate further layer information
regarding the minimum and maximum of the proposed sample path (which
together provide a sample path layer). To achieve this recall (with
reference to Section~\ref{sea} and (\ref{eqdi}), (\ref{eqdili}),
(\ref{eqdiui})) that, having simulated a layer for our proposal Brownian
bridge sample path as per Algorithm~\ref{algbbl}, we know the sample
path is
restricted to the layer $D_\iota$. We can then simply decompose the set
$D_\iota$ into a disjoint union and simulate to which element our
sample path belongs,
%
%e84 #&#
\begin{equation}\label{eqdidecomp}
D_\iota  = L_\iota\cup U_\iota= \underbrace{
(L_\iota\cap U_\iota )}_{D_{\iota,1}}\, \uplus\,\underbrace{
\bigl(U^{\mathrm{C}}_\iota\cap L_\iota
\bigr)}_{D_{\iota,2}}\,\uplus\,\underbrace{ \bigl(L^{\mathrm{C}}_\iota
\cap U_\iota \bigr)}_{D_{\iota,3}}.
\end{equation}
This decomposition corresponds to the sample path attaining the $\iota$th Bessel layer at both its minimum and maximum ($D_{\iota
,1}$) or its minimum ($D_{\iota,2}$) or its maximum ($D_{\iota,3}$). We
can simulate to which set our sample path belongs by application of the
following results and Algorithm~\ref{algint}. Recalling the
definition of a
layer from Definition~\ref{defnlayer}, we term this particular layer
construction the \textit{Intersection Layer}.
% Theorem - Initial Intersection Layer
%

%th8 #&#
\begin{theorem}[(Initial intersection layer)] \label{thmil}
The probability a Brownian bridge sample path is in $D_{\iota,1}$,
given it is in $D_\iota$, can be represented as follows (denoting
$\ell
{\downarrow}:=(x \wedge y) - a_{\iota}$, $\ell{\uparrow}:=(x
\wedge
y) - a_{\iota-1}$, $\upsilon{\downarrow}:=(x \vee y) + a_{\iota-1}$,
$\upsilon{\uparrow}:=(x \vee y) + a_{\iota}$ and $\tilde{\beta
}
(s,t,x,y ) := \beta (s,t,x,y,\ell{\downarrow},\ell
{\uparrow
},\upsilon{\downarrow},\upsilon{\uparrow} )+ \beta
(s,t,x,y,\ell{\downarrow},\ell{\uparrow},(x \vee y),\upsilon
{\downarrow} )+\beta (s,t,x,y,\ell{\uparrow},(x \wedge
y),\upsilon{\downarrow},\upsilon{\uparrow} )$),
%
%e85 #&#
\begin{equation}
p_{D_{\iota,1}}  := \mathbb{P} (D_{\iota,1} | D_\iota,
W_s = x,W_t = y ) = \frac{\beta (s,t,x,y,\ell{\downarrow},\ell
{\uparrow
},\upsilon{\downarrow},\upsilon{\uparrow} )}{\tilde{\beta
}
(s,t,x,y )}.
\end{equation}
\end{theorem}

\begin{pf}
Follows by Bayes rule, Theorem~\ref{thmbetan} and the decomposition of
$D_\iota$ in (\ref{eqdidecomp}).
\end{pf}

%
%
%co8 #&#
\begin{corollary} \label{coril}
Events of probability $p_{D_{\iota,1}}$ can be simulated by retrospective
Bernoulli sampling (as per Corollaries \ref{corlt}, \ref{corris} and
Algorithm~\ref{algris}), noting that $p_{D_{\iota,1}}$ is a function
of $\beta$
probabilities, defining
\begin{eqnarray*}
\tilde{S}^\beta_{k} (s,t,x,y )&:= & S^\beta_{k}
(s,t,x,y,\ell{\downarrow},\ell{\uparrow},\upsilon {\downarrow},\upsilon{\uparrow}
)+S^\beta_{k} \bigl(s,t,x,y,\ell {\downarrow},\ell{
\uparrow},(x \vee y),\upsilon{\downarrow } \bigr)
\\
&&{}+S^\beta_{k} \bigl(s,t,x,y,\ell{\uparrow},(x \wedge y),
\upsilon {\downarrow},\upsilon{\uparrow} \bigr),
\end{eqnarray*}
and using the following sequence,
%
%e86 #&#
\begin{equation}
S^{D({\iota,1})}_k := \frac{S^\beta_k (s,t,x,y,\ell{\downarrow},\ell{\uparrow
},\upsilon{\downarrow},\upsilon{\uparrow} )}{\tilde{S}^\beta
_{k} (s,t,x,y )}.
\end{equation}
\end{corollary}

Noting that by symmetry we have $p_{D_{\iota,2}}:=\mathbb{P}(D_{\iota,2} | D_\iota, W_s = x,W_t = y) =
\mathbb{P}(D_{\iota
,3} | D_\iota, W_s = x,W_t = y)=:p_{D_{\iota,3}}$ and furthermore
$p_{D_{\iota,2}}+p_{D_{\iota,3}}=1-p_{D_{\iota,1}}$ it is possible to
determine to which disjoint set ($D_{\iota,1}$. $D_{\iota,2}$ or
$D_{\iota,3}$) our sample path belongs by direct application of
Theorem~\ref{thmil}, Corollary~\ref{coril} and the following
Algorithm~\ref{algint}.
\begin{algorithm}[t]
\caption{Simulation of an Initial Brownian Bridge Intersection Layer}
\label{algint}
\begin{enumerate}[4.]
\item[1.] Simulate layer $\mathcal{I}=\iota$ as per Algorithm~\ref{algbbl},
simulate $u \sim\mathrm{U}[0,1]$ and set $k=0$.
\item[2.] While $u \in (S^{D({\iota,1})}_{2k+1},S^{D({\iota
,1})}_{2k} )$, $k=k+1$.
\item[3.] If\vspace*{1pt} $u \leq S^{D({\iota,1})}_{2k+1}$, then set $D_{\iota} =
D_{\iota,1}$.
\item[4.] If $u \geq S^{D({\iota,1})}_{2k}$, then with probability $0.5$
set $D_{\iota} = D_{\iota,2}$ else set $D_{\iota} = D_{\iota,3}$.
\end{enumerate}
\end{algorithm}

%s8.2 #&#
\subsection{Simulating
Intersection Layer Intermediate Points} \label{silip}
Having simulated an intersection layer we require a sampling scheme for
simulating the conditional Brownian bridge at some intermediate time $q
\in(s,t)$. As shown in \cite{BBPR12}, the density of the sample path
at the intermediate time point $q$ can be written as follows (where
$\mu
_w$ and $\sigma^2_w$ denote the mean and variance of a Brownian bridge
as in Section~\ref{ssbb}),
%
%e87 #&#
%e88 #&#
\begin{eqnarray}\label{eqilmptarget}
\pi(w) & :=& \mathbb{P} \bigl(W_q = w|W_s,
W_t, \hat{m}_{s,t}\in\bigl[\ell ^{\downarrow}_{s,t},
\ell^{\uparrow}_{s,t}\bigr], \check{m}_{s,t}\in \bigl[
\upsilon ^{\downarrow}_{s,t},\upsilon^{\uparrow}_{s,t}
\bigr] \bigr)
\\
\label{eqilmp}
&\propto & \rho \bigl(s,q,t,x,w,y,\ell^\downarrow_{s,t},\ell
^\uparrow _{s,t},\upsilon^\downarrow_{s,t},
\upsilon^\uparrow_{s,t} \bigr) \cdot\mathrm{N} \bigl(w;
\mu_w,\sigma^2_w \bigr).
\end{eqnarray}
A method of simulating from $\pi(w)$ was outlined in \cite{BBPR12}
based on inversion sampling and numerical methods, however, this scheme
is formally inexact and given particular parameter values can be
computationally extremely inefficient. We provide a number of
alternative schemes which are exact.

In Section~\ref{silipbcs}, we present a method of simulating from
(\ref{eqilmp}) by finding a bound constructed from a mixture of Normal
densities which can be easily simulated from and conducting rejection
sampling. It transpires that this scheme is typically highly
computationally efficient, however, for a small number of parameter
values the acceptance rate of the rejection sampler is very low. As
such, in Section~\ref{siliplip}, we present an alternate rejection sampling
scheme which exploits the known Lipschitz constants of the bounding
sequence in (\ref{eqilmp}) to construct an arbitrarily tight bound of
the target density. This, however, comes at some computational expense,
so we advocate using some mixture of these two approaches (which we
discuss later in Sections~\ref{silipmix} and \ref{sMEA}). Finally,
for completeness, in Section~\ref{sHEA} we construct a third scheme
inspired by the Bessel layer constructions found in Section~\ref{sea3} and
\cite{MCAPBPR08}. This third scheme provides some insight into how the
different layered Brownian bridge constructions of Section~\ref{slbb} and
Section~\ref{snlbb} relate to one another.

%s8.2.1 #&#
\subsubsection{Bounding Cauchy Sequence Approach} \label{silipbcs}
Here we show that it is possible to extend \cite{BBPR12}, and simulate
from $\pi(w)$ exactly by means of composition sampling (see \cite{BKSS}) and rejection sampling. We will begin by considering the upper
convergent bounding sequence of $\rho(w)$ ($\forall k \in\mathbb{Z}_{\geq0}$ we have $\rho(w) \leq S^\rho_{2k}(w)$ and $\lim_{k\to
\infty}S^\rho_{2k}(w)=\rho(w)$), where for conciseness we additionally
denote $\rho(w):=\rho(s,q,t,x,w,y,\ell^\downarrow_{s,t},\ell
^\uparrow
_{s,t}, \upsilon^\downarrow_{s,t},\upsilon^\uparrow_{s,t})$ (as in this
setting we have $s,q,t,x,y,\ell^\downarrow_{s,t},\ell^\uparrow
_{s,t},\upsilon^\downarrow_{s,t}$ and $\upsilon^\uparrow_{s,t}$ are
fixed). Decomposing~(\ref{eqrho1de}) into its elementary form in terms
of $\bar{\varsigma}$ and $\bar{\varphi}$ (see (\ref{eqvarsigrep}) and
(\ref{eqvarphirep}), resp.) yields $K = 64(k+1)^2$ of these
elementary terms.

Recalling that $\varsigma^{\ell,\upsilon}_{s,t}(j;x,y) := \bar
{\varsigma
}^{\ell,\upsilon}_{s,t}(j;x,y) + \bar{\varsigma}^{-\ell,-\upsilon
}_{s,t}(j;-x,-y)$ and $\varphi^{\ell,\upsilon
}_{s,t}(j;x,y)$
$:= \bar{\varphi}^{\ell,\upsilon}_{s,t}(j;x,y) + \bar{\varphi
}^{-\ell
,-\upsilon}_{s,t}(j;-x,-y)$ it can be shown that each of the functions
$\bar{\varsigma}$ and $\bar{\varphi}$ has the structural form $\exp
(a_i +b_i w )$, with known coefficients $a_i$ and $b_i$ (see the
\hyperref[apxecf]{Appendix} for further details). As such, we can find a
bound for
our unnormalised target density (\ref{eqilmp}) as follows (the $c_i
\in\{0,1\}$ determine the sign of each density contribution),
%
%e89 #&#
%e90 #&#
%e91 #&#
\begin{eqnarray}
\pi(w) &\propto & \rho(w) \cdot\mathrm{N} \bigl(w;\mu_w,
\sigma^2_w \bigr) \leq   S^\rho
_{2k}(w)\cdot\mathrm{N} \bigl(w;\mu_w,
\sigma^2_w \bigr)
\\
& = & \sum^{K}_{i=1} \bigl[
(-1)^{c_i}\cdot\exp \{a_i + b_iw \} \cdot
\mathbh{1} \bigl(w\in[\ell_i,\upsilon_i] \bigr)\cdot
\mathrm{N} \bigl(w;\mu _w,\sigma^2_w
\bigr) \bigr]
\\
& =&  \sum^{K}_{i=1} \bigl[
\underbrace{(-1)^{c_i}\cdot\exp \bigl\{ a_i +
\mu_wb_i + b_i\sigma^2_w/2
\bigr\}}_{\upomega_i:=}
\nonumber
\\[-8pt]
\label{rhodominating}
\\[-8pt]
\nonumber
&&\hspace*{16pt} {}\times\mathbh{1} \bigl(w\in[\ell_i,
\upsilon_i] \bigr)\cdot\underbrace{\mathrm{N} \bigl(w; \mu
_w+b_i\sigma^2_w,
\sigma^2_w \bigr)}_{\mathrm{N} (w; \mu
_i,\sigma^2_w
):=} \bigr].
\end{eqnarray}
Here we have a mixture of positively and negatively weighted truncated
Normal densities (with common variance). Although each truncated Normal
in the mixture is unique (due to the truncation points), a large
proportion of them will have common location parameter. We exploit this
by partitioning the interval that provides support for the target
density (\ref{eqilmp}) into sections corresponding to the truncation
points (in particular, we consider the partitioning $ \{[\ell
^{\downarrow}_{s,t},\ell^{\uparrow}_{s,t}],[\ell^{\uparrow
}_{s,t},\upsilon^{\downarrow}_{s,t}],[\upsilon^{\downarrow
}_{s,t},\upsilon^{\uparrow}_{s,t}] \}$ which we denote by $j\in
\{
1,2,3\}$, resp.). As a consequence, the resulting mixture
density has a number of positive and negative elements which cancel
each other out (i.e., they can be \textit{netted} from one another).
Defining $\omega_{i,j}$ as the weight associated with the $j$th partition of the $i$th truncated Normal density of
(\ref{rhodominating}), and $\upomega^+_{i,j} := (\omega_{i,j}\vee0)$,
we can find an upper bound by solely considering the mixture formed
from the components with positive weights,
%
%e92 #&#
%e93 #&#
%e94 #&#
\begin{eqnarray}
\pi(w) & \leq & \sum^{K}_{i=1}
\upomega_i\cdot\mathrm{N} \bigl(w;\mu _i,
\sigma^2_w \bigr)\cdot\mathbh{1} \bigl(w\in[
\ell_i,\upsilon_i] \bigr)
\nonumber
\\[-8pt]
\label{eqrhodom2}
\\[-6pt]
\nonumber
&&\hspace*{15pt}{}\times\bigl[\mathbh{1} \bigl(w\in
\bigl[\ell^{\downarrow}_{s,t},\ell^{\uparrow
}_{s,t}
\bigr] \bigr)+\mathbh{1} \bigl(w\in\bigl[\ell^{\uparrow}_{s,t},\upsilon
^{\downarrow}_{s,t}\bigr] \bigr)+ \mathbh{1} \bigl(w\in\bigl[
\upsilon ^{\downarrow
}_{s,t},\upsilon^{\uparrow}_{s,t}
\bigr] \bigr) \bigr]
\\
& \leq & \sum^{K}_{i=1} \mathrm{N} \bigl(w;
\mu_i,\sigma^2_w \bigr)\mathbh{1} \bigl(w\in[\ell_i,\upsilon_i] \bigr)
\nonumber
\\[-8pt]
\\[-6pt]
\nonumber
&&\hspace*{14pt}{}\times\bigl[
\upomega^+_{i,1}\mathbh{1} \bigl(w\in\bigl[\ell^{\downarrow}_{s,t},
\ell^{\uparrow}_{s,t}\bigr] \bigr)+\upomega^+_{i,2}\mathbh{1} \bigl(w\in\bigl[\ell^{\uparrow
}_{s,t},
\upsilon ^{\downarrow}_{s,t}\bigr] \bigr)+ \upomega^+_{i,3}
\mathbh{1} \bigl(w\in \bigl[\upsilon^{\downarrow}_{s,t},
\upsilon^{\uparrow}_{s,t}\bigr] \bigr) \bigr]\qquad
\\
\label{rhodominating3}
& =:&  S^{\rho,+}_{2k}(w)\cdot\mathrm{N} \bigl(w;
\mu_i,\sigma^2_w \bigr).
\end{eqnarray}
By application of composition sampling (see \cite{BKSS}) we can
simulate from the probability density proportional to $S^{\rho
,+}_{2k}(w)\cdot\mathrm{N}(w;\mu_w,\sigma^2_w)$ by first choosing
one of the
truncated Normal densities partitioned on the interval $[\mathscr{L},
\mathscr{U}]$ with probability proportional to,
%
%e95 #&#
\begin{equation}
\upomega^+_{i,j}\cdot \bigl[\Phi \bigl(\mathscr{U} |
\mu_i,\sigma ^2_w \bigr) - \Phi \bigl(
\mathscr{L} |\mu_i,\sigma^2_w \bigr)
\bigr].
\end{equation}
As $w\sim S^{\rho,+}_{2k}(w)\cdot\mathrm{N} (w;\mu_w,\sigma
^2_w  ) /
Z_D$ and we require $w\sim\rho(w)\cdot\mathrm{N} (w;\mu
_w,\sigma^2_w
) / Z_T$ (where $Z_T$ and $Z_D$ denote the normalising constants of the
target and dominating densities, resp., noting that the rejection
sampling bound $M=Z_D/Z_T$) we accept this draw with probability,
%
%e96 #&#
\begin{equation} \label{eqintpacp}
P=\frac{\rho(w) \cdot\mathrm{N} (w | \mu_w,\sigma^2_w
)/Z_T}{M\cdot S^{\rho,+}_{2k}(w)\cdot\mathrm{N} (w | \mu
_w,\sigma
^2_w  )/Z_D}  = \frac{\rho(w)}{S^{\rho,+}_{2k}(w)} \leq1.
\end{equation}
Events of probability ${{P}}$ can be simulated by retrospective
Bernoulli sampling (as per Corollaries \ref{corlt}, \ref{corris} and
Algorithm~\ref{algris}), noting that ${{P}}$ is a function of $\rho
(w)$. The
complete rejection sampler is presented in Algorithm~\ref{algintb}.
\begin{algorithm}[t]
\caption{Simulation of Intersection Layer Intermediate Points (Bounded
Cauchy Sequence Approach)} \label{algintb}
\begin{enumerate}[4.]
\item[1.] Simulate $u\sim\mathrm{U}[0,1]$ and set $j=1$.
\item[2.] Simulate $w\sim S^{\rho,+}_{2k}(w)\cdot\mathrm{N} (w;\mu
_w,\sigma
^2_w  )/ Z_D$ for some $k\in\mathbb{Z}_{\geq0}$.
\item[3.] While $u \in (\frac{S^{\rho}_{2j+1}(w)}{S^{\rho
,+}_{2k}(w)},\frac{S^{\rho}_{2j}(w)}{S^{\rho,+}_{2k}(w)} )$, $j=j+1$.
\item[4.] If $u \leq\frac{S^{\rho}_{2j+1}(w)}{S^{\rho,+}_{2k}(w)}$ then
accept else reject.
\end{enumerate}
\end{algorithm}

%s8.2.2 #&#
\subsubsection{Lipschitz Approach} \label{siliplip}
Simulating intermediate points as per Algorithm~\ref{algintb} is (typically)
highly efficient as $S^{\rho,+}_{2}(w)\cdot\mathrm{N}(w;\mu
_w,\sigma^2_w)$
typically tightly bounds $\pi(w)$ (as noted in \cite{BBPR12}). If this
is not the case (which occurs for a small number of parameter
configurations), then sampling from the bounding density with $k>1$
isn't usually effective as $S^{\rho,+}_{2k}(w)$ is only formed by the
positive netted components of $S^{\rho}_{2k}(w)$. In this section we
propose an alternative scheme in which we exploit the known Lipschitz
constants of the bounding sequence in (\ref{eqilmp}) to construct a
tight bound of the target density.

If the rejection sampling scheme proposed in Section~\ref{silip} is not
efficient then this implies that $S^{\rho}_{2}(w)\cdot\mathrm
{N}(w;\mu_w,\sigma
^2_w)$ does not tightly bound $\pi(w)$. In this case, the natural
question to ask is at what level the Cauchy sequence approximation
($S^{\rho}_{2k}(w)$) of $\rho(w)$ needs to be evaluated such that
$S^{\rho}_{2k}(w)\cdot\mathrm{N}(w;\mu_w,\sigma^2_w)$ does form a
tight bound
of $\pi(w)$. To address this we note that in analogous form to
Section~\ref{silipbcs} it is possible to also find a \textit{lower}
bound of the
target density,
%
%e97 #&#
\begin{equation}\label{equpperandlower}
S^\rho_{2k+1}(w)\cdot\mathrm{N} \bigl(w;\mu_w,
\sigma^2_w \bigr) \leq\pi(w) \leq S^\rho_{2k}(w)
\cdot\mathrm{N} \bigl(w;\mu_w,\sigma^2_w
\bigr).
\end{equation}
The lower bound of the target density also has the form of a mixture of
positively and negatively weighted Normal densities with known
parameter values (recall the upper bound comprises $K{\uparrow
}=64(k+1)^2$ terms, similarly the lower bound comprises $K{\downarrow
}=64(k+1)^2 - 48$ terms). As such, the normalising constants of the
upper and lower bounds of the target density can be calculated and this
information used to determine whether the upper bound tightly bounds
the target density. In particular, we advocate evaluating the
alternating Cauchy sequence $S^{\rho}_{k}(w)$ until such time that it
exceeds some user specified threshold,
%
%e98 #&#
\begin{eqnarray}
T_Z & \leq & \frac{Z^\rho_{2k+1}(w)}{Z^\rho_{2k}(w)}
\nonumber
\\[-8pt]
\label{eqnormthresh}
\\[-8pt]
\nonumber
& :=&  \biggl[\int^{\upsilon^{\uparrow}_{s,t}}_{\ell^{\downarrow
}_{s,t}}S^\rho_{2k+1}(w)
\cdot\mathrm{N} \bigl(w;\mu_w,\sigma^2_w
\bigr) \,\mathrm{d} w \biggr] \Big/ \biggl[\int^{\upsilon^{\uparrow}_{s,t}}_{\ell
^{\downarrow
}_{s,t}}S^\rho_{2k}(w)
\cdot\mathrm{N} \bigl(w;\mu_w,\sigma^2_w
\bigr) \,\mathrm{d} w \biggr].
\end{eqnarray}
Upon finding an appropriately tight upper bound, a subset of the
positive and negative Normal densities can be netted from one another
leaving the following bounding density form (as argued in Section~\ref{silip} and shown in (\ref{eqrhodom2})),
%
%e99 #&#
\begin{eqnarray}
\pi(w)  &\leq &   \sum^{K{\uparrow}}_{i=1} \mathrm{N}
\bigl(w;\mu _i,\sigma^2_w \bigr)\cdot
\mathbh{1} \bigl(w\in[\ell_i,\upsilon_i] \bigr)
\nonumber
\\
&&\hspace*{16pt}{}\times \bigl[
\upomega _{i,1}\mathbh{1} \bigl(w\in\bigl[\ell^{\downarrow}_{s,t},
\ell ^{\uparrow
}_{s,t}\bigr] \bigr)
+ \upomega_{i,2}\mathbh{1} \bigl(w\in\bigl[\ell
^{\uparrow}_{s,t},\upsilon^{\downarrow}_{s,t}\bigr]
\bigr)
\nonumber
\\[-8pt]
\label{eqcaucgx}
\\[-8pt]
\nonumber
&& \hspace*{118pt}{}+ \upomega _{i,3}\mathbh{1} \bigl(w\in\bigl[
\upsilon^{\downarrow}_{s,t},\upsilon ^{\uparrow}_{s,t}
\bigr] \bigr) \bigr] \\
&=: & g(w).\nonumber
\end{eqnarray}
For any given interval $[q,r]$ (where $q<r$), it is possible to
explicitly calculate for each of the contributing Normal densities
(e.g., $\mathrm{N}(w;\mu_i,\sigma^2_w)$) the local Lipschitz
constant (we
denote $I:=[\mu_i-\sigma_w,\mu_i+\sigma_w]\cap[q,r]$),
%
%e100 #&#
\begin{eqnarray}
\alpha_i(q,r) & :=&  \sup_{w\in[q,r]} \frac{ \mathrm{d}}{ \mathrm{d}w}
\mathrm {N}\bigl(w;\mu_i,\sigma ^2_w\bigr)
\nonumber
\\
& =&  \mathbh{1} (I\neq\varnothing )\cdot\frac{ \mathrm
{d}}{ \mathrm{d}w} \mathrm{N} \bigl(
\mu_i-\sigma_w;\mu_i,\sigma^2_w
\bigr)
\\
&&{}+ \mathbh{1} (I=\varnothing )\cdot\max \biggl\{\frac
{ \mathrm{d}}{ \mathrm{d}w} \mathrm{N}
\bigl(q;\mu_i,\sigma^2_w\bigr),
\frac{ \mathrm{d}}{ \mathrm{d}w} \mathrm{N}\bigl(r;\mu _i,\sigma^2_w
\bigr) \biggr\}.\nonumber
\end{eqnarray}
As such, it is possible to find for the bounding density ($g(w)$ in
(\ref{eqcaucgx})) the local Lipschitz constant for the interval
$[q,r]$ (where $\alpha$ is set to zero when considering an interval of
zero length),
%
%e101 #&#
\begin{eqnarray}
\sup_{u,v\in[q,r]}\frac{g(u) - g(v)}{u-v} & \leq & \sum
^{K{\uparrow}}_{j=1} \bigl[ |\upomega_{j,1}| \alpha
_j \bigl(q \vee \ell^{\downarrow}_{s,t},r \wedge
\ell^{\uparrow
}_{s,t} \bigr) + |\upomega_{j,2}|
\alpha_j \bigl(q \vee \ell^{\uparrow}_{s,t},r \wedge
\upsilon^{\downarrow}_{s,t} \bigr)
 \nonumber\\
 &&\label{eqlipcon}\hspace*{103pt}\qquad{}+ |\upomega_{j,3}| \alpha_j \bigl(q \vee
\upsilon^{\downarrow}_{s,t},r \wedge \upsilon^{\uparrow
}_{s,t}
\bigr) \bigr]\qquad\quad
\\
\nonumber
&=:&  \beta(q,r),
\end{eqnarray}
and consequently, having evaluated the density at $g(q)$ and $g(r)$, we
can find a bound for the upper bound of the target density for the
interval $[q,r]$ (noting that the line $y=g(q)+\beta t$ intersects the
line $y=g(r)+\beta(r-q)-\beta t$ at $t=[g(r)-g(q)+\beta(r-q)]/2\beta
\in[q,r]$),
%
%e102 #&#
\begin{equation}\label{eqlocalbound}
\sup_{w\in[q,r]}g(w) \leq g(r)+\beta(q,r)\cdot t =
\frac{g(r)+g(q)}{2} + \beta(q,r)\cdot\frac{r-q}{2} =: M(q,r).
\end{equation}
As the support of the target density $\pi(w)$ is contained within the
interval $[\ell^{\downarrow}_{s,t},\upsilon^{\uparrow}_{s,t}]$, if we
construct a suitably fine mesh on this interval (for simplicity, we
assume a mesh of size $N$ with regular interval size $\Delta
:=(\upsilon
^{\uparrow}_{s,t}-\ell^{\downarrow}_{s,t})/N$), we can find a piecewise
uniform bound of this density with which to conduct rejection sampling,
%
%e103 #&#
\begin{equation}
g(w) \leq\sum^N_{i=1} \mathbh{1}
\bigl(w\in\bigl[\ell^{\downarrow
}_{s,t}+(i-1)\Delta,
\ell^{\downarrow}_{s,t}+i\Delta\bigr]\bigr)\cdot M \bigl(\ell
^{\downarrow}_{s,t}+(i-1)\Delta,\ell^{\downarrow}_{s,t}+i
\Delta \bigr).
\end{equation}
As in (\ref{eqnormthresh}), we can calculate the normalising constant
of this bounding density, so we advocate choosing the size of the mesh
to be at least as fine as the following user specified threshold,
%
%e104 #&#
\begin{equation}\label{eqboundthresh}
T_M \leq\frac{Z^\rho_{2k}(w)}{Z^{N}_M(w)} := \frac{Z^\rho
_{2k}(w)}{\sum^N_{i=1} \Delta\cdot M (\ell^{\downarrow
}_{s,t}+(i-1)\Delta,\ell^{\downarrow}_{s,t}+i\Delta )}.
\end{equation}
\begin{algorithm}[b]
\caption{Simulation of Intersection Layer Intermediate Points
(Lipschitz Approach)} \label{alglipapproach}
\begin{enumerate}[7.]
\item[1.] Set $k=0$, $N=0$.
\item[2.] While $T_Z \geq\frac{Z^\rho_{2k+1}(w)}{Z^\rho_{2k}(w)}$, $k=k+1$.
\item[3.] While $T_M \geq\frac{Z^\rho_{2k}(w)}{Z^{N}_M(w)}$, increase $N$.
\item[4.] Simulate mesh interval $i$ with probability $\Delta\cdot M
(\ell^{\downarrow}_{s,t}+(i-1)\Delta,\ell^{\downarrow
}_{s,t}+i\Delta
 )/Z^N_M$.
\item[5.] Simulate $w\sim\mathrm{U} [\ell^{\downarrow
}_{s,t}+(i-1)\Delta,\ell
^{\downarrow}_{s,t}+i\Delta ]$, $u\sim\mathrm{U}
[0,M (\ell
^{\downarrow}_{s,t}+(i-1)\Delta,\ell^{\downarrow}_{s,t}+i\Delta
) ]$ and set $j=k$. \label{alglipapproachwsim}
\item[6.] While $u \in (S^{\rho}_{2j+1}(w),S^{\rho}_{2j}(w)
)$, $j=j+1$.
\item[7.] If $u \leq S^{\rho}_{2j+1}(w)$ then accept, else reject and
return to Step~5.
\end{enumerate}
\end{algorithm}
%
%
%f11 #&#
\begin{figure}

\includegraphics{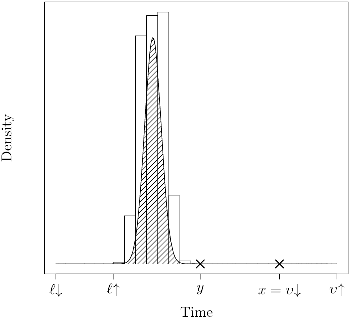}

\caption{Density of intersection layer intermediate point overlaid with
piecewise constant bound calculated using a mesh of size $20$ over the
interval $[\ell{\downarrow},\upsilon{\uparrow}]$ and the corresponding
local Lipschitz constants.} \label{figlipex}
\end{figure}

\noindent We present the synthesis of the above argument in Algorithm~\ref
{alglipapproach}. Clearly the acceptance rate of Algorithm~\ref
{alglipapproach} is at least $T_Z\cdot T_M$ and furthermore is more
robust to different parameter values than the Cauchy sequence approach
outlined in Algorithm~\ref{algintb}, as given sufficient computation an
arbitrarily tight bound of the target density can be found with which
to conduct rejection sampling. In Figure~\ref{figlipex}, we present an
example of a set of parameter values in which the acceptance rate under
the Cauchy sequence approach was less than $10^{-8}$, whereas with the
approach outlined in Algorithm~\ref{alglipapproach} a small mesh of
size $20$
was sufficient to find a tight upper bound of the target density.
%

%s8.2.3 #&#
\subsubsection{Bessel Approach} \label{silipbes}
\label{sHEA}
An alternative scheme to simulate a single intermediate point from
(\ref{eqilmptarget}) is to apply an analogous decomposition of the law of
the sample path as was constructed in the Bessel approach for layered
Brownian bridge outlined in Section~\ref{sea3}. Recall in
Section~\ref{sea3}
that in order to simulate intermediate points from the sample path that
we first simulated the minimum or maximum of the sample path
conditional on the Bessel layer (with probability $1/2$) and then
simulated proposal intermediate points from the law of a Bessel bridge.
The proposal intermediate points were then accepted if the sample path
remained in the appropriate Bessel layer.

We apply the same notion described in Section~\ref{sea3}, however, a
modification has to be made to the acceptance probability as the
intersection layer provides more precise information regarding the
interval in which both the minimum and maximum is contained than the
Bessel layer. In particular, if we have simulated intersection layer
$D_{\iota,1}$ then with probability $1/2$ we propose the auxiliary
minimum (else maximum) in the $\iota$th layer and then only
accept the proposal sample path if the sample path maximum (else
minimum) is contained between the $(\iota-1)$th and $\iota$th Bessel layer. In the case where we have either simulated
intersection layer $D_{\iota,2}$ or $D_{\iota,3}$ then with probability
$1/2$ we propose the auxiliary minimum (else maximum) in the $\iota$th (else $(\iota-1)$th) layer and then only accept
the proposal sample path if the sample path maximum (else minimum) is
contained within the $(\iota-1)$th (else $(\iota-1)$th) Bessel layer. The synthesis of the above argument which is
based on Section~\ref{sea3} can be found in Algorithm~\ref{alghyblayer}.

Although given particular parameter values in (\ref{eqilmptarget}) the
Bessel approach can computationally outperform the Cauchy sequence
approach or Lipschitz approached described in Section~\ref{silipbcs} and
Section~\ref{siliplip}, respectively, as we will discuss in
Section~\ref{silipmix} we advocate a mixture of those two approaches
instead as
the Bessel approach can be particularly inefficient whenever a large
intersection layer is proposed.
\begin{algorithm}
\caption{Simulation of Intersection Layer Intermediate Points (Bessel
Approach)} \label{alghyblayer}
\begin{enumerate}[6.]
\item[1.] Simulate $u_1,u_2\sim\mathrm{U}[0,1]$, set $j=k=0$. \label
{alghyblayerstart}
\item[2.] Simulate Auxiliary Information as per Algorithm~\ref{algbsbm},\label
{alghyblayeraux}
\begin{enumerate}[(a)]
\item[(a)] If $u_1\leq1/2$ simulate minimum $(\tau,X_\tau=\hat{m}\in
[\ell
^{\downarrow}_{s,t},\ell^{\uparrow}_{s,t}])$ setting $c{\downarrow
}:=\upsilon^{\downarrow}_{s,t}$ and $c{\uparrow}:=\upsilon
^{\uparrow}_{s,t}$.
\item[(b)] If $u_1 > 1/2$ simulate maximum $(\tau,X_\tau=\check{m}\in
[\upsilon
^{\downarrow}_{s,t},\upsilon^{\uparrow}_{s,t}])$ setting
$c{\downarrow
}:=\ell^{\downarrow}_{s,t}$ and $c{\uparrow}:=\ell^{\uparrow}_{s,t}$.
\end{enumerate}
\item[3.] Simulate $X_{q}$ from a Bessel bridge conditional on $X_\tau$ as
per Algorithm~\ref{algbbm}.
\item[4.] While $u_2 \in (\prod^{\kappa+2}_{i=1}S^{\delta
}_{2j+1}
(X_\tau,c{\downarrow} ),\prod^{\kappa+2}_{i=1}S^{\delta
}_{2j}
(X_\tau,c{\downarrow} ) )$, $j=j+1$,
\begin{enumerate}[(a)]
\item[(a)] If $u_2 \leq\prod^{\kappa+2}_{i=1}S^{\delta}_{2j+1}
(X_\tau
,c{\downarrow} )$ then reject sample path and return to Step~1.
\end{enumerate}
\item[5.] While $u_2 \in (\prod^{\kappa+2}_{i=1}S^{\delta
}_{2j+1}
(X_\tau,c{\uparrow} ),\prod^{\kappa+2}_{i=1}S^{\delta
}_{2j}
(X_\tau,c{\uparrow} ) )$, $k=k+1$,
\begin{enumerate}
\item[(a)] If $u_2 \geq\prod^{\kappa+2}_{i=1}S^{\delta}_{2k} (X_\tau
,c{\uparrow} )$ then reject sample path and return to Step~1.
\end{enumerate}
\item[6.] Discard Auxiliary Information.
\end{enumerate}
\end{algorithm}

%s8.2.4 #&#
\subsubsection{Implementational Considerations --
Recommended Approach} \label{silipmix}
In Sections~\ref{silipbcs}, \ref{siliplip} and \ref{sHEA}, we have
outlined three separate approaches and algorithms for simulating from
the density of a conditional Brownian bridge at some intermediate time
$q\in(s,t)$~(\ref{eqilmp}). As each of these algorithms is a rejection
sampler in which independent proposals are drawn and then accepted or
rejected, if a proposal is rejected one can change to another of these
algorithms without introducing any bias. As empirically Algorithm~\ref
{algintb} is highly computationally efficient compared to the other
algorithms, but for a small number of parameters values has a very low
acceptance rate, we suggest that on implementation a user specified
threshold number of potential proposals from this algorithm is chosen
(say $N$). If after the first $N$ proposals there has been no
acceptance, then this suggests that the acceptance rate for the
particular parameter configuration is low. As such, at this point we
suggest switching to Algorithm~\ref{alglipapproach} which requires a
significant initial computational effort to find a tight bound to the
target density, but, the acceptance rate will be higher and the
algorithm more robust to different parameter values than Algorithm~\ref
{algintb}. This particular combination of algorithms is advocated as
Algorithm~\ref{alghyblayer} can be inefficient whenever a large intersection
layer is proposed.

%s8.3 #&#
\subsection{Dissecting an
Intersection Layer} \label{silb}
Upon simulating intermediate points of a Brownian bridge sample path
conditional on an intersection layer (e.g., in Section~\ref{silip}), simulating further intermediate points in a sub-interval
between any two existing consecutive points is more complicated as
there is a dependency between all sub-intervals (induced by the
intersection layer). To simplify this problem, we can \textit{dissect}
an intersection layer into separate intersection layers for each pair
of consecutive points by considering all possible dissections and
unbiasedly simulating which one of these occurs.

\begin{figure}[b]

\includegraphics{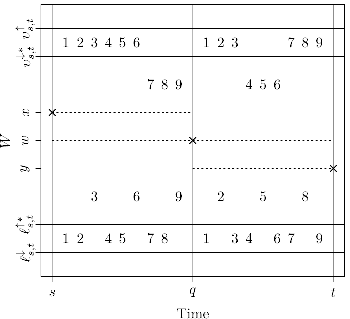}

\caption{Illustration of 9 possible (disjoint) bisections.} \label{figbisect}
\end{figure}

To guide intuition, we first consider the case where we have a single
intermediate point ($W_q=w$) within an existing intersection layer
($W_s=x,W_t=y,\hat{m}_{s,t}\in[\ell^{\downarrow}_{s,t},\ell
^{\uparrow
}_{s,t}],\check{m}_{s,t}\in[\upsilon^{\downarrow}_{s,t},\upsilon
^{\uparrow
}_{s,t}]$) and we want to simulate separate intersection layers for the
intervals $[s,q]$ and $[q,t]$ conditional on the known intersection
layer and the simulated point. We begin by noting that the simulated
point provides further detail on the interval in which the minimum and
maximum lies. In particular, if $w\in[\ell^{\downarrow}_{s,t}, \ell
^{\uparrow}_{s,t}]$ we have that $\hat{m}_{s,t}\in[\ell^{\downarrow
}_{s,t},w]$ and similarly if $w\in[\upsilon^{\downarrow
}_{s,t},\upsilon
^{\uparrow}_{s,t}]$ then we have that $\check{m}_{s,t}\in[w,\upsilon
^{\uparrow
}_{s,t}]$. As such we denote $\ell^{\uparrow*}_{s,t} := (\ell
^\uparrow
_{s,t} \wedge w)$, $\upsilon^{\downarrow*}_{s,t} := (\upsilon
^\downarrow_{s,t} \vee w)$ and we now replace~$D$ with,
%
%e105 #&#
\begin{equation}
D_*  = \bigl\{W_{[s,t]}\dvt \hat{m}_{s,t} \in \bigl[
\ell^\downarrow _{s,t},\ell ^{\uparrow*}_{s,t} \bigr]
\bigr\}\cap \bigl\{W_{[s,t]}\dvt \check {m}_{s,t} \in
\bigl[\upsilon^{\downarrow*}_{s,t},\upsilon^\uparrow_{s,t}
\bigr] \bigr\}.
\end{equation}
The attainment of a particular layer in the interval $[s,t]$ by either
the minimum or the maximum implies that the same layer is attained by
the sample path in at least one of the sub-intervals $[s,q]$ or
$[q,t]$. As such, in our case there are $9$ possible (disjoint)
bisections (which we denote as $B_1$--$B_9$ where $B := D_* = \biguplus
^9_{i=1}B_i$) as illustrated in Figure~\ref{figbisect}. For instance, our
sample path may lie in $B_6$, which more formally has the form,
%
%e106 #&#
\begin{eqnarray}
B_6 &:=&  \bigl( \bigl\{W_{[s,t]}\dvt \hat{m}_{s,q}
\in \bigl[\ell^{\uparrow
*}_{s,t},(x \wedge w) \bigr] \bigr\}\cap
\bigl\{ W_{[s,t]}\dvt \check{m}_{s,q} \in \bigl[
\upsilon^{\downarrow*}_{s,t},\upsilon^{\uparrow
}_{s,t}
\bigr] \bigr\} \bigr)
\nonumber
\\[-9pt]
\\[-9pt]
\nonumber
&&\hspace*{6pt} {}\cap \bigl( \bigl\{W_{[s,t]}\dvt \hat{m}_{q,t} \in
\bigl[\ell ^{\downarrow}_{s,t},\ell^{\uparrow*}_{s,t}
\bigr] \bigr\}\cap \bigl\{ W_{[s,t]}\dvt \check{m}_{q,t}
\in \bigl[(w \vee y),\upsilon ^{\downarrow
*}_{s,t} \bigr] \bigr\}
\bigr).\\[-30pt]\nonumber
\end{eqnarray}
%
%
%f12 #&#

This notion can be extended to the case where we have multiple
intermediate points ($\mathcal{W}:=\{W_{q_1}=w_1,\ldots,W_{q_n}=w_n\}
$), and want to dissect the interval into separate intersection layers.
In particular, we are dissecting a single intersection layer into
$(n+1)$ intersection layers, each with a layer for the minimum and
maximum in their own sub-interval. As the sample path minimum and
maximum must exist in one of the intersection layers there are
$b:=(2^{(n+1)}-1)^2$ possible dissections $B^n_1,\ldots,B^n_b$. We
can simulate which of these dissections our sample path lies in by
application of the following results and Algorithm~\ref{algbis}.
% Theorem - Intersection Layer Multiple Bisection
%

%th9 #&#
\begin{theorem}[(Intersection layer dissection)] \label{thmilmb}
The probability a Brownian bridge sample path is in $B^n_i$ conditional
on $B$ and $\mathcal{W}$ is as follows (denoting by $\mathcal{L}(i)$
and $\mathcal{U}(i)$ the lower and upper layer sets for $B^n_i$),
%
%e107 #&#
\begin{eqnarray}
p_{B^n_i} & :=& \mathbb{P} \bigl(B^n_i |
\hat{m}_{s,t} \in\bigl[\ell^\downarrow _{s,t},
\ell^{\uparrow}_{s,t}\bigr],\check{m}_{s,t}\in\bigl[
\upsilon ^{\downarrow
}_{s,t},\upsilon^{\uparrow}_{s,t}
\bigr],W_s = x,W_t = y,\mathcal {W} \bigr)
\nonumber
\\[-9pt]
\\[-9pt]
\nonumber
& =&  \frac{{}^{(n)}\beta^{\mathcal{L}(i),\mathcal
{U}(i)}_{s,t,x,y}(q_{1:n},\mathcal{W})}{{}^{(n)}\rho^{\ell
{\downarrow},\ell{\uparrow},\upsilon{\downarrow},\upsilon
{\uparrow
}}_{s,t,x,y}(q_{1:n},\mathcal{W})}.
\end{eqnarray}
\end{theorem}

\begin{pf}
Follows directly by Bayes rule, Theorems \ref{thmrhon} and \ref
{thmbetan}.
\end{pf}

%
% Remark - Intersection Layer single Bisection
%
%
%re2 #&#
\begin{remark}[(Intersection layer bisection)] \label{remilb}
In the particular case where we have a single intermediate point then
the probability a Brownian bridge sample path is in $B_i$ (conditional
on $B$ and $W_q=w$) reduces to that in \cite{BBPR12} (denoting $\ell
^{\downarrow,i}_{s,q}$, $\ell^{\uparrow,i}_{s,q}$ $\upsilon
^{\downarrow
,i}_{s,q},\upsilon^{\uparrow,i}_{s,q}$ and $\ell^{\downarrow,i}_{q,t}$,
$\ell^{\uparrow,i}_{q,t}$ $\upsilon^{\downarrow,i}_{q,t},\upsilon
^{\uparrow,i}_{q,t}$ as the bounds for $B_i$ in the interval $[s,q]$
and $[q,t]$, resp.),
%
%e108 #&#
\begin{equation}
p_{B_i}  = \frac{\beta (s,q,x,w,\ell^{\downarrow,i}_{s,q},\ell
^{\uparrow
,i}_{s,q},\upsilon^{\downarrow,i}_{s,q},\upsilon^{\uparrow
,i}_{s,q} )\cdot\beta (q,t,w,y,\ell^{\downarrow
,i}_{q,t},\ell
^{\uparrow,i}_{q,t},\upsilon^{\downarrow,i}_{q,t},\upsilon
^{\uparrow
,i}_{q,t} )}{\rho (s,q,t,x,w,y,\ell^{\downarrow
}_{s,t},\ell
^{\uparrow*}_{s,t},\upsilon^{\downarrow*}_{s,t},\upsilon^{\uparrow
}_{s,t} )}.
\end{equation}
\end{remark}
%
% Lemma - Intersection Layer Bisection Cauchy
%

%co9 #&#
\begin{corollary} \label{corilbcauchy}
Events of probability $p_{B_i}$ can be simulated by retrospective
Bernoulli sampling (as per Corollaries \ref{corlt}, \ref{corris} and
Algorithm~\ref{algris}), noting that it is a function of ${}^{(n)}\beta^{\mathcal{L}(i),\mathcal
{U}(i)}_{s,t,x,y}(q_{1:n},\mathcal
{W})$ and ${}^{(n)}\rho^{\ell{\downarrow},\ell{\uparrow
},\upsilon{\downarrow},\upsilon{\uparrow
}}_{s,t,x,y}(q_{1:n},\mathcal
{W})$ probabilities, using the following sequence,
%
%e109 #&#
\begin{equation}
S^{B(n,i)}_k  := \frac{S^{\beta(n)}_k (s,t,x,y,q_{1:n},\mathcal
{W},\mathcal
{L}(i),\mathcal{U}(i) )}{S^{\rho(n)}_{k+1}
(s,t,x,y,q_{1:n},\mathcal{W},\ell^{\downarrow}_{s,t},\ell^{\uparrow
}_{s,t},\upsilon^{\downarrow}_{s,t},\upsilon^{\uparrow}_{s,t} )}.
\end{equation}
\end{corollary}

\begin{algorithm}[t]
\caption{Dissecting an Intersection Layer} \label{algbis}
\begin{enumerate}[3.]
\item[1.] Simulate $u \sim\mathrm{U}[0,1]$ and set $j=1$ and $k=0$.
\item[2.] While $u \in (C^{B(n,j)}_{2k+1},C^{B(n,j)}_{2k} )$,
$k=k+1$. \label{algbisloop}
\item[3.] If $u \leq C^{B(n,j)}_{2k+1}$ set dissection layer $B=B_j$ else
set $j=j +1$ and return to Step~2.
\end{enumerate}
\end{algorithm}
%
% Remark - Simulating Bisection
Unbiased simulation of the dissection the sample path lies in
can be conducted by inversion sampling and an alternating Cauchy
sequence representation of the CDF of $B$ (\ref{eqbnjcdf}). In
particular, by application of Corollary~\ref{corris}, our sample path lies
in $B^n_j$ if for some $k>0$ and $u\sim\mathrm{U}[0,1]$ we have $u\in
(C^{B(n,j-1)}_{2k+1},C^{B(n,j)}_{2k} )$.
%
%e110 #&#
\begin{equation}\label{eqbnjcdf}
C^{B(n,j)}_k  := \sum^j_{i=1}
S^{B(n,i)}_k.
\end{equation}
%
% Algorithm ilMb
%

%s8.4 #&#
\subsection{Refining an Intersection
Layer}\label{silr}
Suppose we have already simulated layers for the maximum and minimum of
our proposal Brownian bridge sample path ($\hat{m}_{s,t}\in[\ell
^{\downarrow
}_{s,t}, \ell^{\uparrow}_{s,t}]$ and $\check{m}_{s,t}\in[\upsilon
^{\downarrow
}_{s,t},\upsilon^{\uparrow}_{s,t}]$), but we require more \textit
{refined} layer information (i.e., we want a set of narrower layers
$\llvert \ell^{\uparrow^*}_{s,t} - \ell^{\downarrow^*}_{s,t}\rrvert
\leq
\llvert \ell^{\uparrow}_{s,t} - \ell^{\downarrow}_{s,t}\rrvert $ or
$\llvert \upsilon^{\uparrow^*}_{s,t} - \upsilon^{\downarrow^*}_{s,t}\rrvert
\leq\llvert \upsilon_{s,t}^{\uparrow} - \upsilon_{s,t}^{\downarrow
}\rrvert $). This can be achieved by noting that given some $\ell
_{s,t}^\updownarrow \in(\ell_{s,t}^\downarrow,\ell_{s,t}^\uparrow)$
and $\upsilon_{s,t}^\updownarrow \in(\upsilon_{s,t}^\downarrow
,\upsilon_{s,t}^\uparrow)$, the sample path falls in one of the
following $4$ possible (disjoint) intersection layer refinements (where
$R := \biguplus^4_{i=1} R_i$),
%
%e111 #&#
%e112 #&#
%e113 #&#
%e114 #&#
\begin{eqnarray}\label{eqrefpart11}
R_1 & =& \bigl\{W_{[s,t]} \dvt  \hat{m}_{s,t} \in \bigl[
\ell_{s,t}^\downarrow ,\ell _{s,t}^\updownarrow \bigr]
\bigr\}\cap \bigl\{W_{[s,t]}\dvt \check {m}_{s,t}\in
\bigl[\upsilon_{s,t}^\updownarrow,\upsilon_{s,t}^\uparrow
\bigr] \bigr\},
\\[-2pt]
\label{eqrefpart12}
R_2 & =&  \bigl\{W_{[s,t]}\dvt \hat{m}_{s,t} \in \bigl[
\ell _{s,t}^\updownarrow,\ell _{s,t}^\uparrow \bigr]
\bigr\}\cap \bigl\{W_{[s,t]}\dvt \check {m}_{s,t}\in
\bigl[\upsilon_{s,t}^\updownarrow,\upsilon_{s,t}^\uparrow
\bigr] \bigr\},
\\
\label{eqrefpart21}
R_3 & =&  \bigl\{W_{[s,t]}\dvt \hat{m}_{s,t} \in \bigl[
\ell_{s,t}^\downarrow ,\ell _{s,t}^\updownarrow \bigr]
\bigr\}\cap \bigl\{W_{[s,t]}\dvt \check {m}_{s,t}\in
\bigl[\upsilon_{s,t}^\downarrow,\upsilon_{s,t}^\updownarrow
\bigr] \bigr\},
\\
\label{eqrefpart22}
R_4 & =&  \bigl\{W_{[s,t]}\dvt \hat{m}_{s,t} \in \bigl[
\ell _{s,t}^\updownarrow,\ell _{s,t}^\uparrow \bigr]
\bigr\}\cap \bigl\{W_{[s,t]}\dvt \check {m}_{s,t}\in
\bigl[\upsilon_{s,t}^\downarrow,\upsilon_{s,t}^\updownarrow
\bigr] \bigr\},
\end{eqnarray}
and so we can simply unbiasedly simulate which one our sample path lies
in.

In a similar fashion to Section~\ref{silb}, we can simulate unbiasedly
which of the intersection layer refinements our sample path lies in by
application of the following established results and Algorithm~\ref{algref}
(where we denote by $\ell_{s,t}^{\downarrow,i},\ell_{s,t}^{\uparrow
,i},\upsilon_{s,t}^{\downarrow,i},\upsilon_{s,t}^{\uparrow,i}$ with a
superscript $i\in\{1,2,3,4\}$ the corresponding parameter selections
from (\ref{eqrefpart11})--(\ref{eqrefpart22})).
% Theorem - Intersection Layer Refinement
%

\begin{algorithm}[b]
\caption{Refining an Intersection Layer \cite{BBPR12}} \label{algref}
\begin{enumerate}[3.]
\item[1.] Simulate $u \sim\mathrm{U}[0,1]$ and set $j=1$ and $k=0$.
\item[2.] While $u \in (C^{R(j)}_{2k+1},C^{R(j)}_{2k} )$,
$k=k+1$. \label{algrefloop}
\item[3.] If $u \leq S^{R(j)}_{2k+1}$ set layer $R=R_j$ else set $j=j+1$
and return to Step~2.
\end{enumerate}
\end{algorithm}
%

%th10 #&#
\begin{theorem}[(Intersection layer refinement \protect\cite{BBPR12}, Section~5.3)] \label{thmilr}
The probability a Brownian bridge sample path contained within $R$ is
in $R_i$ is as follows,
%
%e115 #&#
\begin{eqnarray}
p_{R_i} & :=&  \mathbb{P} \bigl(R_i |
\hat{m}_{s,t} \in \bigl[\ell _{s,t}^\downarrow,
\ell_{s,t}^\uparrow \bigr],\check{m}_{s,t}\in \bigl[
\upsilon _{s,t}^{\downarrow},\upsilon_{s,t}^{\uparrow}
\bigr],W_s = x,W_t = y \bigr)
\nonumber
\\[-8pt]
\\[-8pt]
\nonumber
& =&  \frac{\beta (s,t,x,y,\ell_{s,t}^{\downarrow,i},\ell
_{s,t}^{\uparrow,i},\upsilon_{s,t}^{\downarrow,i},\upsilon
_{s,t}^{\uparrow,i} )}{\beta (s,t,x,y,\ell
_{s,t}^\downarrow
,\ell_{s,t}^\uparrow,\upsilon_{s,t}^\downarrow,\upsilon
_{s,t}^\uparrow
 )}.
\end{eqnarray}
\end{theorem}
%
% Lemma - ILR Cauchy sequence
%

%co10 #&#
\begin{corollary}[(\protect\cite{BBPR12}, Section~5.3)] \label{corilr}
Events of probability $p_{R_i}$ can be simulated by retrospective
Bernoulli sampling (as per Corollaries \ref{corlt}, \ref{corris} and
Algorithm~\ref{algris}), noting that it is a function of $\beta$
probabilities, using the sequence,
%
%e116 #&#
\begin{equation}
S^{R(i)}_k :=  \frac{S^\beta_k (s,t,x,y,\ell_{s,t}^{\downarrow,i},\ell
_{s,t}^{\uparrow,i},\upsilon_{s,t}^{\downarrow,i},\upsilon
_{s,t}^{\uparrow,i} )}{S^\beta_{k+1} (s,t,x,y,\ell
_{s,t}^\downarrow,\ell_{s,t}^\uparrow,\upsilon_{s,t}^\downarrow
,\upsilon
_{s,t}^\uparrow )}.
\end{equation}
\end{corollary}

% Remark - Simulating Refinement
By application of Corollary~\ref{corris}, unbiased
simulation of
the refinement the sample path lies in can be conducted by inversion
sampling and an alternating Cauchy sequence representation of the CDF
of $R$ (\ref{eqrnjcdf}). In particular, our sample path lies in $R_j$
if for some $k>0$ and $u\sim\mathrm{U}[0,1]$ we have $u\in
(C^{R(j-1)}_{2k+1},C^{R(j)}_{2k} )$, where:
%
%e117 #&#
\begin{equation}\label{eqrnjcdf}
C^{R(j)}_k  := \sum^j_{i=1}S^{R(i)}_k.
\end{equation}
%
% Algorithm - ILR simulation
%

%s8.5 #&#
\subsection{Simulating Layered Brownian Bridges} \label{sSLBB} \label{sMEA}
The \textit{Intersection Layer Approach} for constructing a layered
Brownian bridge is a direct application of the algorithms of Sections~\ref{sil}, \ref{silip} and \ref{silb}. In particular, we simulate
initial intersection layer information for the sample path
(Algorithm~\ref{algauea} Step~2) by application
of Algorithm~\ref{algint}. In
Algorithm~\ref{algauea} Step~4, we iteratively
simulate skeletal
(intermediate) points, then new intersection layer information
conditional on these points. This can be achieved directly by either
Algorithm~\ref{algintb}, \ref{alglipapproach}, \ref{alghyblayer} or
some mixture of these algorithms to simulate the intermediate point (as
discussed in Section~\ref{silip} and in particular Section~\ref{silipmix}) and
Algorithm~\ref{algbis} to bisect the interval.

We present the iterative Algorithm~\ref{algauea} Step~4 in
Algorithm~\ref{alglayer} which can be additionally used to conduct
Algorithm~\ref{algauea} Step~6. $\mathcal{S}$
denotes the set containing all
intersection layer information. The set is composed of $(n-1)$ elements
corresponding to the intervals between $n$ existing time points. In
particular, each element ($\mathcal{S}_{a,b}$) between two successive
time points ($a<b$) contains information regarding the sample path at
the end points and an upper and lower bound for both the minimum and
maximum of the sample path in that interval $ (\mathcal{S}_{a,b}:= \{a,b,X_a,X_b,\ell^{\downarrow}_{a,b},\ell
^{\uparrow
}_{a,b},\upsilon^{\downarrow}_{a,b},\upsilon^{\uparrow}_{a,b}
\}
 )$.
\begin{algorithm}[b]
\caption{Layered Brownian Bridge Simulation (Intersection Layer
Approach)} \label{alglayer}
\begin{enumerate}
\item For each intermediate point required ($q$),
\begin{enumerate}[(a)]
\item[(a)] Select the appropriate existing intersection layer $\mathcal
{S}_{a,b}$ from $\mathcal{S}$ such that $q\in(a,b)$.
\item[(b)] Simulate $X_{q}$ as per Algorithm \ref{algintb} or \ref
{alglipapproach} or \ref{alghyblayer}. \label{alglayerintb}
\item[(c)] Dissect interval as per Algorithm~\ref{algbis} to find new intersection
layers $\mathcal{S}_{a,q}$ and $\mathcal{S}_{q,b}$.
\item[(d)] Set $\mathcal{S} = \mathcal{S}\cup \{\mathcal
{S}_{a,q},\mathcal{S}_{q,b} \}\setminus\mathcal{S}_{a,b}$.
\end{enumerate}
\end{enumerate}
\end{algorithm}

It should be noted that further refinements to Algorithm~\ref
{alglayer} could
be made when considering any particular application, however, we have
omitted the explicit algorithms here. For instance, if the simulation
of intermediate points is required for the AUEA (Algorithm~\ref{algauea}),
then refining the intersection layers as outlined in Section~\ref{silr} and
detailed in Algorithm~\ref{algref} would result in tighter upper and lower
bounds for the sample path. As a consequence tighter upper and lower
bounds for $\phi(X)$ could be computed, resulting in a more efficient
algorithm. Similar notions to this are explored in Sections~\ref{sepss}, \ref{sesea} and \ref{sexampfinal}.

\begin{algorithm}[b]
\caption{Unbiased Estimation of Upper Barrier Crossing} \label{algbndcross}
\begin{enumerate}[4.]
\item[1.] Simulate jump diffusion skeleton $\mathcal{S}_{\mathrm{AUJEA}}
(X ) :=  \{ (\xi_i,X_{\xi_i} )^{\sum_j(\kappa
_j+1)}_{i=0},  (R^{[\xi_{i-1},\xi_i]}_X )^{\sum_j(\kappa
_j+1)}_{i=1} \}$ as per Algorithm~\ref{algaujea}.
\item[2.] Set $\mathcal{S}:=\mathcal{S}_{\mathrm{AUJEA}} (X )$,
$\mathcal{C}:=\varnothing$, $\mathcal{B}:=\varnothing$ and $\mathcal
{U}:= \{ [s^1,t^1 ],\ldots, [s^{\llvert \mathcal
{S}\rrvert },t^{\llvert \mathcal{S}\rrvert } ] \}$.
\item[3.] While $\llvert \mathcal{C}\rrvert =0$ and $\llvert \mathcal
{B}\rrvert <\llvert \mathcal{S}\rrvert $,
\begin{enumerate}[(a)]
\item[(a)] For $i$ in $1$ to $\llvert \mathcal{U}\rrvert $,
\begin{enumerate}[iii.]
\item[i.] If $X^i_s \geq B^i_{s}$ or $X^i_t \geq B^i_{t}$ or $\upsilon
^{i,\downarrow}_{s,t}\geq\sup_{u\in[s(i),t(i)]}B_u$ then $\mathcal
{C}:=\mathcal{C}\cup \{ [s^i,t^i ] \}$ and
$\mathcal{U}:=\mathcal{U}\setminus \{ [s^i,t^i
] \}$.
\item[ii.] If $\upsilon^{i,\uparrow}_{s,t}\leq\inf_{u\in[s(i),t(i)]}B_u$
then $\mathcal{B}:=\mathcal{B}\cup \{ [s^i,t^i
] \}
$ and $\mathcal{U}:=\mathcal{U}\setminus \{ [s^i,t^i
] \}$.
\item[iii.] If $[s^i,t^i]\in\mathcal{U}$ then,
\begin{enumerate}[A.]
\item[A.] Simulate $X_q$ where $q:=(s^{i}+t^{i})/2$ conditional on
$\mathcal
{S}^{i}_{s,t}$ as per Algorithm~\ref{algintb}.
\item[B.] Bisect and refine $\mathcal{S}^{i}_{s,t}$ into $\mathcal
{S}^{i,1}_{s,q}$ and $\mathcal{S}^{i,2}_{q,t}$ as per Algorithm~\ref{algbis}
and Algorithm~\ref{algref}.
\item[C.] Set $\mathcal{S}:=\mathcal{S}\cup\mathcal
{S}^{i,1}_{s,q}\cup
\mathcal{S}^{i,2}_{q,t}\setminus\mathcal{S}^{i}_{s,t}$ and $\mathcal
{U}:=\mathcal{U}\cup \{ [s^i,q ],
[q,t^i
] \}\setminus \{ [s^i,t^i ] \}$.
\end{enumerate}
\end{enumerate}
\end{enumerate}
\item[4.] If $\llvert \mathcal{C}\rrvert >0$ then barrier crossed, else if
$\llvert \mathcal{B}\rrvert =\llvert \mathcal{S}\rrvert $ barrier not crossed.
\end{enumerate}
\end{algorithm}

%s9 #&#
\section{Examples -- Unbiased
Estimation of Irregular Barrier Crossing Probabilities} \label{sexampfinal}
In this section, we present a number of applications of the methodology
developed in this paper. In particular, we show that it is possible to
determine exactly whether a jump diffusion sample path simulated as per
Algorithm~\ref{algaujea} crosses various types of barriers. In
Section~\ref{sesapp1} we consider a nonlinear two sided barrier, in
Section~\ref{sesapp2} we consider the crossing of two jump diffusion
sample paths
from different laws and finally in Section~\ref{sesapp3} we consider the
crossing of a circular barrier by a 2-dimensional jump diffusion sample
path. The flexibility of the methodology developed in this paper allows
us to by extension simulate various non-trivial quantities, for
instance, we can construct an unbiased estimate of the probability that
any barrier is crossed, an unbiased estimate of the probability a
barrier is crossed by any particular time and the killed (or un-killed)
diffusion transition density among many other interesting
possibilities.

In each of our examples we employ variants of Algorithm~\ref{algbndcross}.
For simplicity in Algorithm~\ref{algbndcross} we only consider the crossing
of an upper barrier by a one dimensional jump diffusion, however, as we
will discuss later in this section and in Sections~\ref{sesapp1}--\ref
{sesapp3} this can be straight-forwardly extended. To simplify
notation we define $B_u$ as the evaluation of the upper barrier at time
point~$u$. As before we denote $\mathcal{S}$ as the skeleton comprising
intersection layer information, and further denote $\mathcal{C}$ as the
set of intervals in which the sample path crosses the barrier,
$\mathcal
{B}$ as the set of intervals in which there is no crossing, and
$\mathcal{U}$ as the set in which for each interval crossing is undetermined.

%s9.1 #&#
\subsection{Example~1 -- Nonlinear two sided barrier}\label{sesapp1}
In this section, we consider the simulation of jump diffusion sample
paths which can be represented as solutions to the following SDE,
%
%e117 #&#
\begin{equation}\label{eqexamp1sde7}
\mathrm{d}X_t  = \sin({X_{t-}}) \,\mathrm{d}t +
\mathrm{d}W_t + \mathrm {d}J^{\lambda,\nu}_t,
\end{equation}
where
%
%e118 #&#
\begin{equation}\label{eqexamp1sde8}
X_0 = 1, \qquad t\in[0,2\uppi], \qquad \lambda(X_{t-}) = \llvert
X_{t-}/4\rrvert,\qquad f_{\nu}(\cdot;X_{t-}) =
\mathrm{N}(\cdot;-X_{t-}/2,2),
\end{equation}
determining whether or not they cross the following nonlinear two sided
barrier (where $B^{\downarrow}_u$~and~$B^{\uparrow}_u$ denote the lower
and upper barriers at time point $u$, resp.),
%
%e120 #&#
\begin{equation}\label{eqexamp1bnd}
B^{\downarrow}_u = -2.5 - \cos(u),\qquad B^{\uparrow}_u
= 4 + 0.5\cos(u),\qquad u\in[0,2\uppi].
\end{equation}
In this case, as the jump intensity of (\ref{eqexamp1sde8}) can't be
bounded we simulate sample paths using the AUJEA (see Algorithm~\ref{algaujea}). In
particular, we have $\phi(X_t):=(\sin^2(X_t)+\cos
(X_t))/2\in[-1/2,5/8]$, $\lambda(X_t)| (L_X,U_X )\leq\max
\{
|L_X|,|U_X|\}/4$ and the end point is simulated according as follows,
$X_T\sim h(y; x,T) \propto\exp\{-\cos(y)-y^2/6\}$.

\begin{figure}[t]
\begin{tabular}{@{}cc@{}}

\includegraphics{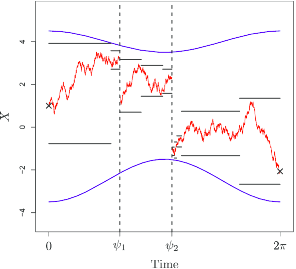}
 & \includegraphics{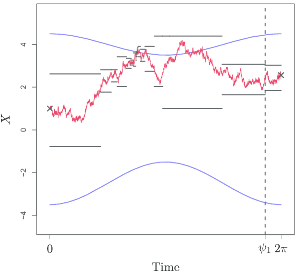}\\
{\fontsize{9}{11}\selectfont{(a) No barrier crossing}} &
{\fontsize{9}{11}\selectfont{(b) Barrier crossed}}
\end{tabular}
\caption{Illustration of the determination of whether a 2-sided
non-linear barrier has been crossed by a sample path using a finite
dimensional sample path skeleton, overlaid with an illustration of the
underlying sample path.}\label{figbarex1pos}
\end{figure}

In Figure~\ref{figbarex1pos}, we present illustrations of whether the two
sided barrier (\ref{eqexamp1bnd}) has been crossed using finite
dimensional realisations of sample paths simulated according to the
measure induced by (\ref{eqexamp1sde7}) and by applying a modified
version of Algorithm~\ref{algbndcross}. This example is motivated by
a number
of possible applications in finance, such as the pay-off of barrier options.
%
%
%f13 #&#

In this example, we simulated 100\,000 sample paths from the measure
induced by (\ref{eqexamp1sde8}) and determined whether the barrier
(\ref{eqexamp1bnd}) was crossed for each sample path. For each sample path
we additionally determined whether one or both barriers were crossed
and if both, which barrier was crossed first. From these simulations,
we calculated unbiased estimates of various barrier crossing
probabilities, the results of which are summarised in Table~\ref{tabbarex1}.
%t1 #&#
\begin{table}[b]
\caption{Nonlinear two sided barrier example: Barrier crossing
probabilities (computed using 100\,000 sample paths)} \label{tabbarex1}
\begin{tabular*}{\tablewidth}{@{\extracolsep{\fill}}lll@{}}
\hline
Crossing type & Empirical
probability & $95\%$ Clopper--Pearson confidence interval \\
\hline
Neither barrier & $18.09\%$ & $[17.85\%,18.36\%]$ \\
Either barrier & $81.91\%$ & $[81.67\%,82.15\%]$ \\
Upper barrier only & $43.98\%$ & $[43.67\%,44.29\%]$ \\
Lower barrier
only & $29.02\%$ & $[28.74\%,29.30\%]$ \\
Both barriers & \phantom{0}$8.92\%$ &
$[8.74\%,9.09\%]$ \\
Upper first given both barriers & $80.45\%$ & $[79.61\%,81.27\%]$ \\
Lower first given both barriers & $19.55\%$ & $[18.73\%,20.39\%]$ \\
\hline
\end{tabular*}
\end{table}
%
%
%f14 #&#
\begin{figure}
\begin{tabular}{@{}p{185pt}@{\quad}p{185pt}@{}}

\includegraphics{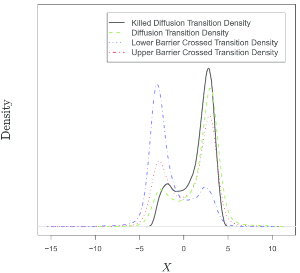}
 & \includegraphics{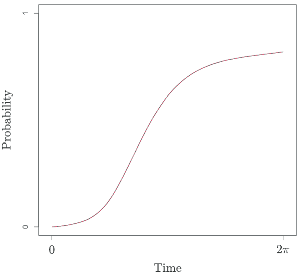}\\
{\fontsize{9}{11}\selectfont{(a) Kernel density estimates of the transition densities of
subsets of sample paths simulated from the measure induced by (\ref
{eqexamp1sde8})}}&
{\fontsize{9}{11}\selectfont{(b) Empirical CDF of barrier crossing probability by time
(crossing time evaluated within interval of length $\varepsilon\leq
10^{-4}$)}}
\end{tabular}
\caption{Nonlinear two-sided barrier example: summary figures computed
using 100\,000 sample paths.}\label{figbarex1extra}
\end{figure}

In Figure~\ref{figbarex1extra}(a), we present kernel density
estimates of
the transition densities of various subsets of the sample paths
simulated, including that for killed diffusions (i.e., sample paths from
the measure induced by (\ref{eqexamp1sde8}) with the restriction that
they remain within the interval between the barriers in (\ref
{eqexamp1bnd})). In Figure~\ref{figbarex1extra}(b), we additionally
determine for each sample path an interval of length $\varepsilon\leq
10^{-4}$, in which the first crossing time occurs (by modifying the
$\varepsilon$-strong algorithms presented in Section~\ref{sepss}) to construct
upper and lower bounds for the empirical CDF of the first barrier
crossing time.
%
%

%s9.2 #&#
\subsection{Example~2 -- Jump diffusion barrier}\label{sesapp2}
In this section, we consider the simulation of jump diffusion sample
paths which can be represented as solutions to the following SDE,
%
%e121 #&#
\begin{equation}\label{eqexamp2sde}
\mathrm{d}X_t  = -X_{t-}\,\mathrm{d}t +
\mathrm{d}W_t + \mathrm{d}J^{\lambda
,\nu}_t,
\end{equation}
where
%
%e122 #&#
\begin{equation}
t\in[0,2], \qquad\lambda(X_{t-}) = \sin^2(X_{t-}),
\qquad f_{\nu
}(\cdot;X_{t-}) = \mathrm{N} (\cdot;-X_{t-}/2,1
).
\end{equation}
We consider sample paths simulated from the measure induced by (\ref
{eqexamp2sde}) initialised at two possible starting values $X^\ell
_0=-2$ and $X^\upsilon_0=2$ (where $X^\ell$ and $X^\upsilon$ denote the
lower and upper diffusions, resp.). In this case, the jump
intensity of (\ref{eqexamp2sde}) is bounded so we simulate sample
paths using the AUEA (see
Algorithm~\ref{algauea}) within the BJEA (see
Algorithm~\ref{algbjea}). Recall that in the BJEA the interval the sample
path is to be simulated over ($t\in[0,2]$), is broken into segments
corresponding to the proposed jump times ($\Psi_1,\ldots$). As such,
if we consider the simulation of a diffusion sample path in the
interval $[\Psi_1,\Psi_2]$ conditional on $X_{\Psi_1}$ then the
proposed end point is simulated as follows, $X_{\Psi_2}\sim h(X_{\Psi
_2};X_{\Psi_1},\Psi_2-\Psi_1) \propto\exp\{-X^2_{\Psi_2}/2-
(X_{\Psi
_2}-X_{\Psi_1})^2/[2(\Psi_2-\Psi_1)]\}$. Furthermore, we have $\phi
(X_t):=(X_t^2-1)/2$, $\phi(X_t)| (L_X,U_X )\in[-1/2,(\max
\{
L^2_X,U^2_X\}-1)/2]$ and $\lambda(X_t)\leq1 =: \Lambda$.

\begin{figure}%[t]
\begin{tabular}{@{}cc@{}}

\includegraphics{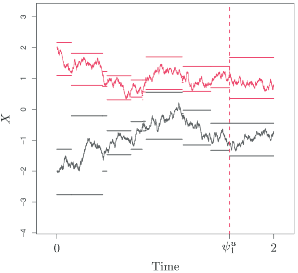}
 & \includegraphics{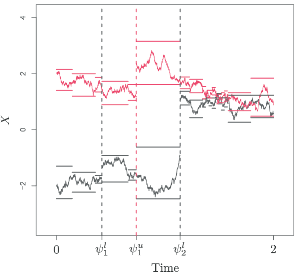}\\
{\fontsize{9}{11}\selectfont{(a) No crossing}}&
{\fontsize{9}{11}\selectfont{(b) Crossing}}
\end{tabular}
\caption{Illustration of the determination of whether two diffusion
sample paths cross one another using finite dimensional sample path
skeletons, overlaid with an illustration of the underlying sample
paths.}\label{figbarex2pos}
\end{figure}

In Figure~\ref{figbarex2pos}, we present illustrations of two sample paths
simulated from the measure induced by (\ref{eqexamp2sde}), initialised
at $X^\ell_0=-2$ and $X^\upsilon_0=2$, which do not cross and cross,
respectively, determined using only a finite dimensional realisation of
the sample paths. This example is motivated by \cite{TRBS12}, in which
(in part) the authors are interested in the probability that two
Brownian motion sample paths cross one another.

%
%
%f15 #&#

\begin{figure}%[t]
\begin{tabular}{@{}p{180pt}@{\quad}p{180pt}@{}}

\includegraphics{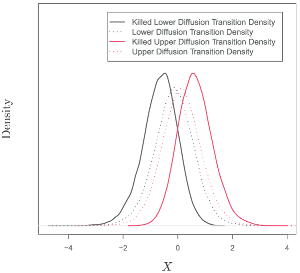}
 & \includegraphics{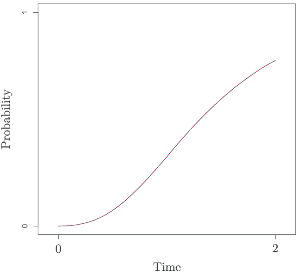}\\
{\fontsize{9}{11}\selectfont{(a) Kernel density estimates of the transition densities of
subsets of sample paths simulated from the measure induced by (\ref{eqexamp2sde})}} &
{\fontsize{9}{11}\selectfont{(b) Empirical CDF of jump diffusion crossing probability
(crossing time evaluated within interval of length $\varepsilon\leq
10^{-4}$)}}
\end{tabular}
\caption{Jump diffusion crossing example: summary figures computed
using 100\,000 sample paths.}\label{figbarex2extra}
\end{figure}

In this example, we simulated 100\,000 pairs of sample paths from the
measure induced by~(\ref{eqexamp2sde}) initialised at $X^\ell_0=-2$
and $X^\upsilon_0=2$ and determined whether or not they crossed. We
present a summary of the unbiased estimates calculated from these
sample paths in Table~\ref{tabbarex2}.
In Figure~\ref{figbarex2extra}(a) we present kernel density estimates of
the transition densities of various subsets of the sample paths
simulated. In Figure~\ref{figbarex2extra}(b), we determine for each sample
path an interval of length $\varepsilon\leq10^{-4}$ in which the first
crossing time occurs in order to construct upper and lower bounds for
the empirical CDF of the first crossing time.

%
%
%t2 #&#
\begin{table}%[b]
\caption{Jump diffusion crossing example: crossing probabilities
(computed using 100\,000 sample paths)} \label{tabbarex2}
\begin{tabular*}{\tablewidth}{@{\extracolsep{\fill}}lll@{}}\hline
Crossing type & Empirical
probability & $95\%$ Clopper--Pearson confidence interval \\
\hline
No crossing & $22.52 \%$ & $[22.26\%, 22.78\%]$ \\
Crossing & $77.48\%$
& $[77.22\%,77.74\%]$ \\
\hline
\end{tabular*}
\end{table}
%%
%
%f16 #&#

%s9.3 #&#
\subsection{Example~3 -- 2-D jump diffusion with circular
barrier}\label{sesapp3}
\begin{figure}[b]
\begin{tabular}{@{}cc@{}}

\includegraphics{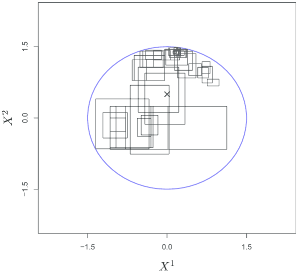}
 & \includegraphics{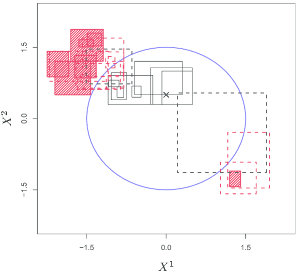}\\
{\fontsize{9}{11}\selectfont{(a) No barrier crossing}}&
{\fontsize{9}{11}\selectfont{(b) Barrier crossed}}
\end{tabular}
\caption{Illustration of the determination of whether a 2-D sample path
crosses a circular barrier using a finite dimensional sample path
skeleton. Inscribed rectangles denote regions where for some time
interval sample paths are constrained. Black and infilled red
rectangles denote intervals constrained entirely within or out-with the
circle, respectively. Dotted black and red rectangles denote intervals
with undetermined or partial crossing, respectively.}\label{figbarex3pos}
\end{figure}

In this section, we consider the simulation of jump diffusion sample
paths which can be represented as solutions to the following SDE,
%
%e123 #&#
\begin{equation}\label{eqexamp3sde1}
X:= \bigl(X^{(1)},X^{(2)} \bigr),
\end{equation}
where
%
%e124 #&#
%e125 #&#
\begin{eqnarray}
\mathrm{d}X^{(1)}_t & =& -X^{(1)}_{t-}
\,\mathrm{d}t + \mathrm {d}W^{(1)}_t + \mathrm{d}J^{\lambda
(X),\nu_1(X{(1)})}_t,
\\
\mathrm{d}X^{(2)}_t & =&  -X^{(2)}_{t-}
\,\mathrm{d}t + \mathrm {d}W^{(2)}_t + \mathrm{d}J^{\lambda
(X),\nu_2(X{(2)})}_t,
\end{eqnarray}
and further denoting by $\theta:= \arctan
(X^{(2)}_t/X^{(1)}_t )$
and $Z\sim\mathrm{U} [0, [ (X^{(1)}_t )^2+
(X^{(2)}_t )^2
]^{1/2} ]$,
%
%e126 #&#
%e127 #&#
\begin{eqnarray}
\label{eqexamp3sde2}
X_0&=& (0,0.5),\qquad t\in[0,3],\qquad \lambda(X_t) = \bigl[
\bigl(X^{(1)}_t \bigr)^2+ \bigl(X^{(2)}_t
\bigr)^2 \bigr]^{1/2},\\
\label{eqexamp3sde3}
f_{\nu_1} \bigl(\cdot;X_{t-}^{(1)} \bigr) &\sim & -\cos(
\theta) Z,\qquad f_{\nu_2} \bigl(\cdot;X_{t-}^{(2)} \bigr)
\sim-\sin(\theta) Z,
\end{eqnarray}
determining whether or not they cross the following circular barriers,
%
%e128 #&#
\begin{equation}\label{eqexamp3bnd}
x^2+y^2=r,\qquad \mbox{where } r=\{0.8,1,\ldots,2.8,3\}.
\end{equation}
The jump intensity of this SDE (\ref{eqexamp3sde1}, \ref{eqexamp3sde2}, \ref{eqexamp3sde3}) can't be bounded so we simulate
sample paths using the AUJEA (see Algorithm~\ref{algaujea}). In Figure~\ref{figbarex3pos}, we present illustrations of whether one particular
circular barrier ($r=1.6$) has been crossed using finite dimensional
realisations of sample paths simulated according to the measure induced
by (\ref{eqexamp3sde1}, \ref{eqexamp3sde2}, \ref{eqexamp3sde3}) and
by applying a modified version of Algorithm~\ref{algbndcross}.
%
%
%f17 #&#
%f18 #&#
\begin{figure}[t]
\begin{tabular}{@{}p{180pt}@{\quad}p{180pt}@{}}

\includegraphics{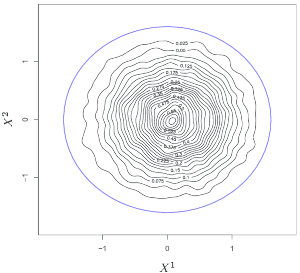}
 & \includegraphics{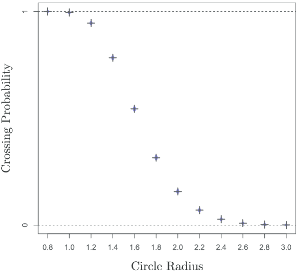}\\[-6pt]
{\fontsize{9}{11}\selectfont{(a) Contour plot of kernel density estimate of killed diffusion
transition density with circular barrier of radius $1.6$}}&
{\fontsize{9}{11}\selectfont{(b) Empirical probabilities of crossing centred circles of
increasing radius overlaid with $95\%$ Clopper--Pearson confidence
intervals}}\\[3pt]

\includegraphics{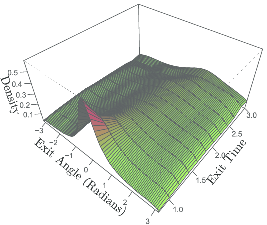}
 & \includegraphics{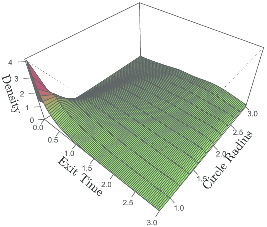}\\[-6pt]
{\fontsize{9}{11}\selectfont{(c) Circle of radius $1.6$ exit angle by exit time (crossing
time and exit angle evaluated within intervals of length $\varepsilon\leq
10^{-3}$ and $\theta\leq10^{-3}$, resp.)}}&
{\fontsize{9}{11}\selectfont{(d) Circle exit time by circle radius (crossing time evaluated
within interval of length $\varepsilon\leq10^{-3}$)}}
\end{tabular}
\caption{2-D jump diffusion with circular barrier example: figures
computed using 50\,000 sample paths.}\label{figbarex3extra}
\end{figure}

In this example we simulated 50\,000 sample paths from the measure
induced by (\ref{eqexamp3sde1}), (\ref{eqexamp3sde2}), (\ref{eqexamp3sde3}), determining for each circular barrier (\ref
{eqexamp3bnd}) whether or not it was crossed. In addition, for each
circular barrier we simulated the time within an interval of length
$\varepsilon\leq10^{-3}$ in which the barrier was first crossed and an
interval of length $\theta\leq10^{-3}$ in which the exit angle lies.
Calculating all circular barriers for every sample path ensures that
the calculated probabilities retain any natural ordering (e.g.,
the first crossing time of a circular barrier of a given radius must
occur before one of larger radius). In Figure~\ref{figbarex3extra}, we
present various results obtained from our simulations which may be of
interest in any given application.
%
%

%sA #&#
\begin{appendix}

\renewcommand{\theequation}{\arabic{equation}}

%sB #&#
\section*{Appendix: Elementary Cauchy Sequence Functions} \label{apxecf}
In Section~\ref{ssbpsp}, we define the functions $\bar{\varsigma}$
and $\bar
{\varphi}$ which form the building blocks for the construction of all
other alternating Cauchy sequences in this paper. In Section~\ref{silip},
we exploit the full representation of $\rho$ found in Theorem~\ref{thmrhon}
and Definition~\ref{dfnrhon} in terms of $\bar{\varsigma}$ and
$\bar{\varphi
}$. In particular, we make the remark that each can be represented in
the form $\exp(a_i+b_iw)$ where each function $a_i$ and $b_i$ can be
explicitly calculated. Furthermore, note that the multiple of any two
such functions can also be represented in the form $\exp(a_j+b_jw)$.

In this \hyperref[apxecf]{Appendix}, we briefly detail the possible functions that can
arise and show an explicit representation for each in terms of $a_i$
and $b_i$. With reference to details in Section~\ref{ssbpsp} and
Section~\ref{ssnbpsp}, we can limit our consideration to the
following four
functional forms (noting that there are also corresponding negative
versions (in which the sign of $x, w$ and $y$ reverses), a set for each
of the possible layer combinations (in which we substitute $[\ell
_i,\upsilon_i]$ for $[\ell{\downarrow},\upsilon{\uparrow}], [\ell
{\uparrow},\upsilon{\uparrow}], [\ell{\downarrow},\upsilon
{\downarrow
}]$ or $[\ell{\uparrow},\upsilon{\downarrow}]$) as well as various
multiples of these functions (which as a consequence of their
exponential form will simply result in the addition of the $a$ and $b$
terms)). Denoting $D:=\llvert \upsilon_i-\ell_i\rrvert $ and $M:=(\ell
_i\wedge\upsilon_i)$ we have,
%
%eB.1 #&#
%eB.2 #&#
%eB.3 #&#
%eB.4 #&#
\begin{eqnarray}
\bar{\varsigma}^{\ell_i,\upsilon_i}_{s,q}(j;x,w) & =& \exp \biggl\{
\underbrace{-\frac{2}{q-s} \bigl(D^2j^2 +
2DMj+M^2-Djx-Mx \bigr)}_{a_{i,1}}
\nonumber
\\[-8pt]
\\[-8pt]
\nonumber
&&\hspace*{96pt}\underbrace{{}-\frac{2}{q-s} (-Dj-M+x )}_{b_{i,1}}w \biggr\} ,
\\
\bar{\varsigma}^{\ell_i,\upsilon_i}_{q,t}(j;w,y) & =&  \exp \biggl\{
\underbrace{-\frac{2}{t-q} \bigl(D^2j^2 +
2DMj+M^2-Djy-My \bigr)}_{a_{i,2}}
\nonumber
\\[-10pt]
\\[-10pt]
\nonumber
&&\hspace*{95pt}{}\underbrace{{}-\frac{2}{t-q} (-Dj-M+y )}_{b_{i,2}}w \biggr\} ,
\\[-3pt]
\bar{\varphi}^{\ell_i,\upsilon_i}_{s,q}(j;x,w) & =&  \exp \biggl\{
\underbrace{-\frac{2j}{q-s} \bigl(D^2j+Dx \bigr)}_{a_{i,3}}
\underbrace{{}+\frac{2j}{q-s}D}_{b_{i,3}}w \biggr\},
\\[-3pt]
\bar{\varphi}^{\ell_i,\upsilon_i}_{q,t}(j;w,y) & =&  \exp \biggl\{
\underbrace{-\frac{2j}{t-q} \bigl(D^2j-Dy \bigr)}_{a_{i,4}}
\underbrace{{}-\frac{2j}{t-q}D}_{b_{i,4}}w \biggr\}.
\end{eqnarray}

\end{appendix}

\section*{Acknowledgements}
M. Pollock would like to thank Dr. Alexandros Beskos and Prof. Omiros Papaspiliopoulos
for stimulating discussion on aspects of this paper. The authors would
like to
thank the anonymous referees for their helpful comments. This work was
supported by the Engineering and Physical Sciences Research Council. M. Pollock was
supported by EPSRC Grant  EP/P50516X/1;  A.M.~Johansen  was partially
supported by
EPSRC Grant  EP/1017984/1; G.O. Roberts was partially supported by EPSRC
Grants
EP/G026521/1, EP/D002060/1.

% imsref loaded by daiva.urboniene, 2014-11-18 09:33:46

\printhistory
\end{document}